\documentclass[aps,prd,superscriptaddress,preprintnumbers,reprint,nofootinbib,showpacs,floatfix]{revtex4-1}
\pdfoutput=1

\usepackage{graphicx}
\usepackage{natbib} 
\usepackage{epsfig}
\usepackage{amssymb}
\usepackage{ragged2e}
\usepackage{amsmath}
\usepackage{amsfonts}
\usepackage{fancyhdr}
 
\usepackage{amsmath}
\usepackage{anysize}
\usepackage{verbatim}
\usepackage{cancel} 
\usepackage{framed}
\usepackage[titletoc]{appendix}
\usepackage{mathrsfs}
\usepackage{amsbsy}
\usepackage{ulem}
\usepackage{soul}
\usepackage{subcaption}
\usepackage{multirow}
\usepackage{vmargin}
\usepackage{bm}
\usepackage[compatibility=false]{caption}
\usepackage{subcaption}

\setpapersize{A4}
\setmargins{2.5cm}       
{1.5cm}                  
{17cm}                   
{23.42cm}                
{10pt}                   
{1cm}                    
{0pt}                    
{2cm}                    

\def\lQ{\Lambda_{\rm QCD}}

\newcommand{\be}{\begin{equation}}
\newcommand{\ee}{\end{equation}}
\newcommand{\bea}{\begin{eqnarray}}
\newcommand{\eea}{\end{eqnarray}}
\newcommand{\nn}{\nonumber}
\newcommand{\ba}{\begin{align}}
\newcommand{\ea}{\end{align}}
\newcommand{\stkout}[1]{\ifmmode\text{\sout{\ensuremath{#1}}}\else\sout{#1}\fi}

\begin{document}

\title{Pion transitions in the Born-Oppenheimer Effective Field Theory: a long distance approach}
\author{Joan Soto} \author{Sandra Tom\`as Valls}
\affiliation{Departament de F\'\i sica Qu\`antica i Astrof\'\i sica and Institut de Ci\`encies del Cosmos, 
Universitat de Barcelona, Mart\'\i $\;$ i Franqu\`es 1, 08028 Barcelona, Catalonia, Spain}

\date{\today}

\preprint{}

\begin{abstract}
We address pion transitions involving heavy quarkonium and heavy exotic states in the Born-Oppenheimer effective field theory. Many of these states have a size similar or larger than the typical hadronic scale, and hence the usual QCD multipole expansion breaks down. Unknown  low energy functions must be introduced, which at short distances must reproduce the known results of the multipole expansion. In order to determine the long distance behavior of these functions, we propose an interaction Lagrangian for pions and the QCD string. By matching the amplitudes obtained with this Lagrangian to the ones of the Born-Oppenheimer effective field theory in the static limit, we obtain the low energy functions in terms of three universal parameters. As a by product, we also obtain the light quark mass dependence of the string tension. Assuming that the low energy functions are long-distance dominated, we calculate several quarkonium-to-quarkonium and hybrid-to-quarkonium transitions. We estimate the universal constants and provide a phenomenological analysis of the most relevant transitions.
\end{abstract}

\maketitle

\section{Introduction}\label{sec:intr}

Exotic hadrons (those beyond mesons and baryons) have been a matter of research since the early days of QCD \cite{Jaffe:1975fd}.
Glueballs, hybrids, tetraquarks, and pentaquarks, among others, have been extensively discussed in the literature (see \cite{Lebed:2016hpi,Brambilla:2019esw,Chen:2022asf} for recent reviews). If they have an exotic flavor structure or an exotic $J^{PC}$, they provide a clear experimental signature. However, many of them may have flavor or $J^{PC}$ quantum numbers like  conventional hadrons. In that case, they may mix with ordinary mesons or baryons and the exotic nature may become difficult to uncover. If they contain hidden heavy flavor, charm or bottom, the mixing with hidden light flavor is suppressed in QCD, and the large charm and bottom mass makes the case much simpler. This is how the first tetraquark $Z_c(3900)$ \cite{BESIII:2013ris,Belle:2013yex} and pentaquark \cite{LHCb:2015yax} were unambiguously identified.
They have masses in the charmonium range but isospin $1$ and baryon number $1$ respectively. Even when there are no exotic flavor or $J^{PC}$ quantum numbers, exotic charmonium or bottomonium may be identified by comparing with quark model predictions, which are supposed to be reliable for these systems. This is the case of the celebrated $X(3872)$ \cite{Belle:2003nnu,CDF:2003cab}. We shall  consider hidden heavy flavor isospin zero mesons, both ordinary and exotic, and focus specifically on the hybrid sector.

An economical approach to address hadrons with two heavy quarks, both ordinary and exotic, is the so-called Born-Oppenheimer effective field theory (BOEFT) \cite{Braaten:2014qka,Berwein:2015vca,Oncala:2017hop,Soto:2017one,Brambilla:2018pyn,Soto:2020xpm}. It exploits the fact that heavy quarks in heavy hadrons move slowly. 
The effect of the light degrees of freedom (LDF), gluons and light quarks, is encoded in a series of potentials organized in a $1/m_Q$ expansion, $m_Q$ being the heavy quark mass. The number of independent potentials at each order depends on the $J^{PC}$ of the LDF, and on its decomposition into irreducible representations of $D_{\infty h}$, the symmetry group of a diatomic molecule. 
The leading potentials ($\mathcal{O}(1/m_Q^0)$) are heavy quark spin and heavy quark mass independent. In the case of hybrids, it has been used to calculate the spin average spectrum \cite{Juge:1999ie,Braaten:2014qka,Berwein:2015vca,Oncala:2017hop}, decays to heavy quarkonium \cite{Braaten:2014ita,Oncala:2017hop,TarrusCastella:2021pld,Brambilla:2022hhi}, and transitions between heavy quarkonium states \cite{Pineda:2019mhw}. The mixing of heavy quarkonium hybrids with heavy quarkonium starts 
at order $1/m_Q$ \cite{Oncala:2017hop}. Heavy-quark spin effects also start at order $1/m_Q$ \cite{Oncala:2017hop,Soto:2017one}, and, in the case of hybrids, have been studied in Refs. \cite{Brambilla:2018pyn,Brambilla:2019jfi,Soto:2023lbh}. We shall address pion transitions within this framework.

\begin{figure*}
    \begin{center}
    \includegraphics[]{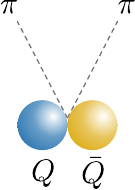}
    \hspace{2cm}
    \includegraphics[]{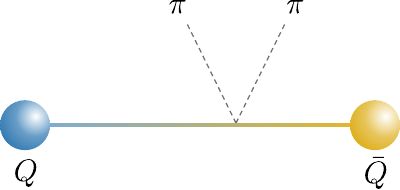}
    \end{center}
    \caption{Multipole expansion emission (left) versus string emission (right)}
\label{multistring}
\end{figure*}

The motivation of this work is twofold. First, to formulate pion transitions within the BOEFT framework, both for ordinary and exotic quarkonium.  And second, to estimate the long distance behavior of the unkown low energy functions (LEF) that arise in this approach.  This is particularly relevant because many exotic states, such as heavy quarkonium hybrids and a number of quarkonium states, are expected to be dominated by long-distance physics.
Traditionally, dipion transitions in heavy quarkonium are usually calculated through a two-step procedure. First, the multipole expansion is used \cite{Gottfried:1977gp}. This is expanding soft gluon operators about the heavy quarkonium center of mass. Next, the soft gluon operators are hadronized in terms of pions. For this approach to be valid, the typical size of the bound states $r$ must be smaller than the typical hadronic scale $1/\lQ$ and smaller than the time scale of the transition $1/\Delta E$, where $\Delta E$ is the energy difference between the initial and final quarkonium states. Often an additional expansion is used, the so-called twist expansion \cite{Voloshin:1978hc}, which localizes soft gluon operators at different times, at the same time. 
This approximation assumes that the typical binding energies are larger than $\lQ$ and $\Delta E$, which is difficult to justify in practice \cite{Pineda:2019mhw}. For quarkonium states with large principal quantum number, even the multipole expansion may fail. For instance, typical values of $1/r$ for $\psi (nS)$ ($\Upsilon (nS)$) are $520$ ($1030$) MeV, $260$ ($430$) MeV, $190$ ($300$) MeV and $150$ ($230$) MeV for $n=1,2,3,4$ \cite{Oncala:2017hop}. Except for the $1S$ states, the remaining figures for $1/r$ are of the order or smaller than $\lQ$. It is then worth exploring the consequences of giving up the multipole expansion. In order to do so, we focus on the opposite limit namely $r\lQ \gg 1$. 
We illustrate this limit and the one corresponding to the multipole expansion in Fig.~\ref{multistring}.
In this regime, the QCD effective string theory (EST) can be used \cite{Luscher:2004ib}. We propose an interaction Lagrangian of the QCD string with pions, which respects both the symmetries of the EST and chiral symmetry.
With this Lagrangian, we then calculate the quark mass dependence of the string tension, elastic pion scattering off the string, and the decay of string excitations into pion pairs. We put forward interaction Lagrangians for quarkonium-to-quarkonium and hybrid-to-quarkonium dipion transitions, which reproduce the string result in the static limit, and obtain from them the long distance behavior of the BOEFT LEF. They are given in terms of three universal constants.

Finally, we carry out a phenomenological analysis. We estimate the universal constants from suitable quarkonium transitions, and once we have obtained them, we use them to calculate dipion invariant mass spectrum of several quarkonium and hybrid-to-quarkonium transitions.

The paper is organized as follows.  Section~\ref{sec:BOEFT} shows the pion interactions in the BOEFT framework. Sec.~\ref{sec:Lpi} provides the interaction between the pions and the QCD string. We obtain the quark mass dependence of the string tension, pion scattering off the string, and the EST expression for the decay of string excitations through pion emission. In Sec.~\ref{sec:ff} we go over the matching of the BOEFT LEF at long distances. Sections~\ref{sec:qtq} and \ref{sec:htq} show the dipion transitions between highly excited quarkonium states, and between exotics with $1^{+-}$ LDF and quarkonium states respectively. Phenomenological applications are worked out for quarkonium and hybrids with $1^{+-}$ LDF in Section~\ref{sec:pheno}. We  discuss our results in Section~\ref{sec:con} and close with a summary and conclusions in Section~\ref{sec:sum}. Several appendices contain technical details and further elaborations.

\section{Pion interactions in the BOEFT} \label{sec:BOEFT}

In the framework of Born-Oppenheimer effective field theory \cite{Oncala:2017hop,Brambilla:2017uyf,Soto:2020xpm,Berwein:2024ztx}, the interaction of pions with heavy quarkonium at leading order (LO) in $1/m_Q$ and the chiral expansion is given by,
\bea
    {\rm L}_{\rm int}&=&\int d^3 \mathbf{R}\int d^3 \mathbf{r}\,\mathrm{Tr}\left[S^\dagger(\mathbf{R},\mathbf{r},t)\left( g_0(r)\partial_0 U^\dagger\partial^0 U\right.\right.\nn\\
    &&\left.\left. + \,  g_1(r)\partial_i U^\dagger\partial^i U+g_2(r){ \hat{r}}^i{ \hat{r}}^j\partial_i U^\dagger\partial_j U\right.\right.\nn\\
    && \left.\left. + \, g_3(r)\left(U^\dagger\mathcal{M}+\mathcal{M}^\dagger U\right)\right)S(\mathbf{R},\mathbf{r},t)\right]
\label{boeftq}
\eea
where $S(\mathbf{R},\mathbf{r},t) =(S_0(\mathbf{R},\mathbf{r},t)\mathbb{I}_2 +  S_1^j(\mathbf{R},\mathbf{r},t)\sigma^j)/\sqrt{2}$, $S_0$ ($S_1^j$) 
is the quarkonium wave function field for the spin 0 (spin 1) 
heavy quarkonium.  $U=U(\mathbf{R},t)=\exp{\left(\vec \pi(\mathbf{R},t)\vec \tau/f_\pi\right)}$, $\pi(\mathbf{R},t)$ are the pion fields, $\vec\tau$ the isospin Pauli matrices, $f_\pi$ the pion decay constant, $\mathcal{M}=m_q \mathbb{I}$, $m_q=(m_u+m_d)/2$, the average of the up quark and down quark masses  ($m_q\simeq m_\pi^2/2B_0$, $B_0\sim \lQ$). $g_s(r)$, $s=0,1,2,3$ are unknown LEF . The trace must be understood to act independently on both spin-symmetry multiplets of quarkonium and on the isospin space of chiral operators. This Lagrangian implicitly assumes that the energies involved in the processes, $\Delta E$, fulfill $\Delta E \ll 1/r \,, \lQ$. The short distance behavior of $g_s(r)$ can be obtained using the QCD multipole expansion \cite{Gottfried:1977gp}. 
It delivers $g_s(r)\sim r^2$. The long distance behavior of $g_s(r)$ is unknown. We are going to estimate it in Sec.~\ref{sec:ff} using the EST results in the following section.

Analogous Lagrangians can be written down when one of the fields represents a heavy quarkonium exotic, namely a quarkonium with non-trivial LDF quantum numbers, rather than a heavy quarkonium ($0^{++}$ LDF). 
We display in Appendix~\ref{ap:op} the lowest order chiral operators relevant for any exotic with LDF of $J\leq 1$.
Let us present here the case with LDF $1^{+-}$  at LO in the $1/m_Q$ and in the chiral expansion 
\bea
\label{boefth}
    {\rm L}_{\rm int}&=&\int d^3 \mathbf{R}\int d^3 \mathbf{r}\,\mathrm{Tr}\left[S^\dagger(\mathbf{R},\mathbf{r},t) \right.\\ 
    &\times &\left.\left(\epsilon_{ijk}\hat{r}^j g_4(\mathbf{r})\partial_0 U^\dagger\partial^k U\right)H^i(\mathbf{R},\mathbf{r},t)\right]\nn
\eea
where $\mathbf{H}(\mathbf{R},\mathbf{r},t)=(\mathbf{H}_0(\mathbf{R},\mathbf{r},t)\mathbb{I}_2+ $\nolinebreak $ \mathbf{H}_1^j(\mathbf{R},\mathbf{r},t)\sigma^j)/\sqrt{2}$, $\mathbf{H}_0$ ($\mathbf{H}_1^j$)
is the wave function field for the spin 0 (spin 1) 
heavy quarkonium
hybrid. The trace must be understood to act independently on both spin-symmetry multiplets of quarkonium and on the chiral operators. 
At short distance, $g_4(r)\sim r$ \cite{TarrusCastella:2021pld}. The  unknown long distance behavior will be obtained in Sec.~\ref{sec:ff}.

\subsection{Pion scattering off static quarkonium}

We compute $\langle \pi(q); S(\mathbf{r}',\mathbf{R}') | S_\text{int} | \pi(p); S(\mathbf{r},\mathbf{0})\rangle$ using the general form of \eqref{boeftq} in the static limit. This results will be used in Sec.~\ref{sec:ff} in order to obtain the matching with EST, hence, we show the result for the special case $r \parallel z$. We obtain,
\bea
\label{eq:BOEFTq}
    &&\langle \pi(\vec q)  ; S(\mathbf{r}',\mathbf{R}') | S_{\rm int} | \pi (\vec p); S(\mathbf{r},\mathbf{0}) \rangle \\
    &&=\delta\left(\mathbf{r}'-\mathbf{r}\right)\delta (\mathbf{R}') \delta (E_q - E_p )\, \frac{8\pi}{f_\pi^2}\, \Big[ g_0(r)E_qE_p \nn \\
    &&  +\, g_1(r)q_ip^i + g_2(r) q_3 p_3 + g_3(r) \frac{m_\pi^2}{2B_0}\Big] \nn  \, ,
\eea
where $E_p=\sqrt{\vec{p}^2+m_\pi^2}$ and $E_q=\sqrt{\vec{q}^2+m_\pi^2}$.

\subsection{Static exotics to static quarkonium pion transitions}

By following the same procedure with the Lagrangian \eqref{boefth} for exotics with $1^{+-}$ LDF we obtain,
\bea\label{eq:BOEFTh}
    &&\langle \pi(\vec q) \pi (\vec p) ; S(\mathbf{r}',\mathbf{R}') | S_{\rm int} | 0_\pi; H^*(\mathbf{r},\mathbf{0}) \rangle \\
    &&=\langle \pi(q) \pi (p) ; S(\mathbf{r}',\mathbf{R}') | S_{\rm int} | 0_\pi; H(\mathbf{r},\mathbf{0}) \rangle^* \nn \\
    &&= \delta\left(\mathbf{r}'-\mathbf{r}\right) \delta (\mathbf{R}')  \delta(E_q+E_{p}+\Delta E) \bigg(-\frac{4\pi}{f_\pi^2}\bigg) \nn \\
     && \times \, i  g_4(r) (E_q \mathrm{p} + E_p \mathrm{q})   \nn \, ,    
\eea 
where $\Delta E<0$ is the static energy of the quarkonium minus the static energy of the exotic. 
$H(\mathbf{r},\mathbf{R})$ and $H^\ast(\mathbf{r},\mathbf{R})$ are defined by the previously introduced $H^1(\mathbf{r},\mathbf{R})= ( H^\ast(\mathbf{r},\mathbf{R}) - H(\mathbf{r},\mathbf{R}) )/\sqrt{2} $ and $H^2(\mathbf{r},\mathbf{R})= i( H^\ast(\mathbf{r},\mathbf{R}) + H(\mathbf{r},\mathbf{R}) )/\sqrt{2} $.

\section{The interaction of pions with the QCD string}\label{sec:Lpi}

The effective QCD string theory provides an accurate description of the static potential at long distances ($r\lQ\gg1$), both for quarkonium and hybrids. This is also the case for a number of $1/m_Q$ suppressed quarkonium potentials. The Nambu-Goto (NG) action provides the leading terms of the EST 
\be
    \label{eq:S_ng}
    S_{NG}=-\sigma \int d^2\xi \sqrt{-\text{det}(\partial_a x^\mu \partial_b x_\mu}) \,,
\ee
where $\sigma\sim \lQ^2$ is the string tension. The energy spectrum in EST, which corresponds to static quark-antiquark energies, can be organized in powers of $1/(r\lQ)$, the leading term being the linear confining potential $\sigma r$ \cite{Luscher:2002qv,Luscher:2004ib}. Bulk corrections to the NG action start at order $1/(r^7\lQ^6)$ \cite{Aharony:2013ipa}, but boundary terms have an impact earlier on, at order $1/(r^4\lQ^3)$ \cite{Billo:2012da,Brandt:2017yzw}. The NG action enjoys reparameterization invariance and Poincar\'e invariance. We choose the frame $\xi^1=x^0=t$, $t\in \mathbb{R}$, $\xi^2=x^3=z$, $z\in \left[-r/2,r/2\right]$. With this choice we can conveniently denote the directions orthogonal to $z$ as $x^\mathsf{i}(t,z)$ with $\mathsf{i}=1,2$, and assume $x^\mathsf{i}\sim 1/\lQ$ and $\partial_t\sim \partial_z \sim 1/r$. For $r\lQ\gg 1$, we then have, 

\bea
    \label{eq:S_g}
    S_{NG}
	&=&-\sigma \int dtdz \left[1-\frac{1}{2}\partial_0 x^\mathsf{i} \partial_0 x^\mathsf{i}  + \frac{1}{2}\partial_3 x^\mathsf{i}  \partial_3  x^\mathsf{i}  + \cdots\right]\nn\\ 
    &=&-\sigma \int dtdz \left[1-\partial_0\varphi^\ast \partial_0 \varphi + \partial_3\varphi^\ast \partial_3 \varphi + \cdots\right]  {\rm ,}
\eea	
where $\varphi=\varphi(z,t)=\left(x^1(z,t)+ix^2(z,t)\right)/\sqrt{2}$ can be written in terms of creation and annihilation operators
\bea
    \label{eq:psi}
    &&\varphi(z,t)=\sum_{n=1}^\infty \frac{1}{2E_n}\left(e^{-iE_nt}\varphi_n(z)\alpha_n + e^{iE_nt}\varphi^*_n(z)\beta_n^\dagger\right) \nn \\
    &&\varphi_n(z)=\frac{1}{\sqrt{2r}}\left(e^{iE_nz}+(-1)^{n+1}e^{-iE_nz}\right) \\
    &&[\alpha_n,\alpha_m^\dagger]=[\beta_n,\beta_m^\dagger]=\frac{2E_n}{\sigma}\delta_{nm} \hspace{0,3cm}, \hspace{0,3cm} E_n=\frac{\pi n}{r} \nn \, .
\eea
The remaining commutators vanish. $\alpha_n^\dagger$ ($\beta_n^\dagger$) on the vacuum creates a state of energy $E_n$, angular momentum $1$ ($-1$) and parity $(-1)^n$. The reflection with respect to the xz plane interchanges $\alpha_n \leftrightarrow\beta_n$.

At low energies
($p\sim m_\pi \ll \lQ$), pion physics is described by the Chiral Lagrangian,
\be
    \mathcal{L}_{\chi}^{LO}=\frac{f_\pi^2}{4} \text{Tr}(\partial_\mu U^\dagger\partial^\mu U)+\frac{f_\pi^2m_\pi^2}{4 m_q} \text{Tr}(U^\dagger\mathcal{M}+\mathcal{M}^\dagger U)
\label{pions}
\ee
which is also Poincar\'e invariant, and approximately invariant under $SU_L(2)\otimes SU_R(2)$ chiral symmetry. The latter symmetry is spontaneously broken by the ground state to the diagonal $SU(2)$ (isospin)  and explicitly broken by the light quark masses, $\mathcal{M}=m_q \mathbb{I}$, $m_q=(m_u+m_d)/2$, which also break isospin at higher orders.

The guidelines for building the interaction of pions with the string in an EFT framework are symmetries and power counting. We shall write down a local effective Lagrangian that respects the symmetries of both  \eqref{eq:S_ng} and \eqref{pions}. For the power counting we are going to assume $1/r$, $p$, $m_\pi\, \ll\, \lQ$, which holds both for the EST and the Chiral Lagrangian. This means an expansion in $\partial x^\mathsf{i}(\xi)/\partial \xi^a \sim 1/r\lQ$ ($a=0,3$), $\partial_\mu/4\pi f_\pi \sim p/\lQ$, $ {\cal M}/4\pi f_\pi \sim m_\pi/\lQ$. Note that in the manifestly Lorentz and reparameterization invariant formulation of the Nambu-Goto string \eqref{eq:S_ng}, the EFT counting is not explicit. 
It contains arbitrary higher orders in the $1/r\lQ$ expansion. This may be remedied by sticking to the physical degrees of freedom ($x^\mathsf{i}(\xi)$), for which the string symmetry group reduces to $SO(1,1)\otimes O(2)$, and implementing the remaining $SO(3,1)$ Lorentz symmetry by a non-linear representation \cite{Isham:1971dv,Volkov:1973vd,Dubovsky:2012sh}. Since we are only aiming at a LO calculation, rather than implementing a non-linear realization of $SO(3,1)$, we will just write down the terms that are invariant under $SO(1,1)\otimes O(2)$ and chiral symmetry and then find a manifestly Lorentz and reparameterization invariant formulation of them.
Since the QCD string does not transform under chiral symmetry, the lowest dimensional local operator we may build in the physical frame reads

\bea
    \mathcal{L}_{\text{ChS}}
    &=&(\lambda +\eta)\text{Tr}(\partial_0 U^\dagger\partial^0 U)+(\lambda+\eta) \text{Tr}(\partial_3 U^\dagger\partial^3 U)\nn\\
    && + \, \lambda \text{Tr}(\partial_\mathsf{i} U^\dagger\partial^\mathsf{i} U)+\lambda' \text{Tr}(U^\dagger\mathcal{M}+\mathcal{M}^\dagger U)
    \label{eq:L_Chs}
\eea
\be
    U=U(t,x^\mathsf{i}(t,z),z) \;,\; \mathsf{i}=1,2\;,\; z=x^3, t=x^0 \label{physicalframe}
\ee

This Lagrangian density is to be integrated over $z\in [-r/2,r/2]$ and $t$ in order to get the LO action. Note that the parameter $\eta$ breaks the linear $SO(3,1)$ symmetry. The terms that respect this symmetry can be obtained from the following manifestly Lorentz and reparameterization invariant
action,
\bea
\label{pionstring}
    S_{\rm int}&=&\int d^2 \xi \sqrt{-\text{det}(\partial_a x^\mu \partial_b x_\mu)}\mathcal{L}_{\text{Ch}}(x(\xi)) \\
    \mathcal{L}_{\text{Ch}}^{LO}&=&\lambda \text{Tr}(\partial_\mu U^\dagger\partial^\mu U)+\lambda' \text{Tr}(U^\dagger\mathcal{M}+\mathcal{M}^\dagger U) \nn \\ 
    \mathcal{L}_{\text{Ch}}^{NLO}&=&\lambda'''\text{Tr}(\mathcal{M}^\dagger\mathcal{M})+\lambda''''\text{Tr}(U^\dagger\mathcal{M}U^\dagger\mathcal{M}+{\rm h.c.})
    \,. \nn
\eea
In the NLO Lagrangian we only display the terms that will be eventually needed for renormalization, see \cite{Gasser:1983yg} for the complete set. In order to obtain the Lorentz invariant terms in \eqref{eq:L_Chs}, we only need the LO Lagrangian above. The terms in \eqref{eq:L_Chs} that break 
the linear $SO(3,1)$ symmetry, can be obtained from the following manifestly Lorentz and reparameterization invariant action,
\bea
\label{delta_pionstring}
    \delta S_{\rm int}&=&\eta\int d^2 \xi \sqrt{-\text{det}(\partial_a x^\mu \partial_b x_\mu)}\partial_a x^\mu \partial^a x^\nu \nn\\
    &\times& \text{Tr}\left(\partial_\mu U^\dagger \partial_\nu U\right) \,  .
\eea

The Fock space on which the Lagrangian above acts on is the tensor product of the string Fock space and the Chiral Lagrangian one. The string coordinates $x^\mathsf{i}(\xi)$ act on the former, but the pion fields act on both. On the latter, through the pion creation and annihilation operators in them, and on the former, through the space dependence on $x^\mathsf{i}(\xi)$. Indeed, consider for instance the charged pion field,
\be
    \pi^+(x)=\int \frac{d^3\Vec{p}}{(2\pi)^3\sqrt{2E_p}}\left(e^{-ipx}a(\Vec{p})+e^{ipx}b^\dagger(\Vec{p})\right)\,,
\ee
$p^0=E_p$, $a(\Vec{p})$ ($b(\Vec{p})$) is the annihilation operators of positively (negatively) charged pions. With $x=x(\xi)$, 
\be
    e^{-ipx}=e^{-ip^0t+ip^3z}\left(1+i(p^\mathsf{i} x^\mathsf{i}(t,z))+\cdots \right)\,.
\label{expexp}
\ee
The expansion is justified since $x^\mathsf{i}(t,z)\sim 1/\lQ$ and $p^\mathsf{i} \ll \lQ$.  It implies that, unlike the usual Chiral Lagrangian, the Lagrangian \eqref{eq:L_Chs} contains arbitrary higher orders in the $p/\lQ$ expansion. Recall finally that $x^\mathsf{i}(t,z)$ with $\mathsf{i}=1,2$, contains creation and annihilation operators of the string excitations,
\bea
    p^\mathsf{i} x^\mathsf{i}(t,z)&=&\mathrm{p}^\ast \varphi(z,t)+ \mathrm{p} \varphi(z,t)^\ast\nn\\
    \mathrm{p}&=&\frac{1}{\sqrt{2}}\left(p^1+ip^2\right) \, .
\label{notation}
\eea

\subsection{The quark mass dependence of the string tension}\label{sec:sm}

As the first application of the Lagrangians \eqref{pionstring} and \eqref{delta_pionstring}, note that in the subspace of zero pions, the Lagrangians \eqref{pionstring} and \eqref{delta_pionstring} reduce to the usual Nambu-Goto action, and hence they redefine the string tension. This redefinition introduces a light quark mass dependence through ${\cal M}$ in \eqref{pionstring} and the pion mass. Indeed,
\be
    \left<\left. 0\right|\right. \mathcal{L}_{\text{ChS}}^{}(x(\xi))\left.\left| 0\right.\right>=\left<\left. 0\right|\right. \mathcal{L}_{\text{ChS}}^{}(0)\left.\left| 0\right.\right>
\ee
\bea
    &&\left<\left. 0\right|\right.\text{Tr}\left(\partial_\mu U^\dagger \partial_\nu U\right)(x(\xi))\left.\left| 0\right.\right>=\\
    &&\frac{g_{\mu\nu}}{4} \left<\left. 0\right|\right.\text{Tr}\left(\partial_\rho U^\dagger \partial^\rho U\right)(0)\left.\left| 0\right.\right>\nn
\eea
At leading order in the light quark mass, there is a tree-level linear contribution from the second term in the LO Lagrangian \eqref{pionstring}. At NLO, there is a one-loop contribution from the first and second terms of the LO Lagrangian and a tree-level contribution from the NLO Lagrangian. The final result reads,
\bea
    \sigma \to \sigma &-&\Bigg(
    4\lambda' m_q + \left[\frac{3B_0}{4\pi^2f_\pi^2}\left(2B_0\left(\lambda+\frac{\eta}{2}\right)-\lambda'\right) \right.
    \nn \\ 
    &\times&
    \left.
    \left(\ln{\frac{2B_0m_q}{\mu^2}} -1\right)
    +2\lambda''\right]m_q^2 
    \Bigg)\,,
\label{sigmam}
\eea
$\lambda''=\lambda'''+2\lambda''''-3B_0\eta/(4\pi f_\pi)^2$ is necessary to absorb the UV divergences of the one-loop calculation. Dimensional regularization in the $\overline{\rm MS}$ scheme has been used. We have also used $ m_\pi^2=2 m_q B_0$ ($B_0\sim \lQ$) in the higher order terms. This result is expected to be useful to understand the differences in the heavy quarkonium spectrum observed in lattice calculations at different pion masses (see for instance  \cite{HadronSpectrum:2012gic} and \cite{Cheung:2016bym} for charmonium), which are more pronounced for hybrids and highly excited quarkonium states. It should also be useful to understand the light quark mass dependence of the string tension recently observed in \cite{Bulava:2024jpj}.

\subsection{Pion scattering off the string}\label{sec:ps}

The simplest process involving pions is the elastic scattering of a pion off any string state.  
The LO contribution arises at tree level, with $\sqrt{-\text{det}(\partial_a x^\mu \partial_b x_\mu)}\simeq 1$. We have
\bea
    \langle \pi(\Vec{q}) |  \mathcal{L}_{\rm ChS}(x) |\pi(\Vec{p}) \rangle
    &=& \frac{4}{f_\pi^2}
    \Big[
    (\lambda+\eta)\left(E_qE_p + q_3 p^3\right)
    \nonumber \\
    + \, \lambda q_\mathsf{i} p^\mathsf{i} &-& \lambda' \frac{m_\pi^2}{2B_0}
    \Big] e^{i(q-p)x}\,,
\label{pionel}
\eea
$x=x(\xi)$ and $\mathsf{i}=1,2$. This expression must be sandwiched between the given string states. At LO, the exponential can be approximated by the first term in \eqref{expexp}, which leads to,
\bea
    &&\langle \pi(\Vec{q}) |  S_{\rm int} + \delta S_{\rm int} |\pi(\Vec{p}) \rangle =\frac{16\pi}{f_\pi^2}\Big[ (\lambda+\eta)\left(E_qE_p + q_3 p^3\right)  \nn \\
    && + \, \lambda q_\mathsf{i} p^\mathsf{i}- \lambda' \frac{m_\pi^2}{2B_0}\Big]\frac{\sin{\left[(q^3-p^3)\frac{r}{2}\right]}}{(q^3-p^3)}\delta(E_q-E_p)\, {\rm .}
\label{ael}
\eea
This LO expression does not depend on the particular string state the pion scatters off. Notice the non-trivial interplay between the third component of the pion momentum transfer $q^3-p^3$ and the string length $r$.

\subsection{Decay of string excitations through pion emission}\label{sec:pe}

The string excitations may decay into the string ground state by a two pion (dipion) emission. Let us calculate the amplitude corresponding to the lowest lying excitations $N=1$, $E_N=\pi N/r$ at LO. They consist of two degenerate $\Pi_u$ states, a clockwise (R) and an anticlockwise (L) rotation of $|L_z|=1$.
\be\label{eq:LR}
    |1\Pi_u^R\rangle=\sqrt{\frac{\sigma}{2E_1}}\beta_1^\dagger|0
    \rangle \quad,\quad |1\Pi_u^L\rangle=-\sqrt{\frac{\sigma}{2E_1}}\alpha_1^\dagger|0
    \rangle
\ee

In order to obtain a non-vanishing transition to the string vacuum, we need terms linear in $x^\mathsf{i}$ in the interaction Lagrangian. Since  $\sqrt{-\text{det}(\partial_a x^\mu \partial_b x_\mu)}$ only produces even powers of $x^\mathsf{i}$, the relevant terms from \eqref{pionstring} can only come from first order of the $p/\lQ$ expansion  in the pion fields \eqref{expexp}. However, from \eqref{delta_pionstring}  there is an additional term at zeroth order of the $p/\lQ$ expansion and first order in the $1/r\lQ$ expansion. It reads,
\bea
    \delta \mathcal{L}_{\rm ChS} &=& 2\eta \big[ \partial_0 x^\mathsf{i} \text{Tr}(\partial_0 U^\dagger \partial_\mathsf{i} U) - \\
    && \partial_3 x^\mathsf{i} \text{Tr}(\partial_3 U^\dagger \partial_\mathsf{i} U) \big] \, , \nn
\eea
with $\mathsf{i}=1,2$.
The part of the calculation involving pions for this term reads,
\bea
    \label{eq:pion_crea}
    &&\langle \pi(\Vec{q}),\pi(\Vec{p})|  \delta\mathcal{L}_{\rm ChS}(x) | 0_\pi \rangle = - \frac{4\eta}{f_\pi^2}
    \Big[
    \partial_0 x^\mathsf{i} \left(E_q p_\mathsf{i} + E_p q_\mathsf{i} \right)
    \nonumber \\
    && - \, \partial_3 x^\mathsf{i} \left(q_3 p_\mathsf{i} + p_3 q_\mathsf{i} \right)
    \Big] e^{i(q+p)x}\,,
\eea
$x=x(\xi)$. We have used $\sqrt{-\text{det}(\partial_a x^\mu \partial_b x_\mu)}\simeq 1$, and the exponential in \eqref{eq:pion_crea} must be taken at zeroth order term in the $p/\lQ$ expansion \eqref{expexp}.  The contribution from \eqref{pionstring} and \eqref{delta_pionstring} at first order in the $p/\lQ$ expansion \eqref{expexp} can be obtained from \eqref{eq:L_Chs}. Putting it together with the contributions from \eqref{eq:pion_crea}, we obtain, 
\bea
\label{eq:ael_Piu}
    && \langle \pi(\Vec{q}),\pi(\Vec{p});0_{\text{EST}}| S_{\text{int}}+\delta S_{\text{int}}|0_\pi;1\Pi_u^{R}\rangle \\
    && =  \langle \pi(\Vec{q}),\pi(\Vec{p});0_{\text{EST}}| S_{\text{int}}+\delta S_{\text{int}}|0_\pi;1\Pi_u^{L}\rangle^\ast \nn\\
	&& = \frac{16\pi\sqrt{\pi}}{f_\pi^2\sqrt{\sigma}r}i \bigg\{ \Big[ (\lambda +\eta)\left(E_qE_p + q_3 p^3\right) + \lambda q_{\mathsf{i}} p^\mathsf{i} \nn \\
    && + \,  \lambda' \frac{m_\pi^2}{2B_0}\Big] \left( \mathrm{q} + \mathrm{p} \right) +\eta \Big[
    \frac{\pi}{r}\big( E_q \mathrm{p} + E_p \mathrm{q} \big)\nn \\ 
    && + \,  (q^3+p^3)\big( q^3 \mathrm{p} + p^3 \mathrm{q} \big)
    \Big] \bigg\} \frac{\cos{\left[(q^3+p^3)\frac{r}{2}\right]}}{(q^3+p^3)^2 - \frac{\pi^2}{r^2}}
    \nonumber \\
    &&\times \, \delta\left(E_q+E_p-\frac{\pi}{r}\right) \nn\,{\rm ,}
\eea
where we used the notation introduced in \eqref{notation} for $\mathrm{q}$ and $\mathrm{p}$ and $\mathsf{i}=1,2$. This is the only expression we will need to calculate the relevant  hybrid decays. Although the lower lying hybrid states are a mixture of $\Pi_u$ ($N=1$) and $\Sigma_u^-$ ($N=3$) components, the latter does not contribute to the pion decays at the order we are working. This is discussed in the Appendix~\ref{ap:stringdecay}. 
For illustration purposes, we also display the outcome for the $N=2$ excitations using the action \eqref{pionstring} in this appendix. The calculation for the $1\Pi_g$ state is similar to the one presented above. For the $1\Delta_g$ state, one needs an extra term in the expansion \eqref{expexp} to get the leading contribution, and for the $2\Sigma_g$ state, there is an additional contribution from the second term in the expansion of $\sqrt{-\text{det}(\partial_a x^\mu \partial_b x_\mu)}$, see \eqref{eq:S_g}.

\section{The long distance behavior of the Low Energy Functions}\label{sec:ff}

In order to establish the long distance interactions of pions with heavy quarkonium and heavy quarkonium hybrids, we just write down interaction Lagrangians that reproduce the amplitudes of the pion scattering off the string and of string excitations decays, calculated in Sec.~\ref{sec:ps} and in Sec.~\ref{sec:pe} respectively, in the static limit.

In the case where only quarkonia are involved, it reads
\bea
    {\rm L}_{\rm int}&=&\int d^3 \mathbf{R}\int d^3 \mathbf{r}\,\mathrm{Tr}\left[S^\dagger(\mathbf{R},\mathbf{r},t)S(\mathbf{R},\mathbf{r},t)\right]\nn\\
    &\times&
    \int_{-r/2}^{r/2} dz\, 
    g(r,z) \mathcal{L}{ '}_{\text{ChS}}(t,\mathbf{R}+z\hat{\mathbf{r}} )\,, 
\label{intQ}
\eea
where  $\mathcal{L}{'}_{\text{ChS}}$ is $\mathcal{L}_{\text{ChS}}$ \eqref{eq:L_Chs} in the frame where $\hat{\mathbf{r}}$ corresponds to the z-direction and the origin is at $\mathbf{R}$, namely,
\be
    \mathcal{L}{'}_{\text{ChS}}= \mathcal{L}_{\text{Ch}}^{LO} 
    + \eta \mathrm{Tr} \big(\partial_0 U^\dagger \partial_0 U - \hat{r}^i\hat{r}^j \partial_i U^\dagger \partial_j U\big)
\label{eq:L'_Chs}\,,
\ee
with $\mathcal{L}_{\text{Ch}}^{LO}$ given in \eqref{pionstring}.
Then we have,
\bea
    &&\langle \pi(\vec q)  ; S(\mathbf{r}',\mathbf{R}') | S_{\rm int} | \pi (\vec p); S(\mathbf{r},\mathbf{0}) \rangle \\
    &&=\delta\left(\mathbf{r}'-\mathbf{r}\right)\delta (\mathbf{R}') \delta (E_q - E_p ) \left(\frac{8\pi}{f_\pi^2}\right) \nn \\
    && \times \, \Big[ (\lambda+\eta)\left(E_qE_p + q_3 p^3\right) + \lambda q_\mathsf{i} p^\mathsf{i} + \lambda' \frac{m_\pi^2}{2B_0}\Big] \nn \\
    && \times  \int_{-r/2}^{r/2} dz g(r,z) e^{-i(q^3-p^3)z} \nn \, .
\eea

This expression matches \eqref{ael} if $g(r,z)=1$ up to factors $\delta\left(\mathbf{r}'-\mathbf{r}\right)\delta (\mathbf{R}')$ that appear due to the normalization of the initial and final states. Note that this is not the expected form of the BOEFT \eqref{boeftq}. In fact, it is a more general expression that also holds when $\Delta E \sim 1/r$. For $r \ll 1/\Delta E \sim 1/m_\pi$,  $\mathcal{L}_{\text{ChS}}(t,\mathbf{R}+z\hat{\mathbf{r}} )$ can be expanded around $\mathbf{R}$ and the expected form in the BOEFT \eqref{eq:BOEFTq} is recovered with 
\bea
    g_0(r)&=& \int_{-r/2}^{r/2} dz\, g(r,z)(\lambda+\eta)=r(\lambda+\eta) \nn\\
    g_1(r)&=& \int_{-r/2}^{r/2} dz\, g(r,z)\lambda=r\lambda \\
    g_2(r)&=&\int_{-r/2}^{r/2} dz\, g(r,z){ (- \eta)}=-r\eta\nn\\
    g_3(r)&=& \int_{-r/2}^{r/2} dz\, g(r,z)\lambda'=r\lambda' \nn
\label{eq:g_ibo}\,.
\eea

Hence, the $r^2$ behavior that $g_s(r)$, $s=0,1,2,3$ show at short distances becomes an $r$ behavior at long distances.
The presence of a tensor term at long distances ($g_2(r)=-r\eta$) is due to the addition of \eqref{delta_pionstring} to the string action \eqref{pionstring}.

In the case of a hybrid with $1^{+-}$ LDF in the initial state, it reads,

\bea\label{intH}
    &&{\rm L}_{\rm int}=\int d^3 \mathbf{R}\int d^3 \mathbf{r}\,\mathrm{Tr}\left[S^\dagger(\mathbf{R},\mathbf{r},t)H^i(\mathbf{R},\mathbf{r},t)\right] \nn \\
    &&\times\int_{-r/2}^{r/2} dz\, 
    h(r,z) \epsilon^{ijk}\hat{r}^j\bigg[ \partial_k\mathcal{L}_{\text{ChS}}(t,\mathbf{R}+z\hat{\mathbf{r}} )    \\
    && + \,  2\eta \Big( \frac{i\pi}{r} \mathrm{Tr} \big(\partial_0 U^\dagger \partial_k U \big) + \hat{r}^l \hat{r}^m \partial_m \mathrm{Tr} \big(\partial_l U^\dagger \partial_k U \big)\Big)
    \bigg]\,, \nn
\eea
and we have,
\bea\label{eq:matchh}
    &&\langle \pi(\vec q) \pi (\vec p) ; S(\mathbf{r}',\mathbf{R}') | S_{\rm int} | 0_\pi; H^*(\mathbf{r},\mathbf{0}) \rangle \\
    &&=\langle \pi(q) \pi (p) ; S(\mathbf{r}',\mathbf{R}') | S_{\rm int} | 0_\pi; H(\mathbf{r},\mathbf{0}) \rangle^* \nn \\
    &&= \delta\left(\mathbf{r}'-\mathbf{r}\right) \delta (\mathbf{R}')  \delta(E_q+E_{p}+\Delta E) \bigg(-\frac{ 8 \pi}{f_\pi^2}\bigg) \nn \\
    &&\times\Bigg\{ \Big[
    (\lambda+\eta)\left(E_qE_p + q_3 p^3\right) + \lambda q_\mathsf{i} p^\mathsf{i} + \lambda' \frac{m_\pi^2}{2B_0}
    \Big](\mathrm{q}+\mathrm{p}) \nn\\
    && + \, \eta \Big[
    \frac{\pi}{r} (E_q \mathrm{p} + E_p \mathrm{q})  + (q^3+p^3) (q^3 \mathrm{p} + p^3 \mathrm{q})
    \Big]\Bigg\} \nn \\
    &&\times\int_{-r/2}^{r/2} dz h(r,z)\,  e^{-i(q^3+p^3)z} \nn \, ,     
\eea
$h(r,z)=i\cos(\pi z/r)/\sqrt{\sigma\pi}$ matches the string theory result \eqref{eq:ael_Piu} up to factors $\delta\left(\mathbf{r}'-\mathbf{r}\right)\delta (\mathbf{R}')$ that appear due to the normalization of the initial and final states. Note that $ H^*(\mathbf{r},\mathbf{0})$ and $ H(\mathbf{r},\mathbf{0})$, as defined below \eqref{eq:BOEFTh}, have the same transformation properties under $D_{\infty h}$ as the states $\Pi_u^R$ and $\Pi_u^L$ \eqref{eq:LR} respectively.
For $r \ll 1/\Delta E \sim 1/m_\pi$, the expected form in the BOEFT \eqref{boefth} is recovered. At LO in this expansion, only the term proportional to $1/r$ in \eqref{intH} and \eqref{eq:matchh} survives and leads to
\be
\label{eq:g4}
    g_4(r)=-\int_{-r/2}^{r/2}dz h(r,z)\left(i2\pi\eta/{ r}\right)=4\eta/\sqrt{\sigma \pi} \,
\ee
in \eqref{eq:BOEFTh}.
Note that the $g_4(r)\sim r$ behavior at short distances \cite{TarrusCastella:2021pld} becomes a constant at long distances.

\section{Dipion transitions between highly excited quarkonium states }\label{sec:qtq}

Highly excited states in quarkonium are dominated by long distance physics. Hence, the formalism developed in the previous section is expected to be relevant for dipion transitions between them.
Once $g(r,z)$ is fixed to $1$, we can calculate the amplitude for these transitions from \eqref{intQ},
\bea
\label{eq:M_QQ}
    &&\mathcal{M}(\vec P n l m\to {\vec P}' n' l' m'\,\pi(q) \pi(p))=\int d^3 \mathbf{r}\, \psi_{n'l'm'}^\ast(\mathbf{r})
    \nn \\
    &&\times \, \psi_{nlm}(\mathbf{r})\left( -\frac{8}{f_\pi^2}\right)\Big[ (\lambda+\eta)\left(E_qE_p + q_3 p^3\right) + \lambda q_\mathsf{i} p^\mathsf{i} \nn\\
    &&+ \, \lambda' \frac{m_\pi^2}{2B_0}\Big] \, \frac{\sin{\left[({\bf q}+{\bf p})\frac{\bf r}{2}\right]}}{({\bf q}+{\bf p})\hat {\bf r}}\,,
\eea
$\vec P n l m$ ($\vec P' n' l' m'$) are the heavy quarkonium center of mass three-momentum, principal quantum number, orbital angular momentum and its third component in the initial (final) state, and $\psi_{nlm}(\mathbf{r})$ is the wave function. It is convenient to take 
$\mathbf{k}_+=\mathbf{q}+\mathbf{p}$ in the $z$-direction. Then, we can define 
\bea
\label{eq:krE}
    \mathbf{k}_-&=&\mathbf{q}-\mathbf{p}=k_-\left(\sin\theta_-\cos\phi_-,     \sin\theta_-\sin\phi_-, \cos\theta_-\right) \nn\\
    \hat{\mathbf{r}}&=&\left(\sin\theta_+\cos\phi_+, \sin\theta_+\sin\phi_+, \cos\theta_+\right) \, ,\nn\\
    E_{\pm}&=&E_q\pm E_p\, ,
\eea
with $\theta_-$ the angle between $\mathbf{k}_+$ and $\mathbf{k}_-$, and $\theta_+$ the angle between $\mathbf{k}_+$ and $\hat{\mathbf{r}}$.
In the following,
we take $\mathbf{k}_+\equiv \mathbf{k}$, $\theta_+\equiv \theta$, and $\phi_+\equiv \phi$ for simplification.
$\psi_{nlm}(\mathbf{r})=R_{n}^l(r) Y_l^m(\theta,\phi)$, where $R_{n}^l(r)$ is the radial wave function and $Y_l^m(\theta,\phi)$ the spherical harmonics. The angular integrals can be carried out in terms of special functions,
\bea
    A_{l'm'lm}^\kappa(r,k) &\equiv&\int d\Omega Y_{l'}^{m'\ast}(\theta,\phi) Y_l^m(\theta,\phi)\mathcal{F}_{2\kappa} (\theta,\phi) \nn \\
    &\times&\frac{\sin\left(k\cos\theta\frac{r}{2}\right)}{k\cos\theta} \,,
\label{eq:Alm}
\eea
where $\mathcal{F}_{2\kappa}(\theta,\phi)$ with $\kappa=0,\pm1,\pm2$ are
defined in Appendix~\ref{ap:ff_q} and $A_{l'm'lm}(r,k)$ corresponds to the trivial case with $\mathcal{F}_{00}(\theta,\phi)=1$. For instance,
\be
    A_{0000}(r,k)= \frac{\mathrm{Si}(\frac{k r}{2})}{k}
\ee
where $\mathrm{Si}(z)$ is the sine integral function. The remaining angular integrals that we need are given in the Appendix~\ref{ap:AlmQQ}. $A_{l'm'lm}^\kappa(r,k)=0$ unless $m'=m+\kappa$ is fulfilled. The radial integrals will be carried out numerically.
\be
\label{eq:Rlm}
    R_{n'l'm'nlm}^\kappa(k)\equiv \int dr r^2 R_{n'}^{l'}(r)R_n^l(r) A_{l'm'ml}^\kappa(r,k)\,,
\ee
and we label $R_{n'l'm'nlm}(k)$ the case corresponding to $A_{l'm'ml}(r,k)$.
The amplitude then reads,
\bea
\label{eq:MQQ}
    &&\mathcal{M}(\vec P n l m\to {\vec P}' n' l' m'\,\pi(q) \pi(p))  \\
    &&= \left( -\frac{8}{f_\pi^2}\right)\Bigg\{\left( c_{\pi\pi}m_{\pi\pi}^2 + c_\pi m_\pi^2 + c_E {E_+}^2\right)R \nn\\
    &&- \,c_E\left({E_+}^2 - m_{\pi\pi}^2\right)R^0 + c_E {k_-}^2 \Bigg[-\frac{1}{2}(R+R^0)  \nn \\
    &&+\, \frac{1}{2}(3 R^0 - R)\cos^2\theta_- + \left(e^{-i\phi_-}R^1 + e^{i\phi_-}R^{-1} \right) \nn \\
    &&\times\cos\theta_-\sin\theta_- + \frac{1}{4}\left(e^{-i2\phi_-}R^2 + e^{i2\phi_-}R^{-2} \right) \nn \\
    && \times\sin^2\theta_- \Bigg]\Bigg\}  \nn \,,
\eea
where we have written $R^\kappa$ instead of $R^\kappa_{n'l'm'nlm}(k)$ to simplify the notation, and
\bea
\label{eq:c_param}
    c_{\pi\pi}&\equiv&\frac{1}{2}\left(\lambda+\frac{\eta}{2}\right), \nn\\
    c_\pi &\equiv& -\lambda -\eta +\frac{\lambda'}{2B_0}, \\
     c_E &\equiv& \frac{\eta}{4}  \, .\nn
\eea

The dipion invariant mass spectrum in the zero-recoil limit is defined as,
\bea
\label{eq:dims0}
    \frac{d\Gamma}{dm_{\pi\pi}^2}&=&\int \frac{d^3 \vec{p}}{(2\pi)^3 2E_p}\frac{d^3 \vec{q}}{(2\pi)^3 2E_q}|\mathcal{M}|^2 \\
    && \times \, 2\pi \delta(\Delta E+E_p+E_q) \nn \\
    &&\times\, \delta\left(m_{\pi\pi}^2-(E_p+E_q)^2+(\mathbf{p}+\mathbf{q})^2\right) \,,\nn
\eea
where $\Delta E<0$ is the mass difference between the final and initial quarkonium state ($\Delta E=E_{n'l'm'}-E_{nlm}$). Notice that the Dirac deltas fix $E_+=E_p+E_q=-\Delta E$ and $k^2=k_+^2=(\mathbf{p}+\mathbf{q})^2=\Delta E^2-m_{\pi\pi}^2$.
Upon changing variables to $\mathbf{k}_\pm$, the integrals over $\phi_-$, $\theta_-$, and $k_-=|\mathbf{k}_-|$ can be carried out analytically. Then, after following the procedure in App.~\ref{ap:dw},
the dipion invariant mass spectrum reads,
\begin{widetext}
\begin{equation} 
\label{eq:dipion_q}
    \begin{aligned}
    &\frac{d\Gamma}{dm_{\pi\pi}^2} =
    \frac{
    \sqrt{\Delta E^2 - m_{\pi\pi}^2}
    }{
    {15}\, m_{\pi\pi}^4\,  \pi^{ 3} f_\pi^4
    }
    \sqrt{1 - \frac{4m_\pi^2}{m_{\pi\pi}^2}}
    \Bigg\{ R\, m_{\pi\pi}^2 \big(c_\pi m_{\pi}^2 + c_{\pi\pi} m_{\pi\pi}^2\big) \Bigg[ 30\, R \, m_{\pi\pi}^2 \big(c_\pi m_{\pi}^2 + c_{\pi\pi} m_{\pi\pi}^2\big) +\, 20\, c_E\,\Big[
    4 \big( R+R^0 \big)  \\
    &\times m_\pi^2 m_{\pi\pi}^2  - \big(R-2R^0\big)m_{\pi\pi}^4 + 2 \big(R-R^0\big)(2m_\pi^2 + m_{\pi\pi}^2)\Delta E^2
    \Big] \Bigg]
    +\, c_E^2 \Bigg[
    16\Big[4\big(R+R^0\big)^2 + \big(R^2\big)^2 + \big(R^{-2}\big)^2\Big]m_\pi^4 m_{\pi\pi}^4 
    \\
    &-8 \Big[2\big(2R-3R^0\big)
    \big(R+R^0\big) + \big(R^2\big)^2 + \big(R^{-2}\big)^2\Big]m_\pi^2 m_{\pi\pi}^6 + \Big[ 4\big(R\big)^2 -12 RR^0 
    + 14\big(R^{0}\big)^2 + \big(R^{2}\big)^2 + \big(R^{-2}\big)^2\Big]m_{\pi\pi}^8   \\
    &+4m_{\pi\pi}^2 \bigg( 16 \Big[\big(R\big)^2 - \big(R^{0}\big)^2 
    + \big(R^{1}\big)^2 + \big(R^{-1}\big)^2\Big]m_\pi^4 + 4 \Big[ 3\big(R\big)^2 + 5RR^0 - 2 \big(4\big(R^{0}\big)^2
    + \big(R^{1}\big)^2 + \big(R^{-1}\big)^2 \big) \Big]m_\pi^2 m_{\pi\pi}^2 \\
    & + \Big[ -4\big(R\big)^2 + 10 RR^0 - 6 \big(R^{0}\big)^2 + \big(R^{1}\big)^2 
    + \big(R^{-1}\big)^2\Big]m_{\pi\pi}^4
    \bigg)\Delta E^2 
    +16 \big(R-R^0\big)^2\big(6m_\pi^4 + 2m_\pi^2m_{\pi\pi}^2 + m_{\pi\pi}^4\big)\Delta E^4
    \Bigg]
    \Bigg\} \, .
\end{aligned}
\end{equation}
\end{widetext}

Heavy quark spin symmetry ensures that the results above hold for each third component of the spin, and hence they can be readily applicable to the spin $0$ case. However, for higher spin, the spin averages are usually carried out with the third component of the total angular momentum rather than with the third component of the spin. Hence, for the   spin $1$ case, further elaborations are needed which we discuss in the next section.

\subsection{Spin 1 case}\label{sec:qtq1}

Heavy quark spin symmetry is manifest in \eqref{intQ}. It implies that the same matrix elements appear both for spin 0 and spin 1. If instead of working with $S_0(\mathbf{R},\mathbf{r},t)$ we have $S_1^j(\mathbf{R},\mathbf{r},t)$, with indices contracted between the initial al final states, the only change in the amplitude is a multiplying factor .
The $S=1$ amplitude $\mathcal{M}(\vec P n J M l$ $\to {\vec P}' n'J'M' l' $ $\pi(q) \pi(p))$ with $J=l \otimes S$ ($J'=l' \otimes S$) and $M$($M'$) its third component in the initial (final) state reads
\bea
     &&\mathcal{M}(\vec P n J M l \to {\vec P}' n'J'M' l'\,  \pi(q) \pi(p))   \\
     &&=\sum_{\mu=0,\pm 1} C(l'1J';M'-\mu \mu)C(l1J;M-\mu \mu) \nn \\
     &&\times\, \mathcal{M}(\vec P n l {M-\mu}\to {\vec P}' n' l'{M'-\mu}\, \pi(q) \pi(p)) \nn  \, ,
\eea
where $C(j_1j_2j;m_1,m_2)$ are the Clebsch-Gordan coefficients. In Appendix~\ref{ap:spin1} we present the results for the spin average of the relevant transitions.

\section{Dipion transitions between hybrids and highly excited quarkonium states }\label{sec:htq}

Heavy quarkonium hybrids are dominated by long distance physics. The static potentials have a minimum at the scale  $r\sim 1/\lQ$, and the energy excitations about the minimum are parametrically smaller than $\lQ$, namely $\sim \sqrt{\lQ^3/m_Q}$. Hence, the formalism in the previous section may also be relevant for dipion transitions between hybrids and highly excited quarkonium states.
Once $h(r,z)=i\cos(\pi z/r)/\sqrt{\pi\sigma}$ is fixed, we can calculate the amplitude for the dipion transitions from the interaction Lagrangian \eqref{intH}. For $r \ll 1/|\Delta E|$ only the second last term in \eqref{intH} is relevant, which leads to the expected BOEFT form of \eqref{boefth}. For simplicity, we stick to this case here and relegate the discussion of the full calculation to Appendix~\ref{ap:NLO}.
\bea
\label{eq:M_HQ}
    &&\mathcal{M}(\vec P n J M l \to {\vec P}' n' l' m'\,\pi(q) \pi(p))  \\
    &&=\int d^3\mathbf{r} 
    \psi_{n'l'm'}^\ast(\mathbf{r}) \left( \vec \psi_{nJMlm}(\mathbf{r}) \times \hat r \right)_k \left(\frac{8\pi\sqrt{\pi}\eta}{f_\pi^2 \sqrt{\sigma}r^2}\right) \nn \\
    &&\times \, \frac{\cos{\left[(\mathbf{q}+\mathbf{p})\frac{\mathbf{r}}{2}\right]}}{[(\mathbf{q}+\mathbf{p})\hat{\mathbf{r}}]^2 - \frac{\pi^2}{r^2}}\big( E_q p^k + E_p q^k \big) \, , \nn
\eea
$\vec P n l m$ ($\vec P' n' l' m'$) were defined in Sec.~\ref{sec:qtq},  $J$ denotes the orbital angular momentum plus the angular momentum of the LDF  ($J=l\otimes 1$) of the hybrid with $M$ its third component. $\psi_{n'l'm'}(\mathbf{r})=R_{n'}^{l'}(r)Y_{l'}^{m'}(\theta,\phi)$ is the quarkonium wave function and $\vec \psi_{nJMlm}(\mathbf{r})=R_{nJ}^l(r)\sum_{\mu=0,\pm1} C(l1J,M-\mu \mu)Y_l^{M-\mu}(\theta,\phi)\vec \chi_\mu$ the hybrid wave function, where $R_{nJ}^l(r)$ is the radial wave function, $Y_l^m(\theta,\phi)$ the spherical harmonics with $m=M-\mu$, $C(j_1j_2j;m_1,m_2)$ are the Clebsch-Gordan coefficients and $\vec \chi_\mu$ the LDF angular momentum $1$ eigenvectors. 
The results we obtain below are readily applicable to spin $0$ hybrids. For spin $1$ hybrids, analogous relations to the ones obtained in Appendix~\ref{ap:spin1} for quarkonium should be worked out.  

Using the form factors $\mathcal{F}_{1\kappa}(\theta, \phi)$ with $\kappa=0,\pm 1$ defined in \eqref{eq:ff_h}, the angular integrals read ($\mu=0,\pm1$),
\bea
\label{eq:Alm_h}
    &&A_{l'm'JMl}^{\kappa\mu}(r,k) \equiv\int d\Omega Y_{l'}^{m'\ast}(\theta,\phi) Y_l^{M-\mu}(\theta,\phi)\mathcal{F}_{1\kappa}(\theta, \phi) \nn \\
    &&\times \, C(l1J,M-\mu\mu)\frac{1}{r^2}\frac{\cos\left(k\cos\theta\frac{r}{2}\right)}{(k\cos\theta)^2 - \frac{\pi^2}{r^2}} \, , 
\eea
and are displayed in Appendix~\ref{ap:AlmHQ}. The radial integrals
\be
\label{eq:Rlm_h}
    R_{n'l'm'nJMl}^{\kappa\mu}(k)\equiv \int drr^2R_{n'}^{l'}(r)R_{nJ}^l(r)  A_{l'm'JMl}^{\kappa\mu}(r,k) \, ,
\ee
are calculated numerically. Then the amplitude reads (see App.~\ref{ap:ff_h}),
\bea
\label{eq:Mh}
    &&\mathcal{M}(\vec P n J M l  \to {\vec P}' n' l' m'\,\pi(q) \pi(p)) \\
    &&=\frac{i16\pi\sqrt{\pi}c_E}{f_\pi^2 \sqrt{2\sigma}}\Bigg[\big(E_+k_+ - E_-k_-\cos\theta_-\big) \nn \\
    && \times \, \big(R^{11}+R^{-1-1}\big) + E_-k_-\sin\theta_- \bigg(e^{i\phi_-}R^{01} \nn \\
    && +  \,  e^{-i\phi_-}R^{0-1} 
    +  \frac{1}{\sqrt{2}}e^{i\phi_-}R^{-10}  
    - \frac{1}{\sqrt{2}}e^{-i\phi_-}R^{10} \bigg) \Bigg] \nn \, .
\eea

With this notation, omitting the $n'l'm'nJMl$ subscripts for simplification purposes, the dipion invariant mass spectrum reads,
\bea 
\label{eq:dipion_h}
    &&\frac{d\Gamma}{dm_{\pi\pi}^2} ={ 4 c_E^2}
    \frac{
    \big(\Delta E^2 - m_{\pi\pi}^2\big)^{3/2}}{{15}\,  f_\pi^4\,  \sigma\,  m_{\pi\pi}^4}
    \sqrt{1 - \frac{4m_\pi^2}{m_{\pi\pi}^2}} \\
    &&\times \,
    \bigg[ 8\Delta E^2 (R^{11}+R^{-1-1})^2 \big( 2m_\pi^2m_{\pi\pi}^2 +6m_\pi^4 +m_{\pi\pi}^4 \big)  \nn \\
    &&+\, m_{\pi\pi}^2 (m_{\pi\pi}^2-4m_\pi^2)^2 \Big( 2(R^{01})^2 + 2\sqrt{2}R^{01}R^{-10}  \nn \\
    &&+\,2(R^{0-1})^2 -
    2\sqrt{2}R^{0-1}R^{10} + (R^{10})^2 + (R^{-10})^2 \Big)
    \bigg] \, , \nn
\eea
with $\Delta E<0$ the mass difference between the final and initial state, once recoil corrections are neglected. See App.~\ref{ap:dw} for details.

\section{Phenomenological analysis}\label{sec:pheno}
Eq.~\eqref{eq:dipion_q} and Eq.~\eqref{eq:dipion_h}  describe the dipion invariant mass spectrum for transitions between highly excited quarkonium states, and between hybrids and highly excited quarkonium states, respectively.  These distributions are parameterized by three low-energy constants, $c_{\pi\pi}$, $c_\pi$ and $c_E$. The decay width of the transition can be obtained by integrating over the dipion invariant mass spectrum. Ideally, the extraction of these constants would rely on experimental data from highly excited quarkonium transitions. However, very limited data are available for those transitions. In practice, we restrict ourselves to the transitions with the larger principal quantum numbers available.

\subsection{Quarkonium}\label{sec:qq}

\subsubsection{The low energy constants}

In order to obtain the low energy constants, we have chosen the following three transitions: $\Upsilon(3s) \to \Upsilon(2s)\pi^+\pi^-$, $\Upsilon(4s) \to \Upsilon(2s)\pi^+\pi^-$ and $\Upsilon(10860) \to \Upsilon(3s)\pi^+\pi^-$. We have also assumed that the $\Upsilon(10860)$ corresponds to the $\Upsilon(5s)$ state. The two first transitions are not expected to have sizable contributions from resonances, and hence the experimental decay width can be directly compared to the one in our calculation.
We use the PDG \cite{ParticleDataGroup:2024cfk} average on $\Gamma\left(\Upsilon(3s) \to \Upsilon(2s)\pi^+\pi^-\right)=0.57 \pm 0.09$ keV and on $\Gamma\left(\Upsilon(4s) \to \Upsilon(2s)\pi^+\pi^-\right)=1.7 \pm 0.4$ keV. 
However, the last transition has sizable contributions from resonances, which must be subtracted before comparing to our results. Hence, we use  \cite{Belle:2014vzn} estimations of the non-resonant amplitude to obtain a decay width which we can identify as $\Gamma\left(\Upsilon(5s) \to \Upsilon(3s)\pi^+\pi^-\right)$. 
This is $\Gamma\left(\Upsilon(10860) \to \Upsilon(3s)\pi^+\pi^-\right)|_{\rm non-resonant}=0.23^{+0.18}_{-0.12}$ keV, see Appendix~\ref{ap:non-resonant}. As all our input data is for transitions to two charged pions, our results are also the ones for two charged pions in their final state. However, our calculations do not depend on whether the pions are charged or not. The contribution for neutral pions is plainly half the one for charged pions.

For each of the above transitions we compute the radial integrals $R^\kappa_{n'l'm'nlm}$ \eqref{eq:Rlm} at each dipion mass using the angular integrals in App.~\ref{ap:Alm} and the LO wave functions of Ref.~\cite{Oncala:2017hop} which uses lattice data of \cite{Juge:2002br}.
After numerically integrating \eqref{eq:dipion_q} with respect of $m_{\pi\pi}^2$, we obtain a system of three equations involving $c_{\pi\pi}$, $c_\pi$ and $c_E$. By solving it we get two sets of solutions $\{\,c_\pi,\; c_{\pi\pi},\; c_E\,\}_1$ and $\{\,c_\pi,\; c_{\pi\pi},\; c_E\,\}_2$
\be 
\label{eq:param1}
    \{\,c_\pi,\; c_{\pi\pi},\; c_E\,\}_1 =
    \begin{Bmatrix}
    c_\pi =\pm 0.15985 \\
    c_{\pi\pi} = \pm 0.00275\\
    c_E =\mp 0.02060
    \end{Bmatrix}   \, ,
\ee

\be 
\label{eq:param2}
    \{\,c_\pi,\; c_{\pi\pi},\; c_E\,\}_2 =
    \begin{Bmatrix}
    c_\pi =\pm 0.16062 \\
    c_{\pi\pi} = \mp 0.01166\\
    c_E =\mp 0.00636
    \end{Bmatrix}   \, ,
\ee
with a global sign ambiguity, namely the signs of the three constants are correlated.

\subsubsection{Uncertanties}\label{sec:uncert}

The uncertainties on the parameters $\{\,c_\pi,\; c_{\pi\pi},\; c_E\,\}$
were estimated using a Monte Carlo procedure that propagates the experimental errors of the decay widths used. For each decay width, pseudo–experimental values were generated assuming Gaussian distributions centered at the measured values with widths given by the corresponding experimental errors from \cite{ParticleDataGroup:2024cfk,Belle:2014vzn}. For each pseudo–dataset, the parameters were re-extracted, yielding empirical probability distributions from which mean values and standard deviations were determined, together with the correlation matrices.
Starting from the two independent solutions,
\eqref{eq:param1} and \eqref{eq:param2}, the Monte Carlo analysis yields the following four solutions (quoted at the $1\sigma$ level):
\bea
    \text{From }\eqref{eq:param1}: && 
    \begin{Bmatrix}
        c_\pi = \pm (0.160\pm 0.007)\\
        c_{\pi\pi} = \pm (0.0025 \pm 0.0015)\\
        c_E = \mp (0.0204 \pm 0.0009)
    \end{Bmatrix}
    \, , \\
    \text{From }\eqref{eq:param2}: && 
    \begin{Bmatrix}
        c_\pi = \pm (0.160 \pm 0.007)\\
        c_{\pi\pi} = \mp (0.0114 \pm 0.0011)\\
        c_E = \mp (0.0066 \pm 0.0020)
    \end{Bmatrix}
    \, ,
\eea

For set $1$ ($2$), $c_\pi$ shows a mild anti-correlation (correlation) with $c_{\pi\pi}$ and a very weak (strong) anti-correlation with $c_E$, whereas $c_{\pi\pi}$ and $c_E$ present a mild (very strong) anti-correlation.

The theoretical uncertainties include contributions from the next order in both the chiral and EST expansions. While the relative chiral uncertainty can be reasonably estimated as $\Delta E/(4\pi f_\pi)$, the next-order correction in the string expansion, $\langle 1/r \rangle / (\sigma \langle r \rangle)$, for the transitions between quarkonium states used to determine the parameter sets \eqref{eq:param1}–\eqref{eq:param2}, leads to a very large uncertainty. 
This suggests that short-distance effects remain relevant for the considered transitions, and that an interpolation between the short- and long-distance regimes would likely provide a more reliable description (see \cite{Oncala:2017hop,Soto:2020pfa,Soto:2023lbh} for examples of interpolations).
Therefore, in the statistical analysis of our results, we include only the experimental uncertainties, and consider the theoretical results as order of magnitude estimates. From \eqref{eq:param1} and \eqref{eq:param2} we obtain the corresponding
\be 
    \bigg\{\,\lambda,\; \eta,\; \frac{\lambda'}{2B_0}\,\bigg\}_1 =
    \begin{Bmatrix}
    \lambda =\pm (0.046 \pm 0.004) \\
    \eta = \mp (0.082 \pm 0.004) \\
    \frac{\lambda'}{2B_0} =\pm (0.124 \pm 0.005)
    \end{Bmatrix}   \, ,
\ee

\be 
    \bigg\{\,\lambda,\; \eta,\; \frac{\lambda'}{2B_0}\,\bigg\}_2 =
    \begin{Bmatrix}
    \lambda =\pm (0.010 \pm 0.006) \\
    \eta = \pm (0.026 \pm 0.008) \\
    \frac{\lambda'}{2B_0} =\mp (0.124 \pm 0.005)
    \end{Bmatrix}   \, .
\ee

\subsubsection{Dipion invariant mass spectrum}

With \eqref{eq:dipion_q}, \eqref{eq:param1}, and \eqref{eq:param2} we obtain predictions for the dipion invariant mass spectrum as well as the dipion decay width for some relevant dipion transitions between excited quarkonium states. The allowed transitions by the angular integrals are those between states with initial (final) orbital angular momentum $l$ ($l'$) that fulfill $l'=l+0,\pm 2, \dots$ (see App.~\ref{ap:Alm}). 
We compute those that have data in the PDG \cite{ParticleDataGroup:2024cfk} or are relevant to identify XYZ resonances. For charmonium, we obtain, in particular, the spin average result of the transition $\chi_{c1}(2p) \to \chi_{c1}(1p)\pi^+\pi^-$, which is identified as ${ \chi_{c1}}(3872)\to \chi_{c1}(1p)\pi^+\pi^-$. For bottomonium, we identify the states $\Upsilon_1(3d)$, $\Upsilon(5s)$ and $\Upsilon_1(4d)$ as $\Upsilon(10753)$, $\Upsilon(10860)$ and $\Upsilon(11020)$ respectively, according to Ref.~\cite{Oncala:2025mqj}. 

\begin{figure*}[htbp]
\includegraphics[height=0.25\textheight,width=0.48\textwidth]{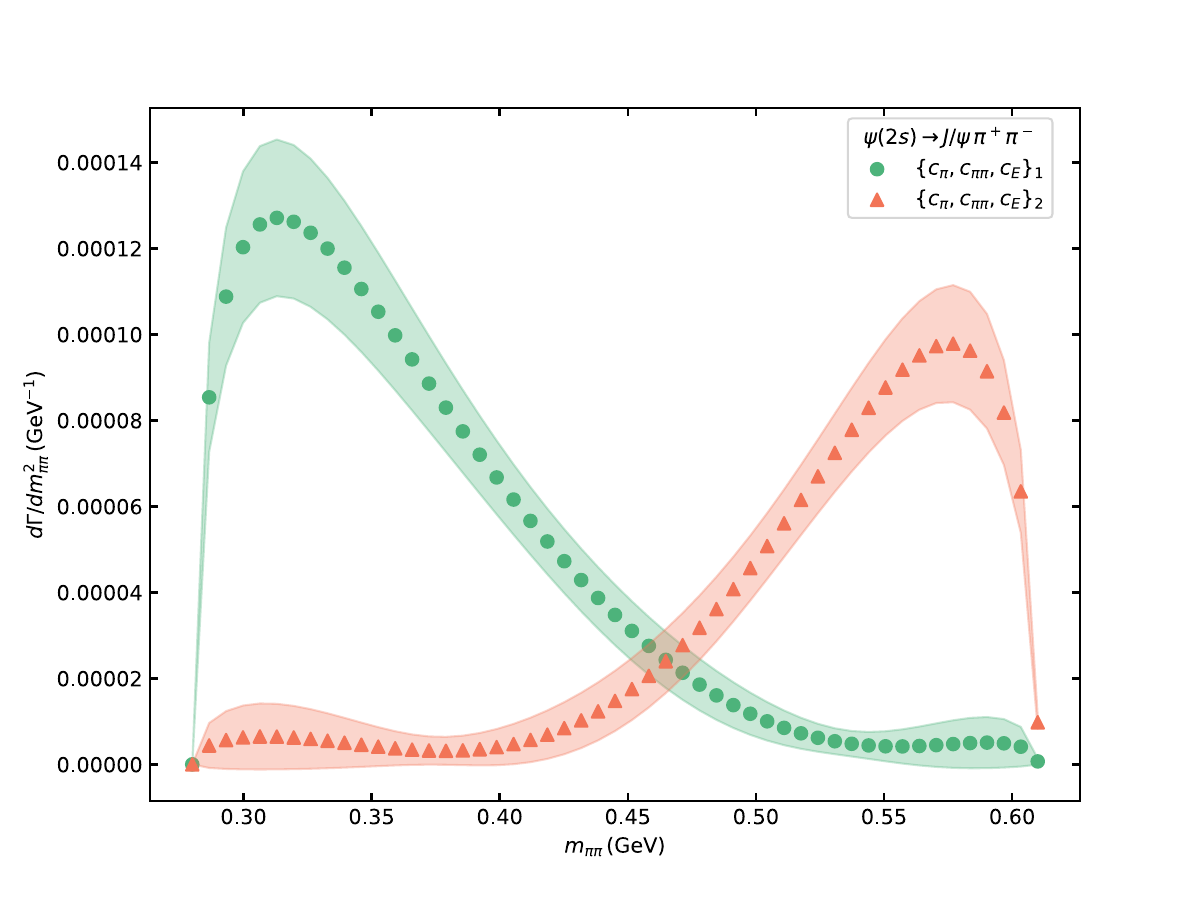} 
\includegraphics[height=0.25\textheight,width=0.48\textwidth]{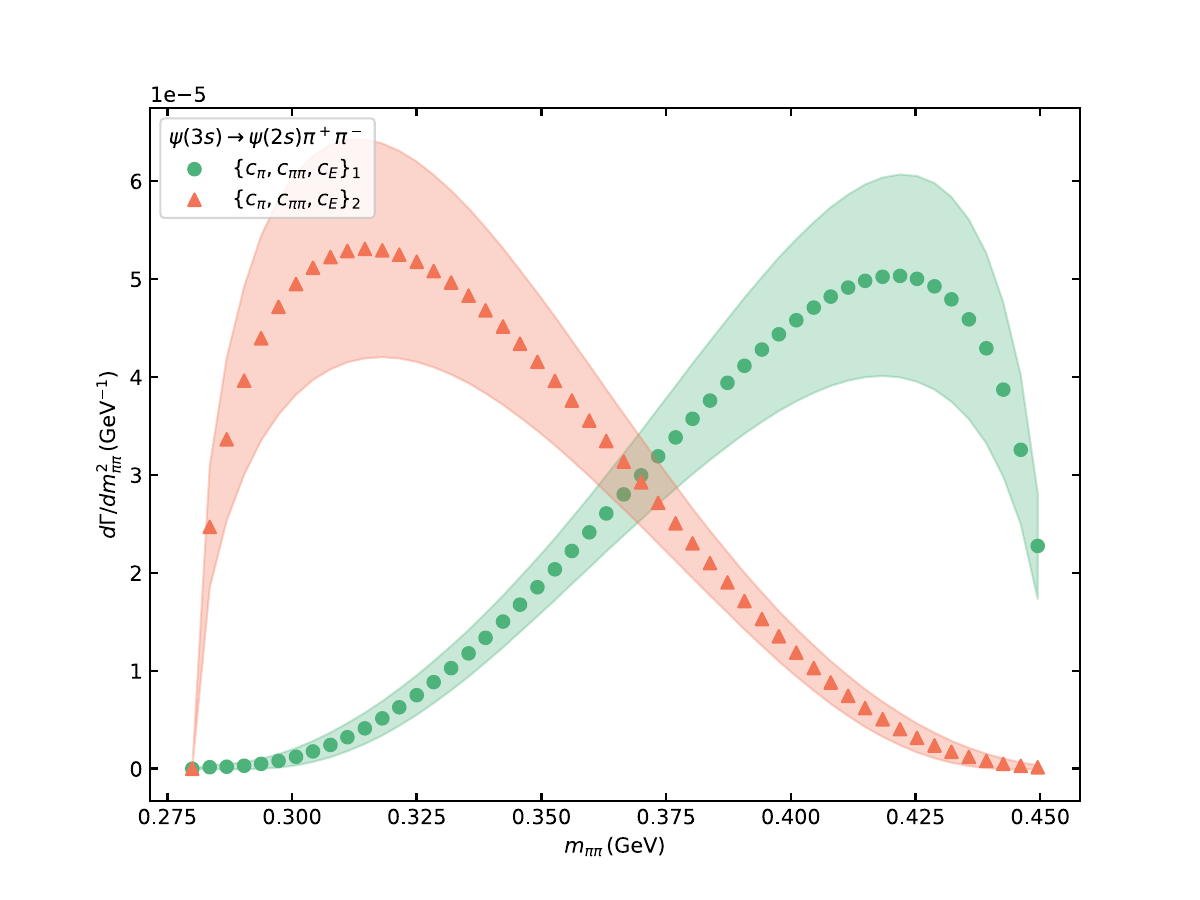} 
\caption{\justifying Dipion invariant mass spectrum obtained for charmonium transitions with $l'=l=0$ using \eqref{eq:dipion_q}. The green circular points correspond to the results for the set $\{\,c_\pi,\; c_{\pi\pi},\; c_E\,\}_1$ \eqref{eq:param1} and the triangular red points to the ones for the set $\{\,c_\pi,\; c_{\pi\pi},\; c_E\,\}_2$ \eqref{eq:param2}. We include the uncertainties generated by the errors on the parameters $c_\pi,c_{\pi\pi},c_E$. For completeness, we have also included a decay involving the ground state, but it is not expected to be reliable in our approach.} 
  \label{fig:StoS-c}
\end{figure*}

\begin{figure}[htbp]
\includegraphics[height=0.25\textheight,width=0.48\textwidth]{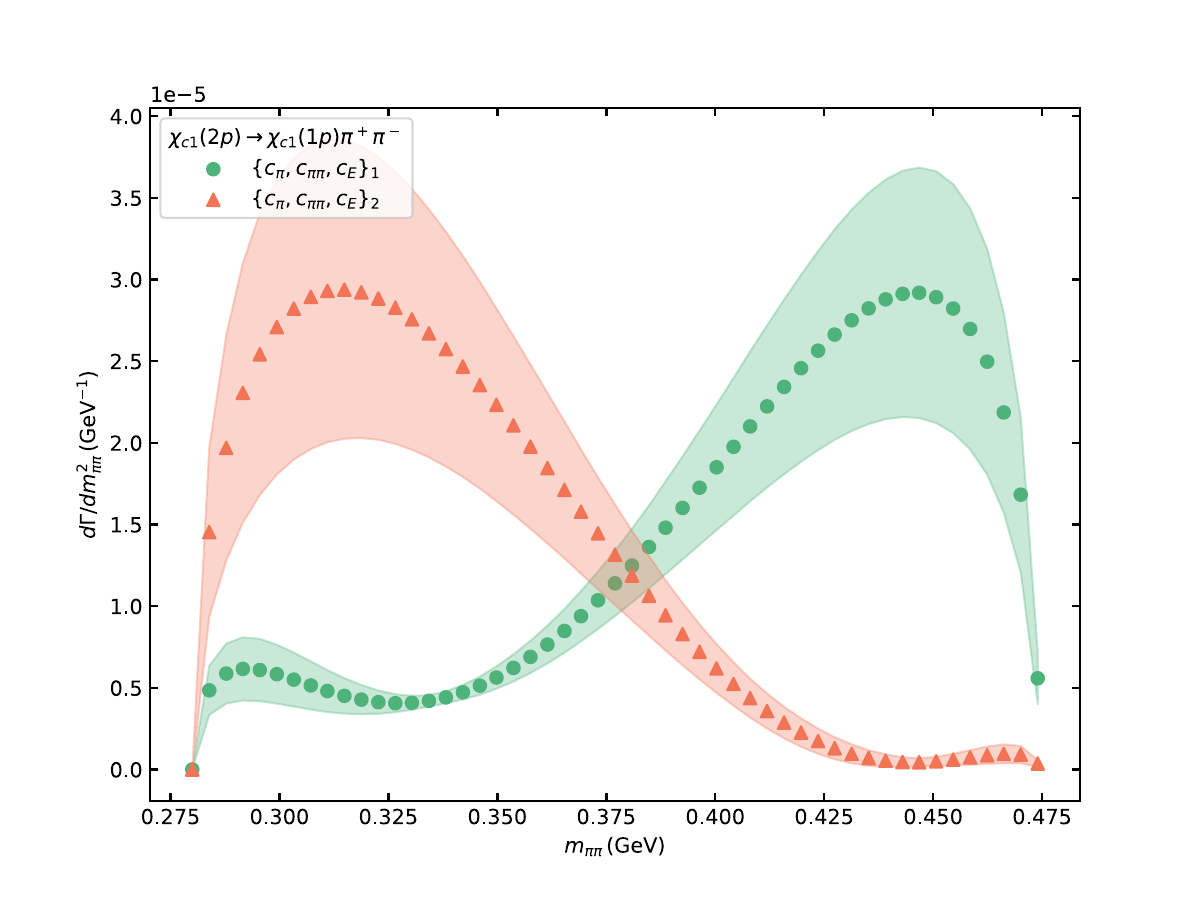} 
\caption{\justifying Dipion invariant mass spectrum obtained for charmonium transitions with $l'=l=1$ using \eqref{eq:dipion_q}. The green circular points correspond to the results for the set $\{\,c_\pi,\; c_{\pi\pi},\; c_E\,\}_1$ \eqref{eq:param1} and the triangular red points to the ones for the set $\{\,c_\pi,\; c_{\pi\pi},\; c_E\,\}_2$ \eqref{eq:param2}. We include the uncertainties generated by the errors on the parameters $c_\pi,c_{\pi\pi},c_E$.}
\label{fig:PtoP-c}
\end{figure}

\begin{figure*}[htbp]
\begin{center}
\includegraphics[height=0.25\textheight,width=0.48\textwidth]{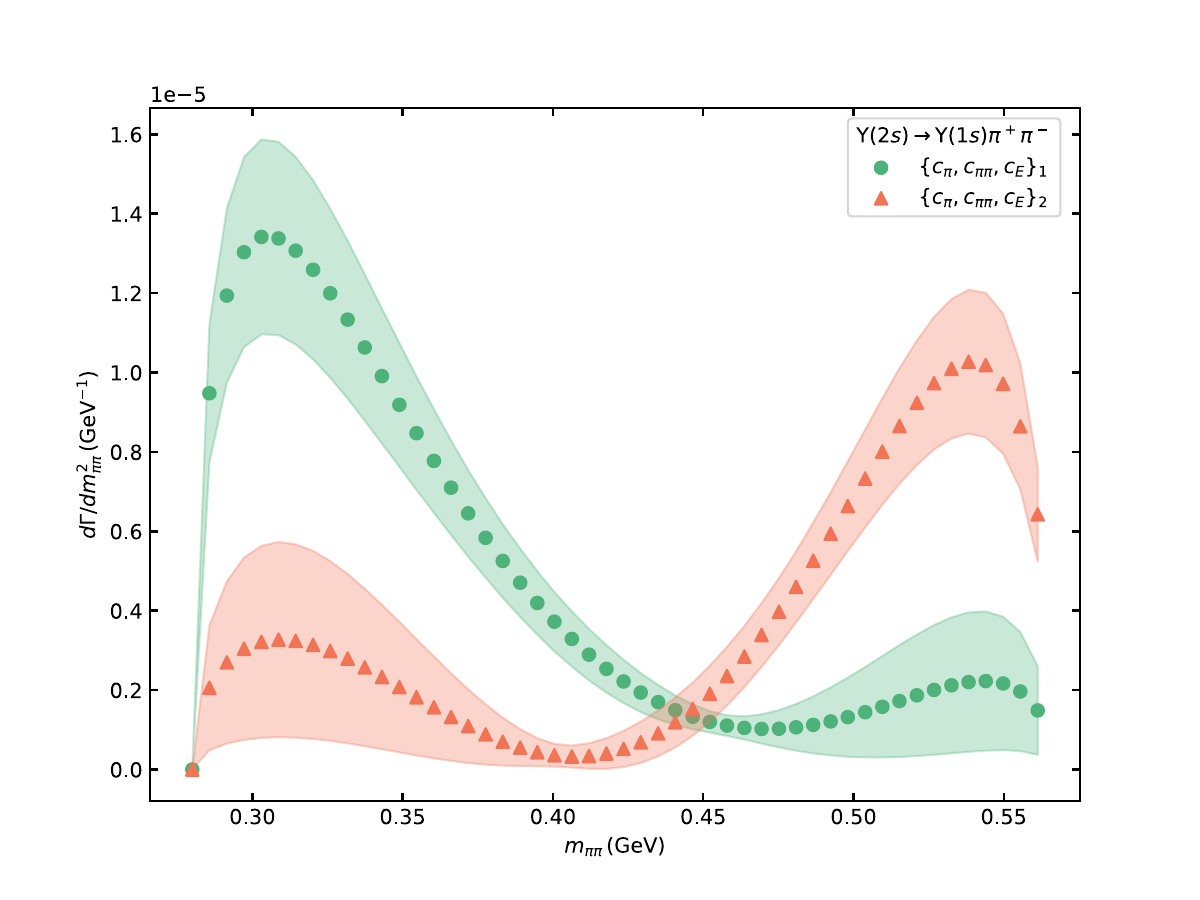}
\includegraphics[height=0.25\textheight,width=0.48\textwidth]{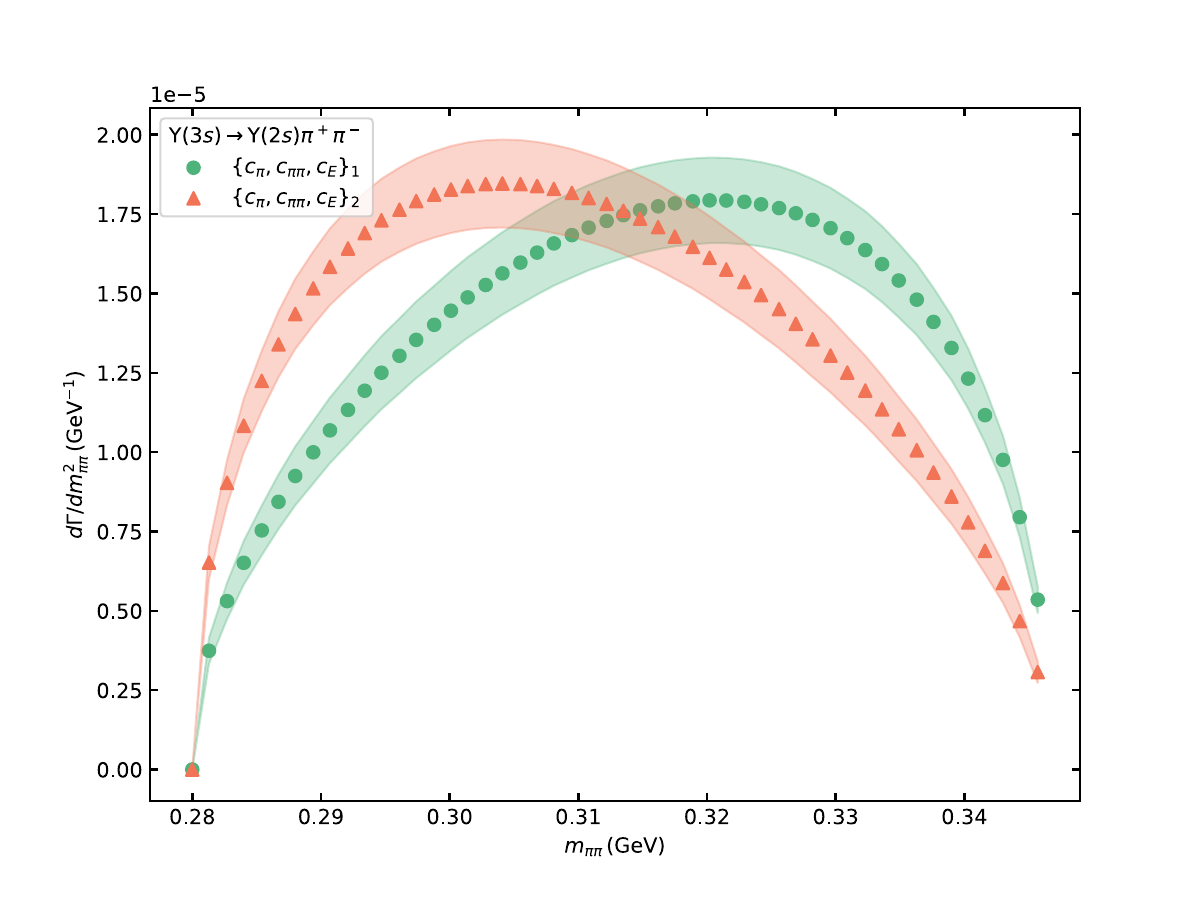}
\includegraphics[height=0.25\textheight,width=0.48\textwidth]{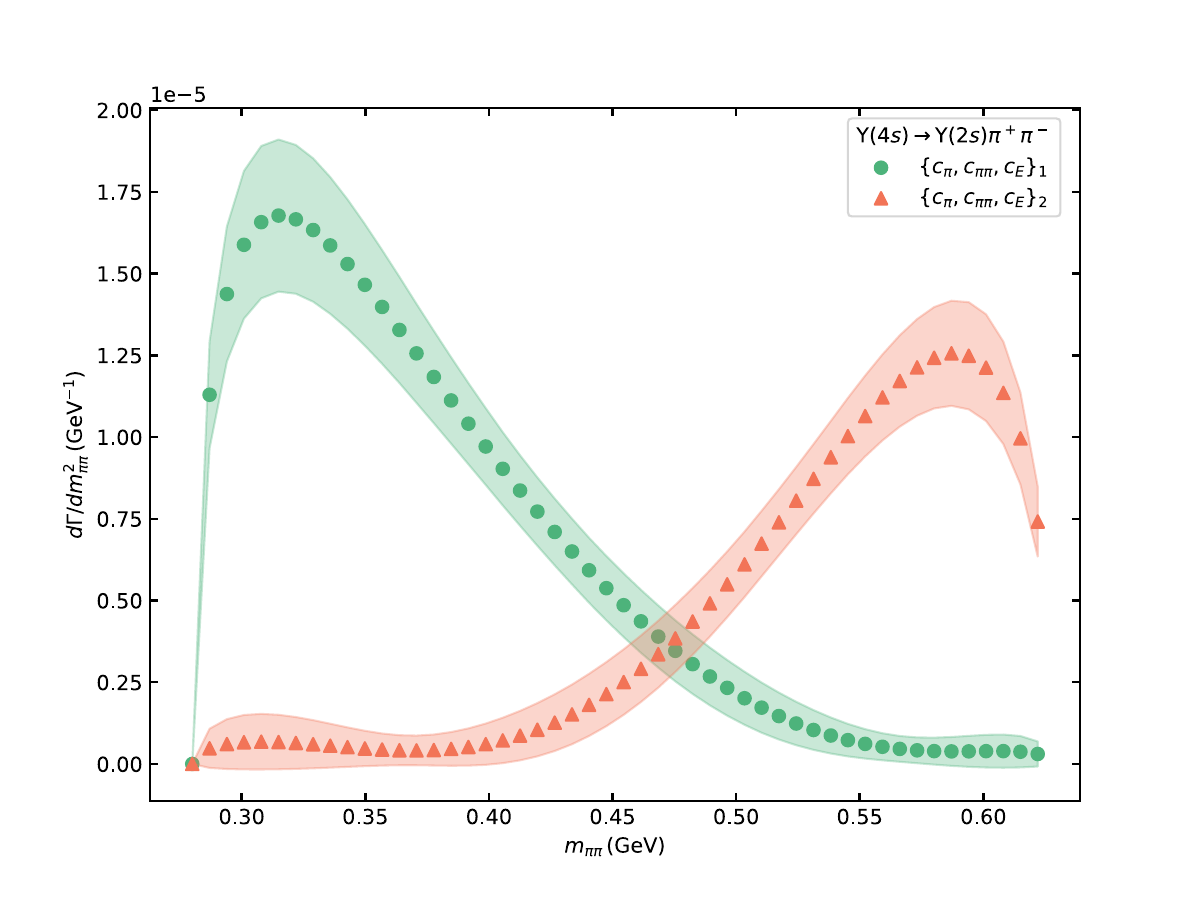}
\includegraphics[height=0.25\textheight,width=0.48\textwidth]{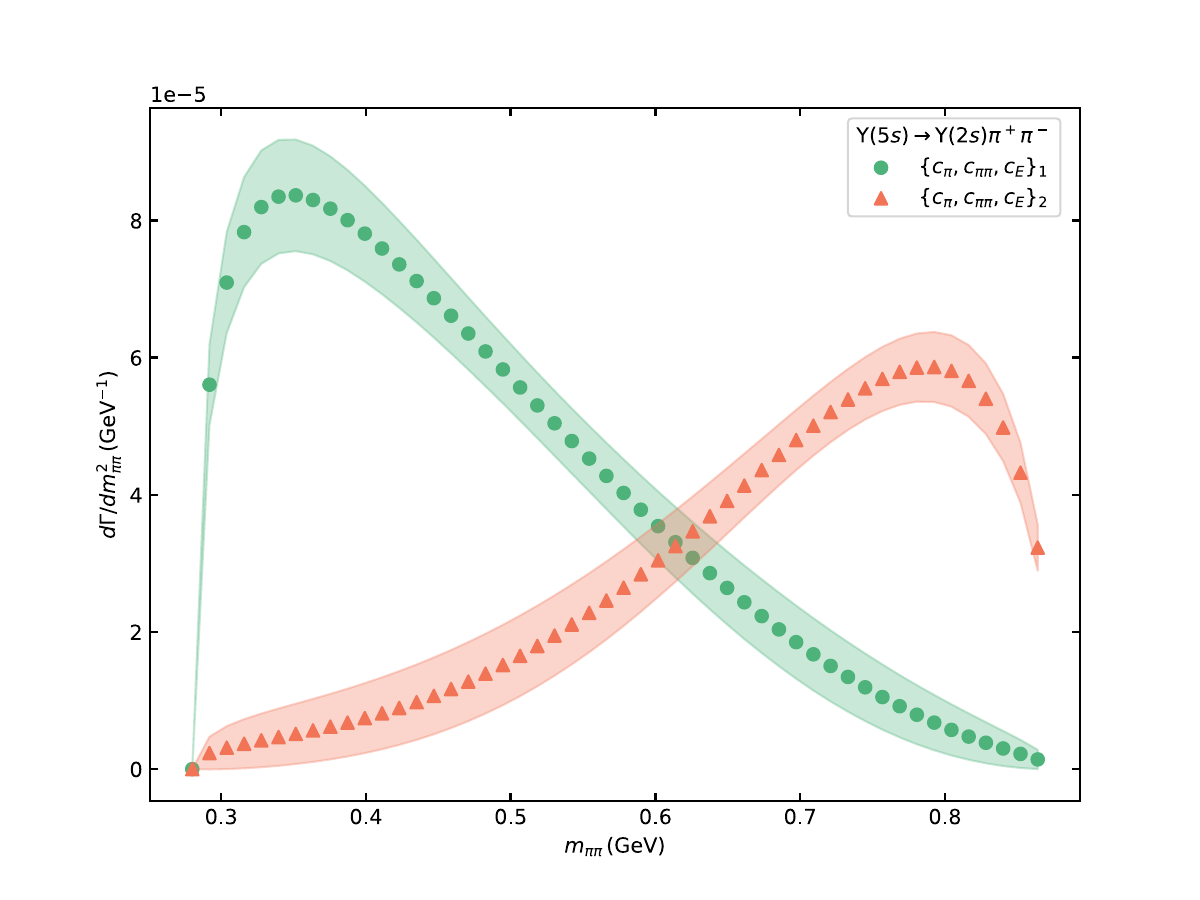} 
\includegraphics[height=0.25\textheight,width=0.48\textwidth]{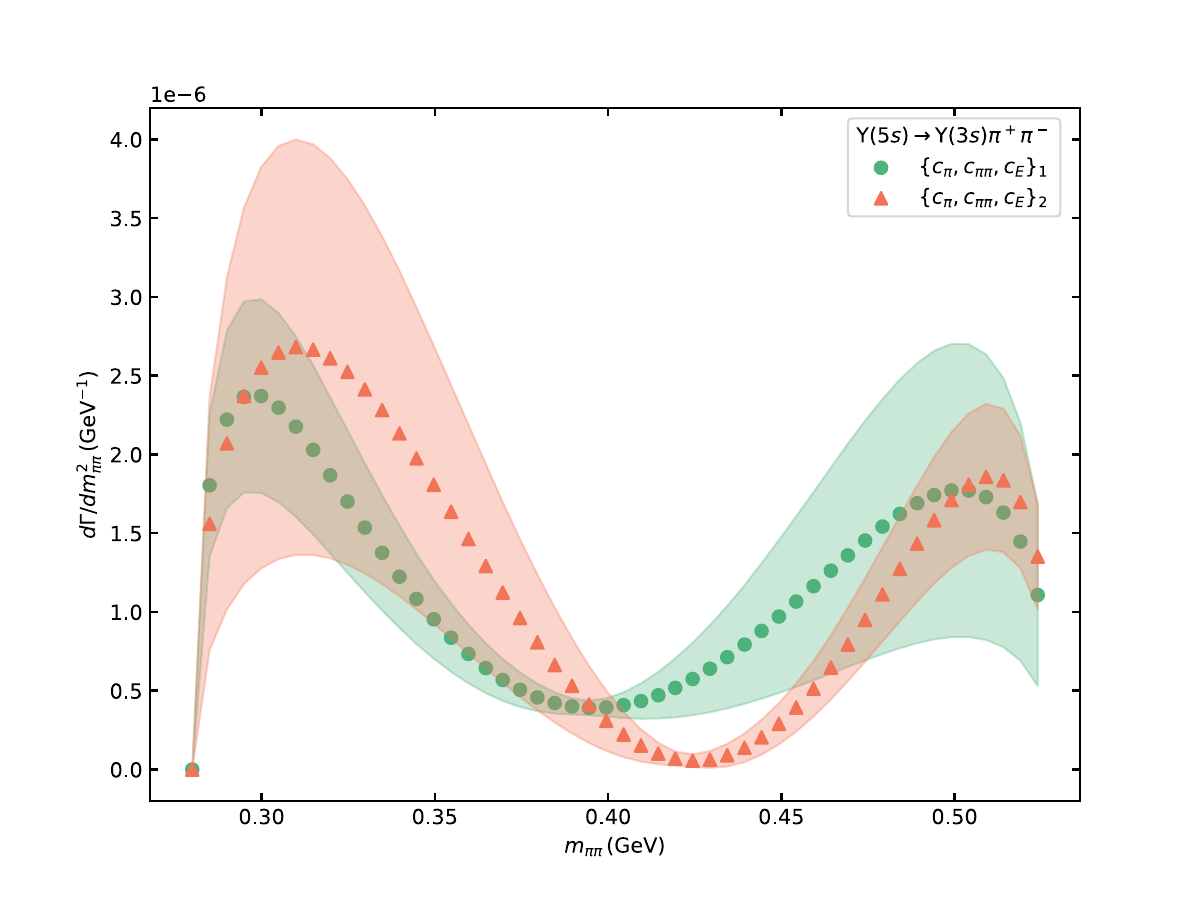}
\caption{ \justifying Dipion invariant mass spectrum obtained for bottomonium transitions with $l'=l=0$ using \eqref{eq:dipion_q}. The green circular points correspond to the results for the set $\{\,c_\pi,\; c_{\pi\pi},\; c_E\,\}_1$ \eqref{eq:param1} and the triangular red points to the ones for the set $\{\,c_\pi,\; c_{\pi\pi},\; c_E\,\}_2$ \eqref{eq:param2}. The plots correspond to the non-resonant contributions for those states that could have sizable contributions from resonances ($\Upsilon(5s)$). We include the uncertainties generated by the errors on the parameters $c_\pi,c_{\pi\pi},c_E$. For completeness, we have also included decays involving the ground states, but those are not expected to be reliable in our approach.}
\label{fig:StoS-b}
\end{center}
\end{figure*}

\begin{figure*}
\begin{center}
\includegraphics[height=0.25\textheight,width=0.48\textwidth]{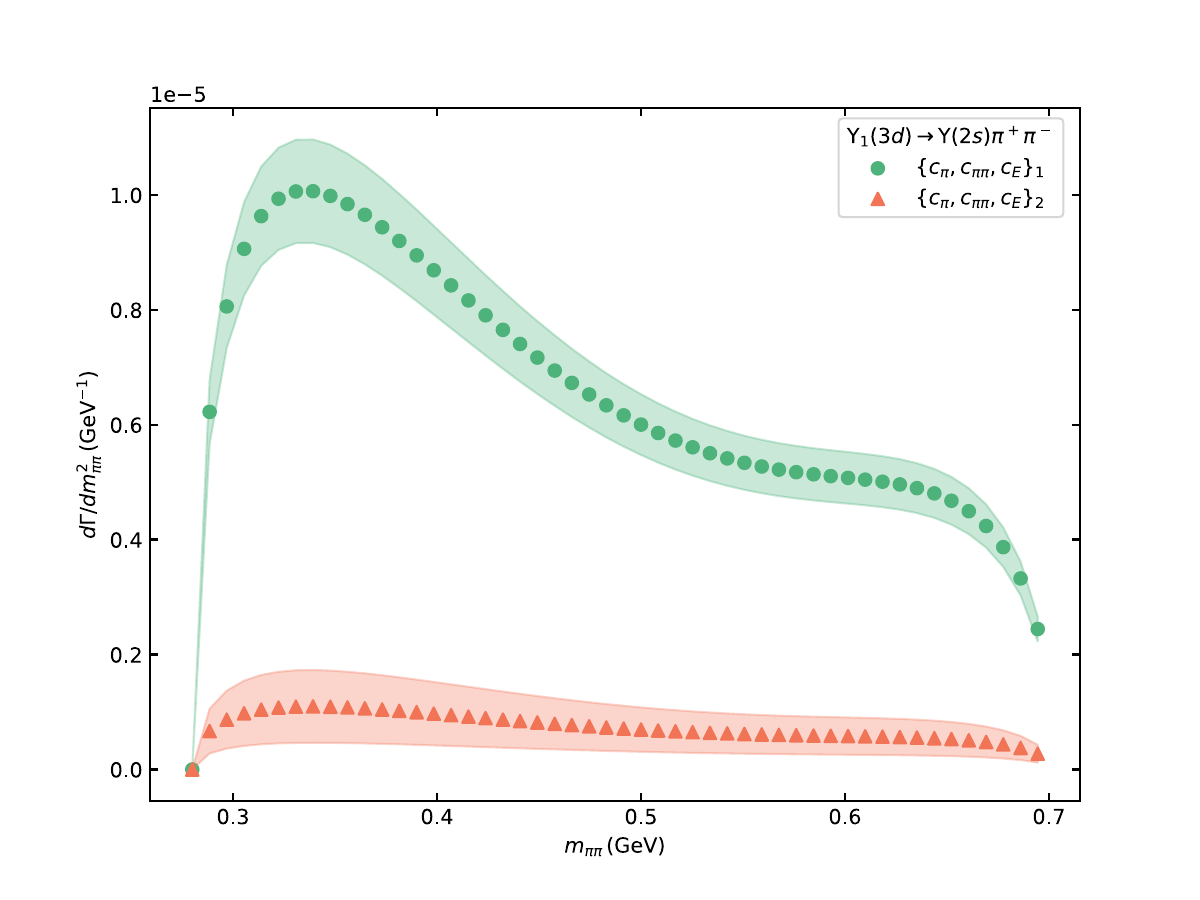}
\includegraphics[height=0.25\textheight,width=0.48\textwidth]{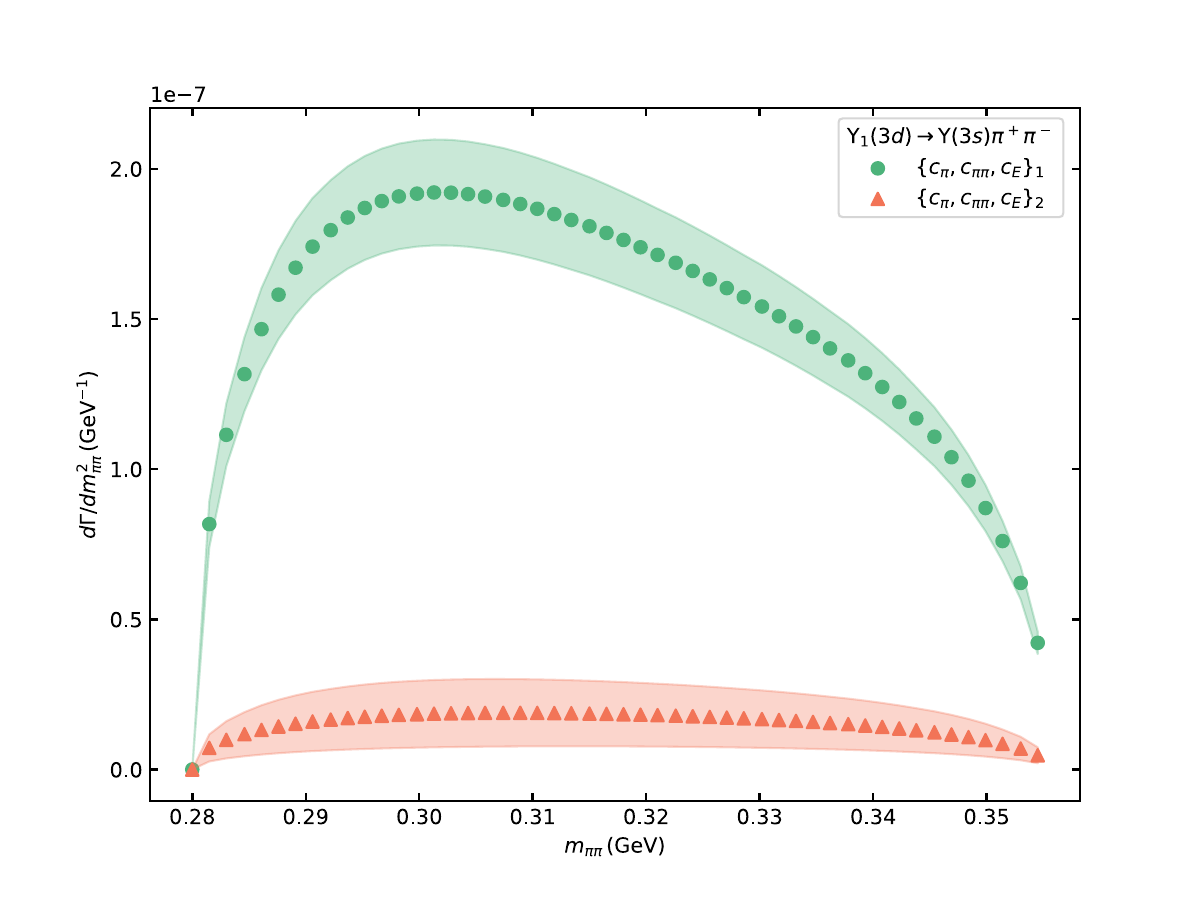}
\includegraphics[height=0.25\textheight,width=0.48\textwidth]{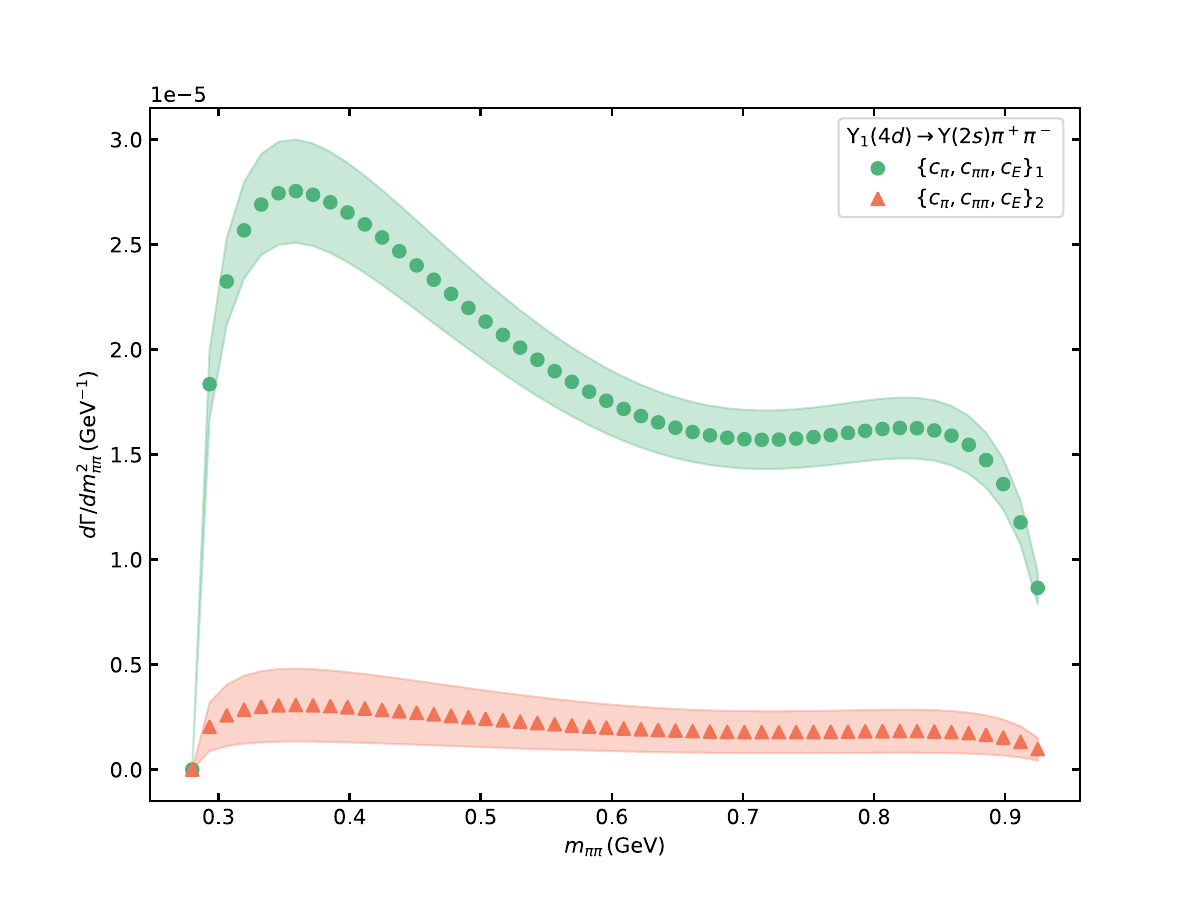} 
\includegraphics[height=0.25\textheight,width=0.48\textwidth]{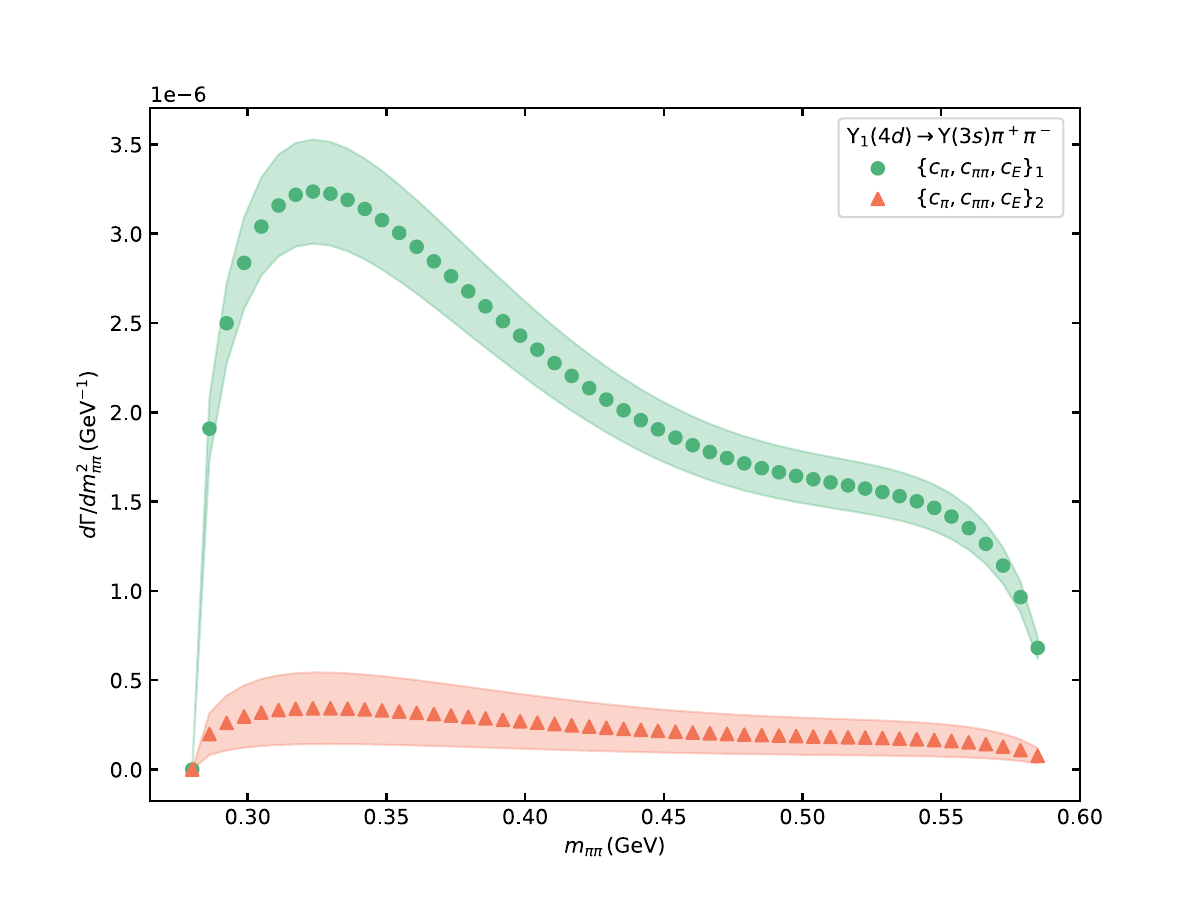}
\caption{\justifying Dipion invariant mass spectrum obtained for bottomonium transitions with $l'=0,l=2$ using \eqref{eq:dipion_q}. The green circular points correspond to the results for the set $\{\,c_\pi,\; c_{\pi\pi},\; c_E\,\}_1$ \eqref{eq:param1} and the triangular red points to the ones for the set $\{\,c_\pi,\; c_{\pi\pi},\; c_E\,\}_2$ \eqref{eq:param2}. The plots correspond to the non-resonant contributions for those states that could have sizable contributions from resonances ($\Upsilon_1(4d)$). We include the uncertainties generated by the errors on the parameters $c_\pi,c_{\pi\pi},c_E$. }
\label{fig:DtoS-b}
\end{center}
\end{figure*}

\begin{figure}[htbp]
\includegraphics[height=0.25\textheight,width=0.48\textwidth]{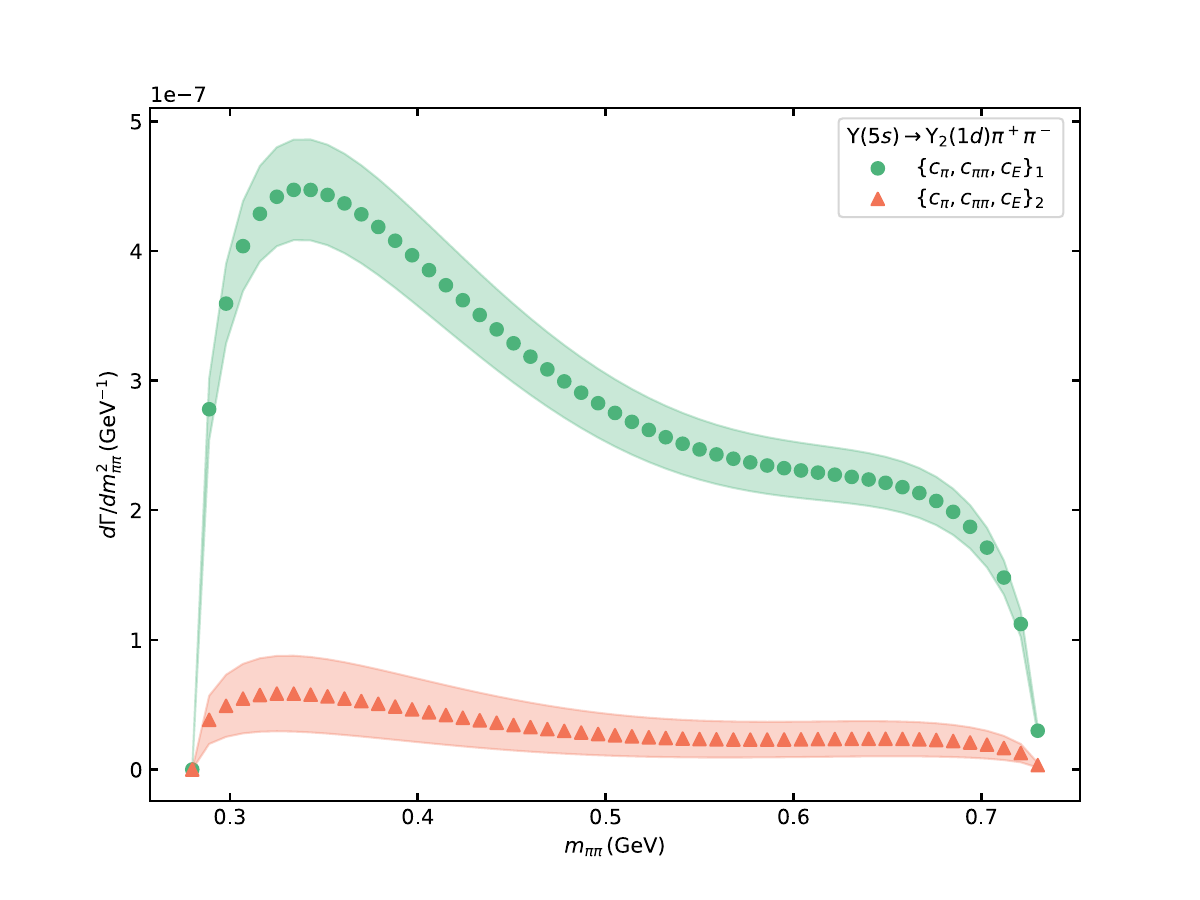} 
\caption{\justifying Dipion invariant mass spectrum obtained for bottomonium transitions with $l'=2,l=0$ using \eqref{eq:dipion_q}. The green circular points correspond to the results for the set $\{\,c_\pi,\; c_{\pi\pi},\; c_E\,\}_1$ \eqref{eq:param1} and the triangular red points to the ones for the set $\{\,c_\pi,\; c_{\pi\pi},\; c_E\,\}_2$ \eqref{eq:param2}. The plot correspond to the non-resonant contribution of the dipion invariant mass spectrum. We include the uncertainties generated by the errors on the parameters $c_\pi,c_{\pi\pi},c_E$. }
\label{fig:StoD-b}
\end{figure}

Table~\ref{t:dwquark} shows our predictions for the charmonium (up) and bottomonium (down) dipion decay widths. In Fig.~\ref{fig:StoS-c} and \ref{fig:StoS-b}, we show our results for the dipion invariant mass spectrum for charmonium and bottomonium $s\to s + \pi^+\pi^-$  transitions respectively. In Fig.~\ref{fig:PtoP-c},  we show the $\chi_{c1}(2p) \to \chi_{c1}(1p)\pi^+\pi^-$ dipion invariant mass spectrum.
In Fig.~\ref{fig:DtoS-b}, we show the results for the bottomonium $d \to s + \pi^+\pi^-$ and in Fig.~\ref{fig:StoD-b}, for the $\Upsilon(5s) \to \Upsilon_2(1d)\pi^+\pi^-$ transition, which is yet to be seen. 

\begin{table}[htbp]
\centering
\begin{tabular}{ |c||c|c| } 
 \hline
 \multicolumn{3}{|c|}{$\Gamma_i$ (keV)} \\
 \hline
 Transitions + $\pi^+\pi^-$ & $\{\,c_\pi,\; c_{\pi\pi},\; c_E\,\}_1$ & $\{\,c_\pi,\; c_{\pi\pi},\; c_E\,\}_2$ \\
 \hline
 $\psi(2s)\to J/\psi$   & $11.8 \pm 1.5$    & $11.7 \pm 1.5$ \\
 $\chi_{c1}(2p) \to \chi_{c1}(1p)$ & $2.2\pm 0.5$    & $1.7 \pm 0.5$ \\
 $\psi(3s)\to \psi(2s)$   & $3.4 \pm 0.6$    & $3.2 \pm 0.5$ \\
 \hline
 $\Upsilon(2s)\to \Upsilon(1s)$   & $1.00 \pm 0.20$     & $0.97 \pm 0.19$  \\
 $\Upsilon(3s)\to \Upsilon(2s)$   & $0.57 \pm 0.04$     & $0.57 \pm 0.04$  \\
 $\Upsilon(4s)\to \Upsilon(2s)$   & $1.68 \pm 0.20$     & $1.68 \pm 0.20$  \\
 $\Upsilon_1(3d)\to \Upsilon(2s)$   & $2.52 \pm 0.22$     & $0.28 \pm 0.16$  \\
 $\Upsilon_1(3d)\to \Upsilon(3s)$   & $ \left( 7.1 \pm 0.7 \right) \times 10^{-3} $     & $ \left( 7 \pm 4 \right) \times 10^{-4}$  \\
 $\Upsilon_1(4d)\to \Upsilon(2s)$   & $14.0 \pm 1.2 $     & $1.6 \pm 0.9 $  \\
 $\Upsilon_1(4d)\to \Upsilon(3s)$   & $0.53 \pm 0.05$     & $0.06 \pm 0.03$  \\
 $\Upsilon(5s)\to \Upsilon(2s)$   & $22.8 \pm 2.4$    & $22.7 \pm 2.4$  \\
 $\Upsilon(5s)\to \Upsilon_2(1d)$   & $0.123 \pm 0.011$    & $0.014 \pm 0.008 $  \\
 $\Upsilon(5s)\to \Upsilon(3s)$   & $0.23 \pm 0.08$    & $0.23 \pm 0.08$ \\
 \hline
\end{tabular}
\caption{\justifying Predictions of the dipion decay widths for charmonium (up) and bottomonium (down) relevant dipion transitions between highly excited states using the two sets of parameters \eqref{eq:param1} and \eqref{eq:param2} and dipion invariant mass spectrum \eqref{eq:dipion_q}. We include the uncertainties generated by the errors on the parameters $\{\,c_\pi,\; c_{\pi\pi},\; c_E\,\}$. For completeness, we have also included decays involving the ground states, but those are not expected to be reliable in our approach.}
\label{t:dwquark}
\end{table}

\subsection{Hybrids}\label{sec:cc}

\begin{figure*}
\begin{center}
\includegraphics[height=0.25\textheight,width=0.48\textwidth]{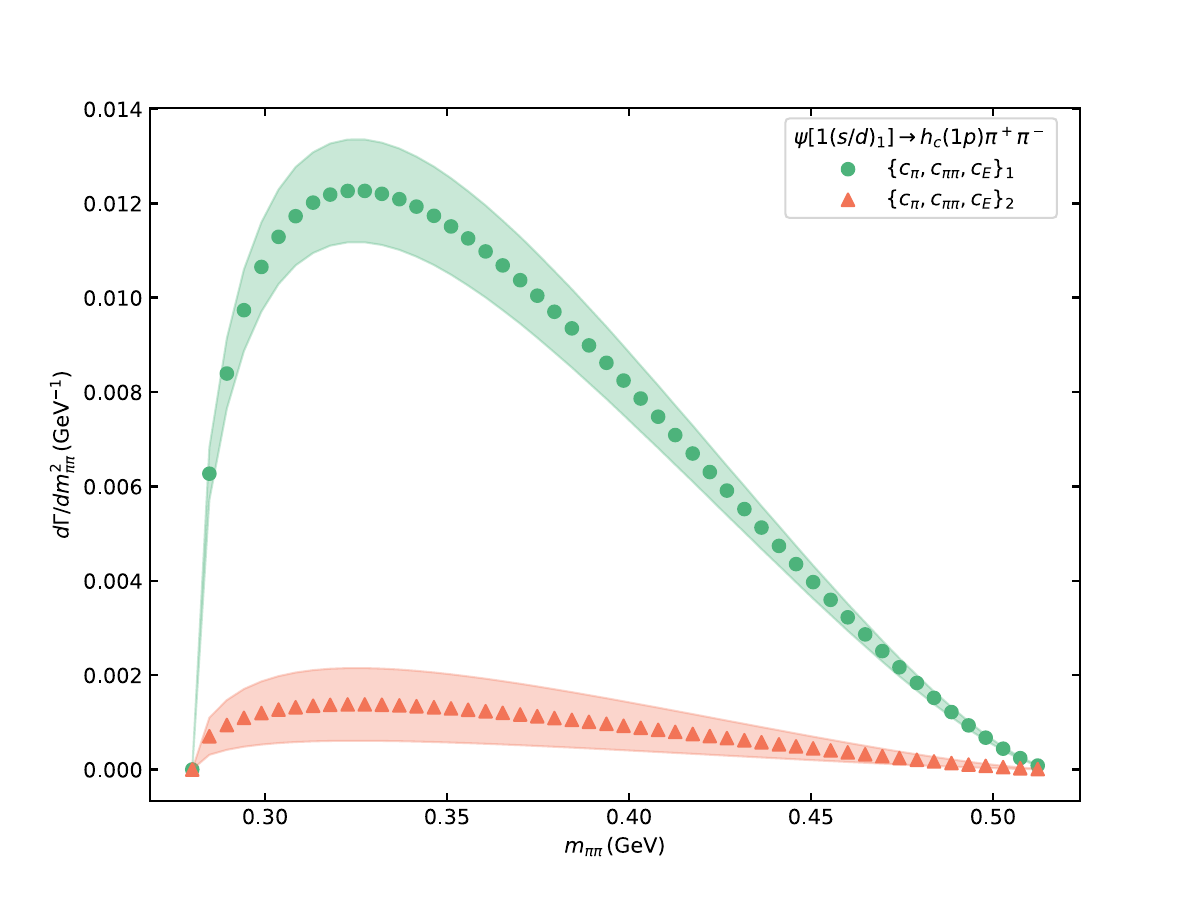}
\includegraphics[height=0.25\textheight,width=0.48\textwidth]{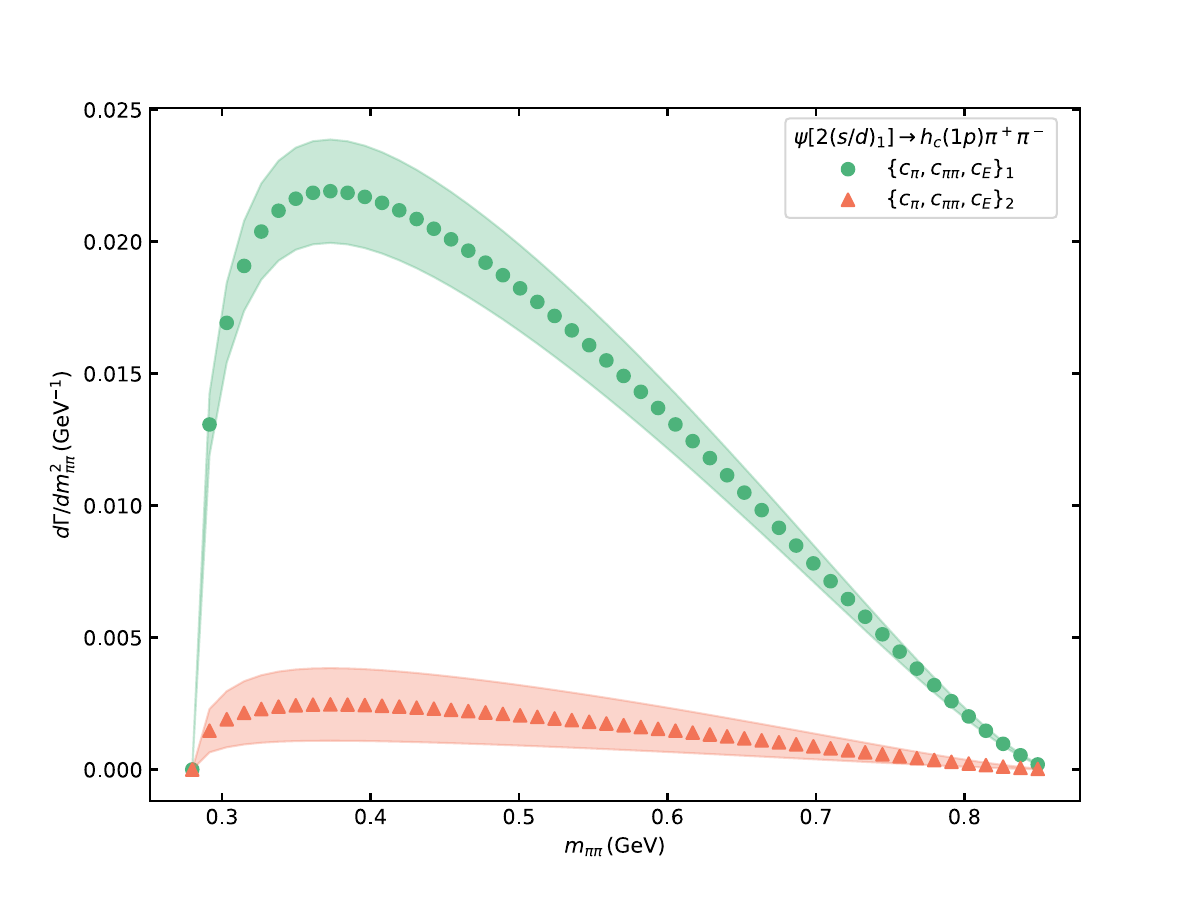} 
\caption{\justifying Dipion invariant mass spectrum obtained for transitions between charmonium spin-0 $n(s/d)_1$ hybrids to $1p$ quarkonium using \eqref{eq:dipion_h}. The green circular points correspond to the results for the set $\{\,c_\pi,\; c_{\pi\pi},\; c_E\,\}_1$ \eqref{eq:param1} and the triangular red points to the ones for the set $\{\,c_\pi,\; c_{\pi\pi},\; c_E\,\}_2$ \eqref{eq:param2}. We include the uncertainties generated by the errors on the parameters $c_\pi,c_{\pi\pi},c_E$. }
\label{fig:SDtoP-c}
\end{center}
\end{figure*}

\begin{figure*}
\begin{center}
\includegraphics[height=0.25\textheight,width=0.48\textwidth]{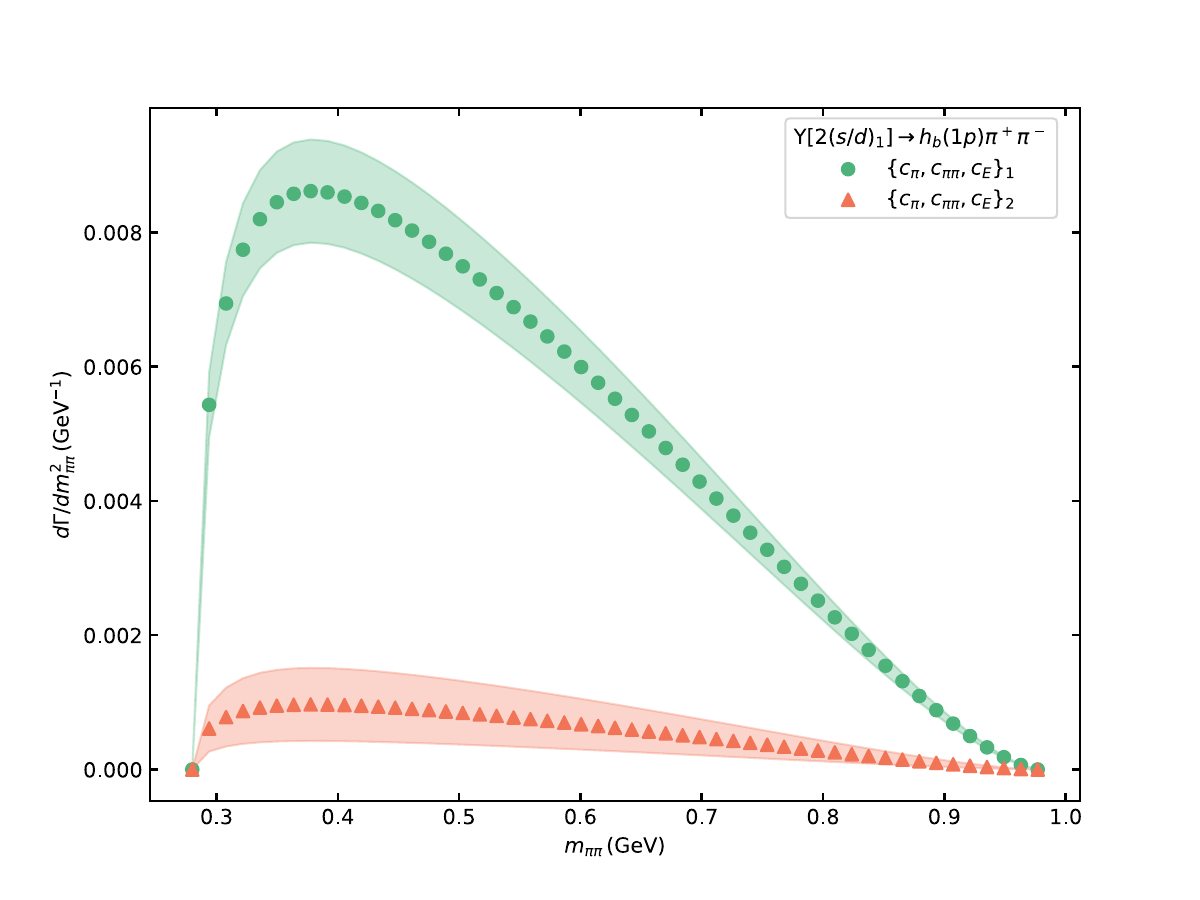}
\includegraphics[height=0.25\textheight,width=0.48\textwidth]{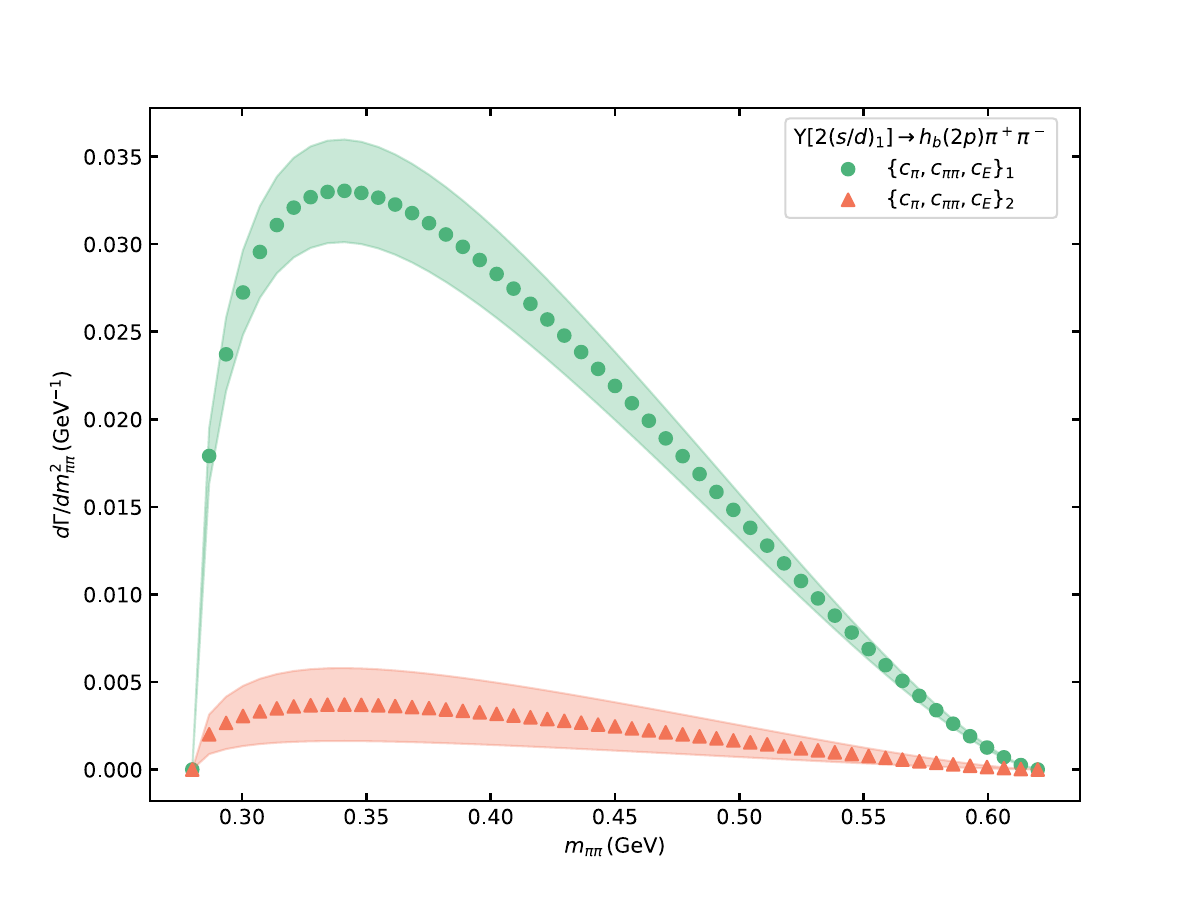} 
\caption{\justifying Dipion invariant mass spectrum obtained for transitions between bottomonium spin-0 $2(s/d)_1$ hybrids to $np$ quarkonium using \eqref{eq:dipion_h}. The green circular points correspond to the results for the set $\{\,c_\pi,\; c_{\pi\pi},\; c_E\,\}_1$ \eqref{eq:param1} and the triangular red points to the ones for the set $\{\,c_\pi,\; c_{\pi\pi},\; c_E\,\}_2$ \eqref{eq:param2}. We include the uncertainties generated by the errors on the parameters $c_\pi,c_{\pi\pi},c_E$.  }
\label{fig:SDtoP-b}
\end{center}
\end{figure*}

\begin{figure}[htbp]
\includegraphics[height=0.25\textheight,width=0.48\textwidth]{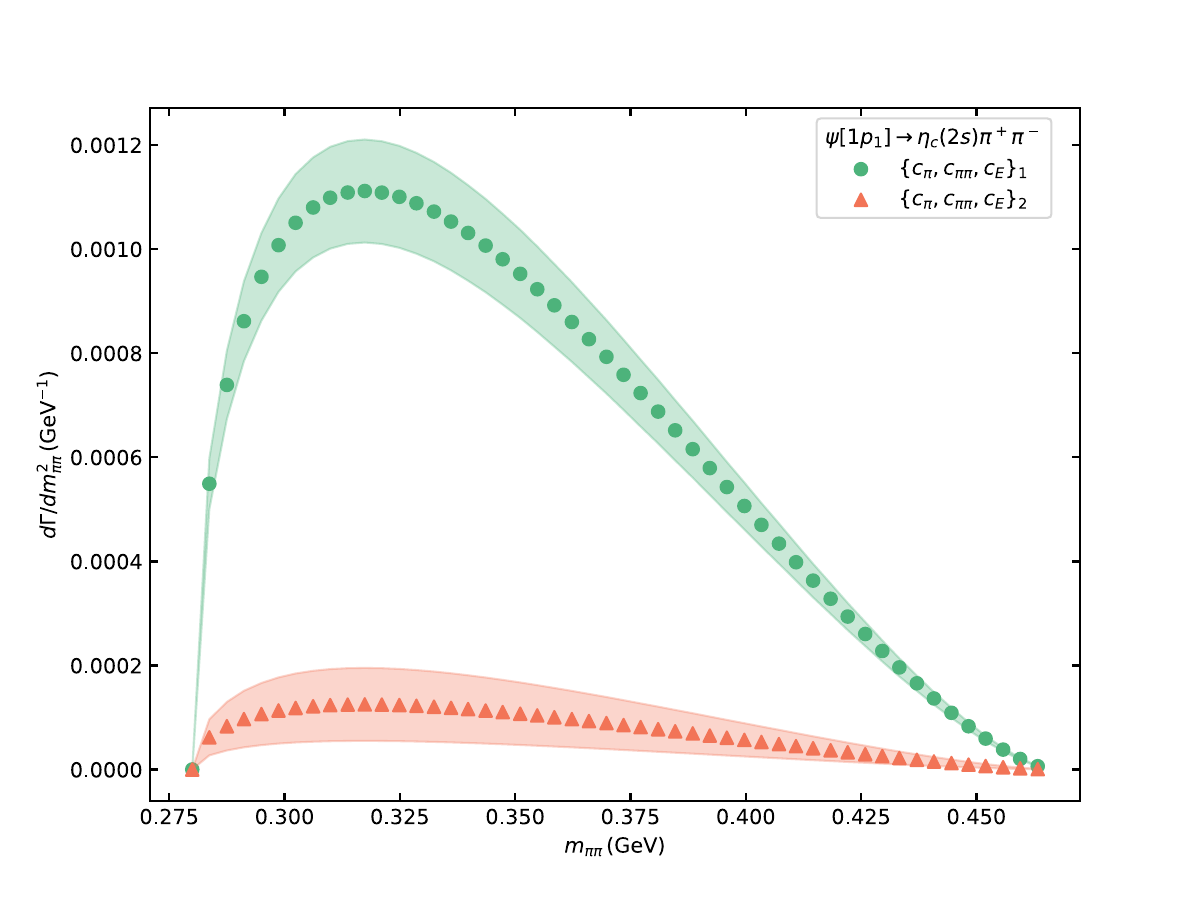} 
\caption{\justifying Dipion invariant mass spectrum obtained for the charmonium spin-0 $1p_1 \to 2s + \pi^+\pi^-$ transition using \eqref{eq:dipion_h}. The green circular points correspond to the results for the set $\{\,c_\pi,\; c_{\pi\pi},\; c_E\,\}_1$ \eqref{eq:param1} and the triangular red points to the ones for the set $\{\,c_\pi,\; c_{\pi\pi},\; c_E\,\}_2$ \eqref{eq:param2}. We include the uncertainties generated by the errors on the parameters $c_\pi,c_{\pi\pi},c_E$. }
\label{fig:P1toS}
\end{figure}

The permitted dipion transitions between hybrids and highly excited quarkonium are $(s/d)_1 \to p + \pi\pi$, $p_1 \to s + \pi\pi$, and $p_1 \to d + \pi\pi$\footnote{ The $p_0$ states contain the $\Sigma_u^-$ representation of the $D_{\infty h}$ group only \cite{Oncala:2017hop}, which gives zero in the EST calculation \eqref{eq:Su-}.}. We restrict our analysis to XYZ resonances that have a hybrid candidate, and focus on the ones that pion transitions have already been observed. Since there is no experimental evidence for the $p_1$ transitions yet, most of the transitions we will deal with are $(s/d)_1 \to p + \pi\pi$.
In addition since almost all hybrid candidates correspond to spin $0$ states, we restrict ourselves to this case. Using Eq.~\eqref{eq:dipion_h}, \eqref{eq:param1}, and \eqref{eq:param2} we can compute the dipion invariant mass spectrum and the decay width. Consistently with the quarkonium case, the wave functions used are the ones in \cite{Oncala:2017hop}.
Fig.~\ref{fig:SDtoP-c} shows our results for charmonium transitions $\psi \left[ 1(s/d)_1 \right] \to h_c(1p)  \pi^+\pi^-$ and $\psi \left[ 2(s/d)_1 \right] \to h_c(1p)  \pi^+\pi^-$. Where $ 1(s/d)_1$ is identified as $\psi(4040)$ and $ 2(s/d)_1$ as $\psi(4360)$ \cite{Oncala:2025mqj}, both with seen transitions to $h_c\pi\pi$. 
Fig.~\ref{fig:SDtoP-b} shows our results for bottomonium transitions with initial state $\Upsilon \left[ 2(s/d)_1 \right]$ identified as $\Upsilon(10860)$ \cite{Oncala:2025mqj} to $h_b(1p) \pi^+\pi^-$ and $h_b(2p) \pi^+\pi^-$ final states.
For completeness we include in Fig.~\ref{fig:P1toS} our prediction for the transition $\psi(1p_1) \to \eta_c(2s) \pi^+\pi^-$ which has no experimental evidence but could be identified as $\chi_{c1}(4140)$ or $X(4160)$ \cite{Oncala:2025mqj}.
The dipion decay widths predicted for the transitions above appear in Table~\ref{t:dwhybrid}.

In Appendix \ref{ap:NLO} we present analogous results obtained from the full Lagrangian \eqref{intH}. 

\begin{table}[!htbp]
\centering
\begin{tabular}{ |c||c|c| } 
 \hline
 \multicolumn{3}{|c|}{$\Gamma_i$ (keV)} \\
 \hline
 Transitions + $\pi^+\pi^-$ & $\{\,c_\pi,\; c_{\pi\pi},\; c_E\,\}_1$ & $\{\,c_\pi,\; c_{\pi\pi},\; c_E\,\}_2$ \\
 \hline
 $\psi \left[ 1(s/d)_1 \right] \to h_c(1p)$   & $1210 \pm 107$    & $136 \pm 76$ \\
 $\psi\left[ 1p_1 \right] \to \eta_c(2s)$   & $83.7 \pm 7.4$    & $9.3 \pm 5.2$ \\
 $\psi \left[ 2(s/d)_1\right] \to h_c(1p)  $   & $7300 \pm 650$    & $821 \pm 460$ \\
 \hline
 $\Upsilon \left[2(s/d)_1 \right] \to h_b(1p)$   & $3568 \pm 320$    & $401 \pm 220$ \\  
 $\Upsilon \left[ 2(s/d)_1 \right] \to h_b(2p)$   & $5144 \pm 460$    & $578 \pm 320$ \\  
 \hline
\end{tabular}
\caption{\justifying Predictions of the decay widths for relevant dipion transitions between charmonium hybrids (up) and bottomonium hybrids (down) and highly excited quarkonium states using the two sets of parameters \eqref{eq:param1} and \eqref{eq:param2} and dipion invariant mass spectrum \eqref{eq:dipion_h}. We include the uncertainties generated by the errors on the parameters $\{\,c_\pi,\; c_{\pi\pi},\; c_E\,\}$.}
\label{t:dwhybrid}
\end{table}

\section{Discussion}\label{sec:con}
We have focused on quarkonium ($0^{++}$) and lowest lying hybrid ($1^{+-}$) transitions. Generalization to other isospin zero exotics is straightforward in the BOEFT limit. We have listed the relevant operators in the Appendix~\ref{ap:op}. Different exotics with the same LDF quantum numbers would be described by the same interaction Lagrangian with different functions for the LEF. For instance, for  $1^{+-}$ tetraquarks, the Lagrangian \eqref{boefth}
still holds but the estimate \eqref{eq:g4} for $g_4(r)$ need not do it\footnote{The EST that leads to \eqref{eq:g4} may have to be modified including fermionic degrees of freedom, see \cite{Soto:2021cgk} for a proposal.}.

By comparing with the PDG \cite{ParticleDataGroup:2024cfk} we can see that our results for the dipion decay widths for the transitions $\psi(2s)\to J/\psi \, \pi^+\pi^-$ ($99 \pm 7$ keV) and $\Upsilon(2s)\to\Upsilon(1s)\pi^+\pi^-$ ($5.7 \pm 0.6$ keV) are one order of magnitude smaller for charmonium and of the same order of magnitude but far smaller for bottomonium. This is not surprising as the $1s$ states are very compact and not suitable for long-distance treatment. 
One may argue that this could also be the case for the transition $\chi_{c1}(2p)\to\chi_{c1}(1s)\pi^+\pi^-$, however if we identify $\chi_{c1}(2p)$ as $\chi_{c1}(3872)$ and compare with the decay width constraint given by the PDG ($< 9.5\pm 1.7 \, {\rm keV}$), we see that for both sets $\{\,c_\pi,\; c_{\pi\pi},\; c_E\,\}_{1,2}$ the experimental constraint is fulfilled. 

The experimental constraint in the transition involving $\psi(4040)$ is $\Gamma_{\psi(4040)\to h_c(1p)\pi^+\pi^-}^{\rm exp}< 252 \pm 36$ keV. By identifying the state with the $1(s/d)_1$ hybrid only the $\{\,c_\pi,\; c_{\pi\pi},\; c_E\,\}_{2}$ set gives compatible results.

Both $\chi_{c1}(4140)$ and $X(4160)$ can be identified as charmonium hybrid $1p_1$. None of them have data on the PDG involving dipion decays to a $s$-state.
An alternative assignment for $\psi(4160)$ is $\psi(3s)$ \cite{Oncala:2025mqj}, PDG gives a constraint on the decay width of $\Gamma_{\psi(4160)\to\psi(2s)\pi^+\pi^-}^{\rm exp}< 276\pm 40$ keV which is fulfilled for both sets of parameters.

Our charmonium hybrid $2(s/d)_1$, can be identified as both $\psi(4360)$ and $\psi(4415)$. Only the former has a seen transition to $h_c\pi^+\pi^-$.

In the case of $\Upsilon(10753)$ (our $\Upsilon_1(3d)$) transitions, the PDG gives $\Gamma_{\Upsilon(10753)\to\Upsilon(3s)\pi^+\pi^-}/\Gamma_{\Upsilon(10753)\to\Upsilon(2s)\pi^+\pi^-} = 0.27 \pm 0.14 $ which is two orders of magnitude higher than the one we obtain independently of the set of parameters used. 

The description of $\Upsilon(10860)$ as $\Upsilon(5s)$ and $\Upsilon[2(s/d)_1]$ at spin-0 are both possible. 
For the quarkonium identification, we cannot use directly the PDG data to compare with our results, due to the large contributions from resonances \cite{Belle:2014vzn}. We use the non-resonant contributions  given in \cite{Bondar:2013xxx} $\Gamma_{\Upsilon(5s)\to \Upsilon(2s)\pi^+\pi^-}\vert_\mathrm{non-resonant}=0.82^{+0.26}_{-0.23}$ keV, but this figure remains an order of magnitude smaller than our predictions. This disfavors the interpretation of $\Upsilon(10860)$ as a pure quarkonium state. $\Upsilon(5s)\to\Upsilon_2(1d)\pi^+\pi^-$ has no experimental data in the literature. PDG has data for both of our computed transitions from $\Upsilon[2(s/d)_1]$ initial state, $\Gamma_{\Upsilon[2(s/d)_1]\to h_c(1p)\pi^+\pi^-}^{\rm exp} = 129.5^{+4.0}_{-5.2}$ keV and $\Gamma_{\Upsilon[2(s/d)_1]\to h_c(2p)\pi^+\pi^-}^{\rm exp} = 210.9^{+6.8}_{-8.4}$ keV. These values are of the same order of magnitude as our results for set 2 in Tab.~\ref{t:dwhybrid}. For both transitions we get even better comparisons with our results in Tab.~\ref{t:dwhybrid-full} for the same set.
This strongly suggests that $\Upsilon(10860)$ is mainly a $2(s/d)_1$ hybrid with a small component of $5s$ quarkonium \cite{Oncala:2017hop}.

Finally, regarding $\Upsilon(11020)$ transitions, identified as $\Upsilon_1(4d)$, the PDG value for the fraction $\Gamma_{\Upsilon(11020)\to\Upsilon(3s)\pi^+\pi^-}^{\rm exp}/\Gamma_{\Upsilon(11020)\to\Upsilon(2s)\pi^+\pi^-}^{\rm exp} = 0.5 \pm 0.7 $ is about an order of magnitude higher than for our sets:
$\Gamma_{\Upsilon_1(4d)\to\Upsilon(3s)\pi^+\pi^-}/\Gamma_{\Upsilon_1(4d)\to\Upsilon(2s)\pi^+\pi^-} |_1 = 0.038 \pm 0.007 $ and $\Gamma_{\Upsilon_1(4d)\to\Upsilon(3s)\pi^+\pi^-}/\Gamma_{\Upsilon_1(4d)\to\Upsilon(2s)\pi^+\pi^-} |_2=0.04 \pm 0.04$.

In Ref.~\cite{Surovtsev:2015hna}, the authors present dipion transition spectra incorporating all relevant resonances, as well as separate plots isolating the combined contributions of $f_0(500)$, $f_0(980)$, and $f_0^\prime(1500)$. By evaluating the difference between the full resonant results and the spectrum containing only these $f_0$ resonances, we obtain an approximation for the non-resonant background. 
A qualitative comparison suggests that the
$\Upsilon(4s)\to\Upsilon(2s)\pi^+\pi^-$ and $\Upsilon(5s)\to\Upsilon(2s)\pi^+\pi^-$ transitions favor the parameter set $1$. For the $\Upsilon(5s)\to\Upsilon(3s)\pi^+\pi^-$ transition, both parameter sets provide a similar description. Furthermore, experimental data from Ref.~\cite{CLEO:2007rbi} suggest a slight preference for set 1 in the $\Upsilon(3s)\to\Upsilon(2s)\pi^+\pi^-$ transition.

Given the limited experimental data currently available and the large uncertainties in our results, we cannot identify either of the two parameter sets $\{\,c_\pi,\; c_{\pi\pi},\; c_E\,\}$ as clearly preferred. We have also tried to obtain them from analysis based on hadronic effective theories \cite{Chen:2016mjn,Baru:2020ywb}, but it has not been possible due to the incomplete set of operators used in those theories, see Appendix~\ref{ap:comparison}.
Several improvements to our procedure are possible. If more experimental data on the dipion spectrum for highly excited quarkonium transitions become available, one could fit to the dipion invariant mass spectrum to extract the parameters $\{\,c_\pi,\; c_{\pi\pi},\; c_E\,\}$, rather than relying on the less informative approach of fitting the decay widths. This would also require properly taking into account final state interactions between the emitted pions \cite{Surovtsev:2015hna,TarrusCastella:2021pld,Chen:2025jip} and threshold effects of intermediate heavy-light meson pairs \cite{Chen:2016mjn,Baru:2020ywb}, which we have either neglected or subtracted according to Ref.~\cite{Bondar:2013xxx}. 
In addition, the large theoretical uncertainties that were not included in our predictions suggest that the transitions considered involve states that are not large enough to completely neglect short-distance effects. An interpolation between the short- and long-distance behavior of the LEF would therefore provide a more appropriate description.

We have also restricted ourselves to the study of pion transitions  and hence we have used $SU(2)$ ChPT in the isospin limit. The generalization to $SU(3)$ in order to include, for instance $\eta$ transitions, or to strong isospin breaking effects, in order to include, for instance $\pi^0$ transitions, is straightforward.

\section{Summary and Conclusions}\label{sec:sum}

We have addressed pion transitions in the BOEFT at leading order in the $1/m_Q$ and chiral expansions both for quarkonium and exotics. 
We have established the interaction Lagrangian for quarkonium and isospin-zero exotics of several $J^{PC}$ of the LDF. The interaction Lagrangian depends on a few  low energy functions. Their short-distance behavior was known, since it can be obtained from the QCD multipole expansion, but their long-distance behavior was unknown. We have estimated the latter in the case of quarkonium and $1^{+-}$ hybrids. In order to do so, we have proposed an interaction Lagrangian of pions with the QCD string that is compatible with the chiral and string symmetries. We have matched the results for the amplitudes obtained with this Lagrangian to those obtained by an interaction Lagrangian of pions with quarkonium and hybrids.
The latter contains non-localities in the direction separating the heavy quark and the heavy antiquark, and hence it is a generalization of the BOEFT one, reducing to it in the limit $r\Delta E \ll 1$. We find that the long-distance LEF can be described in terms of three unknown constants. We proceed to estimate them using quarkonium transitions and assuming that they are dominated by long-distance effects. We use PDG and Belle data on the decay width as an input and obtain two sets of parameters.
We have also matched our Lagrangian to hadronic theories and compared our sets with the sets of parameters used there.
With those parameters, we have calculated the dipion invariant mass spectrum for a number of quarkonium and hybrid transitions of phenomenological interest. We obtain qualitative agreement with experimental results for most of the transitions, except for those involving d-waves either in the initial or final states. A particularly nice feature of our results is that they support the interpretation of $\Upsilon(10860)$ as a hybrid meson with a small mixture of quarkonium \cite{Oncala:2017hop,Oncala:2025mqj}.  

\section*{Acknowledgments}

We thank Roman Mizuk, Alex Bondar and, especially, Alexey Garmash for making Ref.~\cite{Bondar:2013xxx} available and providing clarifications on it, and Christoph Hanhart and Vadim Baru for clarifications on Refs. \cite{Chen:2016mjn,Baru:2020ywb}.
We also thank Llu\'\i s Garrido for discussions.
We acknowledge financial support from Grant No. 2021-SGR-249 from the Generalitat de Catalunya and from projects No. PID2022-136224NB-C21, PID2022-139427NB-I00 and No. CEX2024-001451-M from Ministerio de Ciencia, Innovaci\'on y Universidades (MICIU/AEI/10.13039/501100011033).
S.T.V. also acknowledges financial support from Grant PREP2022-000803 funded by MICIU/AEI/10.13039/501100011033 and, as
appropriate, by “ESF Investing in your future”, by “ESF+” or by “European Union
NextGenerationEU/PRTR”.

\appendix
\section{Operators} \label{ap:op}

Note that our entire analysis has been carried out with hybrid states in mind. However, the same considerations apply to any exotic quarkonium state with LDF quantum numbers $1^{+-}$, provided that the function $h(r,z)$ is modified accordingly.
For other exotic hadrons with different LDF quantum numbers $\kappa^{PC}$, the procedure remains the same, using the corresponding operators in the construction of the effective Lagrangian.
Here we present the leading-order operators for each $\kappa^{PC}$, see Table~\ref{t:op}. The isospin-violating operators of lower or equal order for each $\kappa^{PC}$ are listed in Table~\ref{t:op_iso}.

\begin{table}[htbp]
\centering
\begin{tabular}{ |c|c|c| } 
\hline
 $\kappa^{PC}$   & Operator & Order \\
 \hline
 \hline
 \multirow{4}{*}{$0^{++}$} & ${\rm Tr} (\partial_0 U^\dagger \partial^0 U)$ & \multirow{4}{*}{$\mathcal{O}(p^2)$} \\
                          & ${\rm Tr} (\partial_i U^\dagger \partial^i U)$ &  \\
                          & ${\rm Tr} (\mathcal{M} U^\dagger + U\mathcal{M}^\dagger)$ &  \\
                          & $r^i r^j{\rm Tr} (\partial_i U^\dagger \partial_j U)$ &  \\
\hline
$0^{+-}$ &  $r^i{\rm Tr} (\partial_0 U^\dagger \partial_i U )$ & $\mathcal{O}\left(p^2\right)$ \\
\hline
\multirow{3}{*}{$0^{-+}$} &  ${\rm Tr} ( \partial_i U^\dagger \partial^i U \partial_0 U^\dagger U )$  & \multirow{3}{*}{$\mathcal{O}\left(p^3\right)$} \\
                        &   $r^i r^j {\rm Tr} ( \partial_i U^\dagger \partial_j U \partial_0 U^\dagger U )$  &  \\
                        &  $ \epsilon_{ijk} r^i r^l {\rm Tr } \partial^j ( \partial^k U^\dagger \partial_l U )$  &  \\
\hline
\multirow{3}{*}{$0^{--}$} & $r^i{\rm Tr} (\partial_0 U^\dagger \partial^0 U\partial_i U^\dagger U)$ & \multirow{3}{*}{$\mathcal{O}\left(p^3\right)$} \\
                          & $r^i{\rm Tr} (\partial_j U^\dagger \partial^j U\partial_i U^\dagger U)$ &  \\
                           & $\epsilon_{ijk}r^i{\rm Tr} \left[ \partial^j(\partial^k U^\dagger \partial^0U ) \right] $ &  \\
\hline
\multirow{1}{*}{$1^{++}$} & $ \epsilon_{ijk} r^i r^l {\rm Tr }( \partial_l U^\dagger \partial_j U ) $ & \multirow{1}{*}{$\mathcal{O}\left(p^2\right)$} \\
\hline
$1^{+-}$ & $\epsilon_{ijk}r^i{\rm Tr}(\partial^j U^\dagger \partial^0 U)$ & $\mathcal{O}\left(p^2\right)$  \\
\hline
\multirow{2}{*}{$1^{-+}$} & ${\rm Tr}(\partial_0U^\dagger\partial_iU )$ & \multirow{2}{*}{$\mathcal{O}\left(p^2\right)$} \\
                        &  $r^ir^j{\rm Tr}(\partial_j U^\dagger\partial_0 U)$ &  \\
\hline
\multirow{5}{*}{$1^{--}$} & $r^i{\rm Tr}(\partial_0 U^\dagger\partial^0 U)$ & \multirow{5}{*}{$\mathcal{O}\left(p^2\right)$} \\
                        & $r^i{\rm Tr}(\partial_j U^\dagger\partial^j U)$ &  \\
                        & $r^i{\rm Tr}(\mathcal{M} U^\dagger + U\mathcal{M}^\dagger)$ &  \\
                        & $r^i{\rm Tr}(\partial_i U^\dagger\partial_j U)$ &  \\
                          & $r^ir^jr^k{\rm Tr}(\partial_j U^\dagger\partial_k U)$ &  \\
\hline
\end{tabular}
\caption{\justifying Lowest order operators for each LDF quantum numbers $\kappa^{PC}$ with explicit chiral order $\mathcal{O}\left(p^n\right)$ where $p$ is the pion momentum.}
\label{t:op}
\end{table}

\begin{table}[htbp]
\centering
\begin{tabular}{ |c|c|c| } 
\hline
 $\kappa^{PC}$   & Operator & Order \\
 \hline
 \hline
$0^{-+}$ & ${\rm Tr} (\mathcal{M} U^\dagger - U\mathcal{M}^\dagger)$ & $\mathcal{O}(p^2)$ \\
\hline
$0^{--}$ & $r^i {\rm Tr} ( \mathcal{M}^\dagger \partial_i U - \partial_i U^\dagger \mathcal{M} )$ & $\mathcal{O}\left(p^3 \right)$ \\
\hline
$1^{+-}$ & $r^i {\rm Tr} ( \mathcal{M}^\dagger U - U^\dagger \mathcal{M} )$ & $\mathcal{O}\left(p^2\right)$  \\
\hline
\end{tabular}
\caption{\justifying Isospin-violating lowest order operators for each LDF quantum numbers $\kappa^{PC}$ with explicit chiral order $\mathcal{O}\left(p^n\right)$ where $p$ is the pion momentum. }
\label{t:op_iso}
\end{table}

\section{Decay of higher string excitations}\label{ap:stringdecay}
The $N=2$ excitations consist of five states. A $\Sigma_g^+$ state, two degenerate $\Pi_g$ states, and two degenerate $\Delta_g$ states corresponding to the clockwise (R) and anticlockwise (L) rotations of $|L_z|=1$ and $|L_z|=2$ respectively. 

\bea
\label{eq:N2_states}
    |1\Pi_g^R\rangle &=& \sqrt{\frac{\sigma}{2E_2}}\beta_2^\dagger|0\rangle\hspace{0,2cm},\hspace{0,2cm} |1\Pi_g^L\rangle= -\sqrt{\frac{\sigma}{2E_2}}\alpha_2^\dagger|0\rangle \nn\\
    |2\Sigma_g^+\rangle &=& -\frac{\sigma}{2E_1}\alpha_1^\dagger\beta_1^\dagger|0 \rangle \\
    |1\Delta_g^R\rangle &=& \frac{\sigma}{2E_1}\beta_1^\dagger\beta_1^\dagger|0\rangle\hspace{0,2cm},\hspace{0,2cm} |1\Delta_g^L\rangle=\frac{\sigma}{2E_1}\alpha_1^\dagger\alpha_1^\dagger|0\rangle \nn
\eea

We compute their amplitudes using the manifest Lorenz and reparameterization invariant action from Eq.~\eqref{pionstring}. 
We will need  the expansion \eqref{expexp} and
\be
\label{eq:metricexp}
    \sqrt{-\text{det}(\partial_a x^\mu \partial_b x_\mu)}\simeq 1 - \frac{1}{2}\partial_ax^\mathsf{i}(z,t)\partial^ax^\mathsf{i}(z,t) \, ,
\ee
with $a=0,3$ and $\mathsf{i}=1,2$. 

For the $\Pi_g$ states, we need 
the first order term of the expansion \eqref{expexp} and the leading order term in the expansion \eqref{eq:metricexp}, as we did for the $\Pi_u$ states. We obtain,
\bea
  &&  \langle \pi(\Vec{q}),\pi(\Vec{p});0_{\text{EST}}|S_{\text{int}}|0_\pi;1\Pi_g^{R}\rangle\\
  &&=  \langle \pi(\Vec{q}),\pi(\Vec{p});0_{\text{EST}}|S_{\text{int}}|0_\pi;1\Pi_g^{L}\rangle^\ast\nn\\
  &&= \frac{16\pi\sqrt{2\pi}}{f_\pi^2\sqrt{\sigma}r}\Big[
  (\lambda+\eta)\left(E_qE_p + q_3 p^3\right) +  \lambda q_\mathsf{i} p^\mathsf{i} + \lambda' \frac{m_\pi^2}{2B_0}\Big]\nn \\
  &&\times \, \frac{\sin{\left[(q^3+p^3)\frac{r}{2}\right]}}{\frac{4\pi^2}{r^2}-(q^3+p^3)^2}i\left(\mathrm{q}+\mathrm{p}\right)\delta\left(E_q+E_p-\frac{2\pi}{r}\right) \nn\, ,
\eea

For the $\Delta_g$ states, we need to use the second order term in the expansion \eqref{expexp} and the leading order term in the expansion \eqref{eq:metricexp}. We obtain,
\bea
  &&  \langle \pi(\Vec{q}),\pi(\Vec{p});0_{\text{EST}}|S_{\text{int}}|0_\pi;1\Delta_g^{R}\rangle\\
  &&=  \langle \pi(\Vec{q}),\pi(\Vec{p});0_{\text{EST}}|S_{\text{int}}|0_\pi;1\Delta_g^{L}\rangle^\ast\nn\\
  &&= -\frac{32\pi^2}{f_\pi^2\sigma r^2}\Big[
  (\lambda+\eta)\left(E_qE_p + q_3 p^3\right) +  \lambda q_\mathsf{i} p^\mathsf{i} + \lambda' \frac{m_\pi^2}{2B_0}\Big]\nn \\
  &&\times \, \frac{\sin{\left[(q^3+p^3)\frac{r}{2}\right]}}{\frac{4\pi^2}{r^2}-(q^3+p^3)^2}\frac{\left(\mathrm{q}+\mathrm{p}\right)^2}{\left(q^3+p^3\right)}\delta\left(E_q+E_p-\frac{2\pi}{r}\right) \nn\, ,
\eea

Finally, for the $\Sigma_g^{+\prime}$ state there are two contributions. The first one arises from the leading order term in the expansion \eqref{eq:metricexp}  and the second order term in the expansion \eqref{expexp}, as in the previous case. The second one comes from the leading order term in the expansion \eqref{expexp} and the next-to-leading order term in the expansion \eqref{eq:metricexp}. We obtain,
\bea
  &&  \langle \pi(\Vec{q}),\pi(\Vec{p});0_{\text{EST}}|S_{\text{int}}|0_\pi;2\Sigma_g^+\rangle \\
  &&= \frac{16\pi^2}{f_\pi^2\sigma r^2}\Big[(\lambda+\eta)\left(E_qE_p + q_3 p^3\right) +  \lambda q_\mathsf{i} p^\mathsf{i} + \lambda' \frac{m_\pi^2}{2B_0}\Big]\nn \\
  &&\times \, \frac{\sin{\left[(q^3+p^3)\frac{r}{2}\right]}}{\left(q^3+p^3\right)}\Bigg[1-\frac{2|\mathrm{q}+\mathrm{p}|^2}{\frac{4\pi^2}{r^2}-(q^3+p^3)^2}\Bigg] \nn \\
  &&\times \, \delta\left(E_q+E_p-\frac{2\pi}{r}\right) \nn \, .
\eea

Unlike the previous $N=2$ string excitations, the $\Sigma_u^-$ $N=3$ string excitation is relevant for the lower lying hybrids \cite{Brambilla:2014eaa,Oncala:2017hop}. It corresponds to the state,
\be
    |\Sigma_u^-\rangle= \frac{\sigma}{2\sqrt{2E_1E_2}}\left(\alpha_1^\dagger\beta_2^\dagger-\beta_1^\dagger\alpha_2^\dagger\right)|0\rangle \, .
\ee

In order to generate a non-vanishing matrix element with the vacuum, we need terms that contain the same number of $\varphi$ and $\varphi^\ast$ and that they are odd under the exchange $\varphi\leftrightarrow\varphi^\ast$. This is not possible from the action \eqref{pionstring}, since all the terms with the same number of $\varphi$ and $\varphi^\ast$ arise from the expansion of the exponential in \eqref{expexp} and hence are even  under the  $\varphi\leftrightarrow\varphi^\ast$ exchange. A non-vanishing matrix element may arise from the action \eqref{delta_pionstring}. But in order to have the same number of $\varphi$ and $\varphi^\ast$ we need to expand the exponential \eqref{expexp} at least at first order, and hence the matrix element is $p/\lQ$ suppressed with respect  to the one of $\Pi_u$. We will not consider them in this work. Hence, at the order we are working,
\be
\label{eq:Su-}
    \langle \pi(\Vec{q}),\pi(\Vec{p});0_{\text{EST}}|S_{\text{int}}|0_\pi;\Sigma_u^-\rangle=0
\ee

\section{Form factors}\label{ap:ff}
As explained in Sec.~\ref{sec:qtq} it is convenient to take the z-axis in the $\mathbf{k}_+=\mathbf{q}+\mathbf{p}$ direction. Then, we can define $\mathbf{k}_-, \hat{\mathbf{r}}, E_\pm$ using \eqref{eq:krE}. In the rest of the paper we have called $\mathbf{k}_+\equiv \mathbf{k}$, $\theta_+\equiv \theta$, and $\phi_+\equiv \phi$ but, here, we make the subscripts explicit for clarification.

\subsection{Between excited quarkonium states}\label{ap:ff_q}

With this notation, \eqref{eq:M_QQ} reads
\bea
    &&\mathcal{M}(\vec P n l m\to {\vec P}' n' l' m'\,\pi(q) \pi(p))=\int d^3 \mathbf{r}\, \psi_{n'l'm'}^\ast(\mathbf{r})
    \nn \\
    &&\times \, \psi_{nlm}(\mathbf{r})\left( -\frac{8}{f_\pi^2}\right)\Bigg[ \frac{\lambda}{2}m_{\pi\pi}^2 + \left(-\lambda-\eta+\frac{\lambda'}{2B_0}\right)m_\pi^2 \nn\\
    &&+ \, \frac{\eta}{2}\left[ {E_+}^2 - \frac{1}{2}\left({k_+}^2 + {k_-}^2\right) \right] - \frac{\eta}{4}\Big[ {k_+}^2\cos^2\theta_+ - {k_-}^2 \nn \\
    &&\times \,\left(\cos\theta_+\cos\theta_- + \cos{(\phi_+-\phi_-)\sin\theta_+\sin\theta_-}\right)^2 \Big]\Bigg] \nn \\
    &&\times\, \frac{\sin{\left[k_+\cos\theta_+\frac{r}{2}\right]}}{k_+\cos\theta_+}\,,
\eea
with $E_+\equiv k_+^0$. For simplification purposes we introduce the following form factors,

\bea
    \label{eq:ff_q}
    &&\mathcal{F}_{00}= 1 = \sqrt{4\pi}Y_0^0(\theta_+,\phi_+) \\
    &&\mathcal{F}_{20} = \cos^2\theta_+ = \frac{1}{3}\left(\sqrt{\frac{16\pi}{5}}Y_2^0(\theta_+,\phi_+)+1\right) \nn \\
    &&\mathcal{F}_{2\pm1} = \cos\theta_+ \sin\theta_+ e^{\pm i\phi_+} = \mp\sqrt{\frac{8\pi}{15}}Y_2^{\pm1}(\theta_+,\phi_+) \nn \\
    &&\mathcal{F}_{2\pm2} = \sin^2\theta_+ e^{\pm i2\phi_+} = \sqrt{\frac{32\pi}{15}}Y_2^{\pm2}(\theta_+,\phi_+)  \nn \, ,
\eea
we obtain,

\bea
\label{eq:Mff_q}
    &&\mathcal{M}(\vec P n l m\to {\vec P}' n' l' m'\,\pi(q) \pi(p)) \\
    &&=\int d^3 \mathbf{r}\, \psi_{n'l'm'}^\ast(\mathbf{r})\psi_{nlm}(\mathbf{r}) \left( -\frac{8}{f_\pi^2}\right)\Bigg\{\big( c_{\pi\pi}m_{\pi\pi}^2 \nn \\
    && +\, c_\pi m_\pi^2 + c_E {E_+}^2\big)\mathcal{F}_{00} - c_E\left({E_+}^2 - m_{\pi\pi}^2\right)\mathcal{F}_{20} \nn \\
    &&   +\, c_E {k_-}^2 \Bigg[-\frac{1}{2}(\mathcal{F}_{00}+\mathcal{F}_{20}) +\frac{1}{2}(3\mathcal{F}_{20} - \mathcal{F}_{00})\cos^2\theta_- \nn \\
    &&  + \, \left(e^{-i\phi_-}\mathcal{F}_{21} + e^{i\phi_-}\mathcal{F}_{2-1} \right)\cos\theta_-\sin\theta_-  + \frac{1}{4}\sin^2\theta_- \nn \\
    && \times \, \left(e^{-i2\phi_-}\mathcal{F}_{22} + e^{i2\phi_-}\mathcal{F}_{2-2} \right) \Bigg]\Bigg\}\frac{\sin{\left[k_+\cos\theta_+\frac{r}{2}\right]}}{k_+\cos\theta_+} \, \nn \, ,
\eea
where $c_{\pi\pi}$, $c_{\pi}$ and $c_E$ are defined in \eqref{eq:c_param}. 
From here, and using \eqref{eq:Rlm},

\bea
\label{eq:Mq}
    &&\mathcal{M}(\vec P n l m\to {\vec P}' n' l' m'\,\pi(q) \pi(p))  \\
    &&= \left( -\frac{8}{f_\pi^2}\right)\Bigg\{\left( c_{\pi\pi}m_{\pi\pi}^2 + c_\pi m_\pi^2 + c_E {E_+}^2\right)R \nn\\
    &&- \, c_E\left({E_+}^2 - m_{\pi\pi}^2\right)R^0 + c_E {k_-}^2 \Bigg[-\frac{1}{2}(R+R^0)  \nn \\
    &&+ \, \frac{1}{2}(3 R^0 - R)\cos^2\theta_- + \left(e^{-i\phi_-}R^1 + e^{i\phi_-}R^{-1} \right) \nn \\
    &&\times\, \cos\theta_-\sin\theta_- + \frac{1}{4}\left(e^{-i2\phi_-}R^2 + e^{i2\phi_-}R^{-2} \right) \nn \\
    && \times\, \sin^2\theta_- \Bigg]\Bigg\}  \nn \,,
\eea
where we have written $R^\kappa$ instead of $R^\kappa_{n'l'm'nlm}(k_+)$ to simplify the notation.

\subsection{Between hybrids and excited quarkonium states}\label{ap:ff_h}

To compute $\left( \vec \psi_{nJMlm}(\mathbf{r}) \times \hat r \right)_k \propto \left( \vec \chi_\mu \times \hat r \right)_k$ we write $\hat r = \sum_{\nu=0,\pm1} (-1)^\nu \mathrm{r}_\nu \vec \chi_{-\nu}$ obtaining 
\be
\label{eq:chir}
    \big(\vec \chi_\mu \times \hat r\big)_k=i\sqrt{2}\sum_\nu (-1)^\nu \mathrm{r}_\nu C(111,\mu-\nu) \chi_{\mu-\nu, k}
\ee
where $\mathrm{r}_{\nu=\pm1}=\mp \frac{1}{\sqrt{2}}\sin\theta_+ e^{\pm i \phi_+}$, $\mathrm{r}_{\nu=0}=\cos\theta_+$ and  
\be
\label{eq:chi}
    \boldsymbol{\chi}_{\pm 1} = \mp \frac{1}{\sqrt{2}} \begin{pmatrix}
        1 \\
        \pm i\\
        0
    \end{pmatrix} 
    \hspace{0.5cm} , \hspace{0.5cm}
    \boldsymbol{\chi}_0 = \begin{pmatrix}
        0\\
        0\\
        1
    \end{pmatrix} \, .
\ee

Using $\mathbf{k}_\pm$ notation \eqref{eq:M_HQ} reads
\bea
\label{eq:Mff_h}
    &&\mathcal{M}(\vec P n J M l \to {\vec P}' n' l' m'\,\pi(q) \pi(p))   \\
    &&=\int dr r^2 R_{n'}^{l'}(r)R_{nJ}^l(r) \int d\Omega_+ Y_{l'}^{m'\ast}\left( 
    \frac{8\pi\sqrt{\pi}\eta}{f_\pi^2 \sqrt{\sigma}r^2}\right) \nn \\
    &&\times \, \frac{\cos{\left[k_+\cos\theta_+\frac{r}{2}\right]}}{k_+^2\cos^2\theta_+ - \frac{\pi^2}{r^2}}  \frac{i}{\sqrt{2}}\sum_{\mu ,\nu}C(l1J,M-\mu \mu)Y_l^{M-\mu} \nn \\
    &&\times \, (-1)^\nu \mathrm{r}_\nu C(111,\mu-\nu) \big( E_+ k_+^k - E_- k_-^k \big) \chi_{\mu -\nu, k} \nn \, .
\eea

By introducing the following form factors,
\bea
\label{eq:ff_h}
    &&\mathcal{F}_{10} = \cos\theta_+ = \sqrt{\frac{4\pi}{3}}Y_1^0(\theta_+,\phi_+)  \\
    &&\mathcal{F}_{1\pm 1} = \sin\theta_+e^{\pm i\phi_+} = \mp\sqrt{\frac{8\pi}{3}}Y_1^{\pm 1}(\theta_+,\phi_+) \, , \nn
\eea
and using the definitions \eqref{eq:Alm_h} and \eqref{eq:Rlm_h} the amplitude becomes
\bea
    &&\mathcal{M}(\vec P n J M l  \to {\vec P}' n' l' m'\,\pi(q) \pi(p)) \\
    &&=\frac{i16\pi\sqrt{\pi}c_E}{f_\pi^2 \sqrt{2\sigma}}\Bigg[\big(E_+k_+ - E_-k_-\cos\theta_-\big) \nn \\
    && \times \, \big(R^{11}+R^{-1-1}\big) + E_-k_-\sin\theta_- \bigg(e^{i\phi_-}R^{01} \nn \\
    && +  \,  e^{-i\phi_-}R^{0-1} 
    +  \frac{1}{\sqrt{2}}e^{i\phi_-}R^{-10}  
    - \frac{1}{\sqrt{2}}e^{-i\phi_-}R^{10} \bigg) \Bigg] \nn \, ,
\eea
with $R^{\kappa \mu}\equiv R_{n'l'm'nJMl}^{\kappa\mu}(k_+)$.

\section{Angular integrals}\label{ap:Alm}

\subsection{Between excited quarkonium states}\label{ap:AlmQQ}
The angular integrals $A_{l'm'lm}^\kappa(r,k)$ \eqref{eq:Alm} for the allowed dipion transitions between highly excited quarkonium states depending on the form factors in \eqref{eq:ff_q} are the following. Where $\kappa=0,\pm1,\pm2$ and the trivial case with form factor $\mathcal{F}_{00}=1$ is labeled without $\kappa$, $A_{l'm'lm}(r,k)$.

\be
A_{0000}= \frac{\mathrm{Si}(\frac{k r}{2})}{k}
\ee
\be
A_{0000}^0= \frac{-2 r k \cos\left(\frac{r k}{2}\right) + 4 \sin\left(\frac{r k}{2}\right)}{r^{2} k^{3}}
\ee
\be
A_{1010}=-\frac{6 \left(r k \cos\left(\frac{r k}{2}\right) - 2 \sin\left(\frac{r k}{2}\right)\right)}{r^{2} k^{3}}
\ee
\bea
A_{1010}^0&=&\frac{1}{r^{4} k^{5}}\Bigg[-6 r k \left(-24 + r^{2} k^{2}\right) \cos\left(\frac{r k}{2}\right) \nn \\
&+& 36 \left(-8 + r^{2} k^{2}\right) \sin\left(\frac{r k}{2}\right)\Bigg]
\eea
\bea
A_{1111}&=&A_{1-11-1}  \\
&=&\frac{3}{2 r^{2} k^{3}}\Bigg[2 r k \cos\left(\frac{r k}{2}\right) - 4 \sin\left(\frac{r k}{2}\right) \nn \\
&+& r^{2} k^{2} \mathrm{Si}\left(\frac{r k}{2}\right)\Bigg] \nn
\eea
\bea
A_{1111}^0&=&A_{1-11-1}^0  \\
&=&-\frac{12}{r^{4} k^{5}}\Bigg[6 r k \cos\left(\frac{r k}{2}\right) + \left(-12 + r^{2} k^{2}\right) \nn \\
&\times& \sin\left(\frac{r k}{2}\right)\Bigg] \nn
\eea
\bea
A_{1110}^1&=&A_{1011}^{-1}=-A_{101-1}^{1}=-A_{1-110}^{-1}  \\
&=&\frac{12 \sqrt{2}}{r^{4} k^{5}} \Bigg[ 6 r k \cos\left(\frac{r k}{2}\right) + \left(-12 + r^{2} k^{2}\right) \nn \\
&\times&\sin\left(\frac{r k}{2}\right)  \Bigg] \nn
\eea
\bea
A_{111-1}^2&=&A_{1-111}^{-2}  \\
&=&-\frac{3}{2 r^{4} k^{5}} \Bigg[ 2 r k \left(24 + r^{2} k^{2}\right) \cos\left(\frac{r k}{2}\right) \nn \\
&+& 4 \left(-24 + r^{2} k^{2}\right) \sin\left(\frac{r k}{2}\right) + r^{4} k^{4} \mathrm{Si}\left(\frac{r k}{2}\right)  \Bigg] \nn
\eea
\bea
A_{2000}&=&A_{0020}  \\
&=&-\frac{\sqrt{5}}{2 r^{2} k^{3}} \Bigg[6 r k \cos\left(\frac{r k}{2}\right) - 12 \sin\left(\frac{r k}{2}\right) \nn \\
&+& r^{2} k^{2} \mathrm{Si}\left(\frac{r k}{2}\right)\Bigg] \nn
\eea
\bea
A_{2000}^0&=&A_{0020}^0  \\
&=&-\frac{2\sqrt{5}}{r^{4} k^{5}} \Bigg[r k\left(-36 + r^{2} k^{2}\right)\cos\left(\frac{r k}{2}\right) \nn \\
&-& 8\left(-9 + r^{2} k^{2}\right)\sin\left(\frac{r k}{2}\right)
\Bigg] \nn
\eea
\bea
A_{2100}^1&=&A_{0021}^{-1}=-A_{002-1}^1=-A_{2-100}^{-1} \\
&=&\frac{4\sqrt{30}}{r^{4} k^{5}}\Bigg[6 r k \cos\left(\frac{r k}{2}\right) + \left(-12 + r^{2} k^{2}\right) \nn \\
&\times&\sin\left(\frac{r k}{2}\right)
\Bigg] \nn
\eea
\bea
A_{2200}^2&=&A_{0022}^{-2}=A_{002-2}^2=A_{2-200}^{-2} \\
&=&\frac{1}{2 r^{4} k^{5}}\sqrt{\frac{15}{2}}\Bigg[2 r k\left(24 + r^{2} k^{2}\right)\cos\left(\frac{r k}{2}\right) \nn \\
&+& 4\left(-24 + r^{2} k^{2}\right)\sin\left(\frac{r k}{2}\right) + r^{4} k^{4}\mathrm{Si}\left(\frac{r k}{2}\right)\Bigg] \nn
\eea
\bea
A_{2020}&=&\frac{5}{4 r^{4} k^{5}}\Bigg[-6 r k\left(-72 + r^{2} k^{2}\right)\cos\left(\frac{r k}{2}\right) \\
&+& 12\left(-72 + 7 r^{2} k^{2}\right)\sin\left(\frac{r k}{2}\right) + r^{4} k^{4}\mathrm{Si}\left(\frac{r k}{2}\right)\Bigg] \nn
\eea
\bea
A_{2020}^0&=&-\frac{10}{r^{6} k^{7}}\Bigg[r k \left(4320 - 144 r^{2} k^{2} + r^{4} k^{4}\right) \cos\left(\frac{r k}{2}\right) \nn\\
&-& 2 \left(4320 - 504 r^{2} k^{2} + 7 r^{4} k^{4}\right) \sin\left(\frac{r k}{2}\right)
\Bigg] 
\eea
\bea
A_{2120}^1&=&A_{2021}^{-1}=-A_{202-1}^1=-A_{2-120}^{-1}  \\
&=&\frac{20\sqrt{6}}{r^{6} k^{7}}\Bigg[18 r k \left(-40 + r^{2} k^{2}\right) \cos\left(\frac{r k}{2}\right)\nn \\
&+& \left(1440 - 156 r^{2} k^{2} + r^{4} k^{4}\right)\sin\left(\frac{r k}{2}\right)
\Bigg] \nn
\eea
\bea
A_{2121}&=&A_{2-12-1}  \\
&=&-\frac{60}{r^{4} k^{5}}\Bigg[6 r k \cos\left(\frac{r k}{2}\right) + \left(-12 + r^{2} k^{2}\right) \nn \\
&\times&\sin\left(\frac{r k}{2}\right)
\Bigg] \nn
\eea
\bea
A_{2121}^0&=&A_{2-12-1}^0  \\
&=&-\frac{60}{r^{6} k^{7}}\Bigg[2 r k \left(-240 + 7 r^{2} k^{2}\right) \cos\left(\frac{r k}{2}\right) \nn \\
&+& \left(960 - 108 r^{2} k^{2} + r^{4} k^{4}\right) \sin\left(\frac{r k}{2}\right)
\Bigg] \nn
\eea
\bea
A_{212-1}^2&=&A_{2-121}^{-2}  \\
&=&-\frac{480}{r^{6} k^{7}}\Bigg[r k \left(-60 + r^{2} k^{2}\right) \cos\left(\frac{r k}{2}\right) \nn \\
&-& 12 \left(-10 + r^{2} k^{2}\right) \sin\left(\frac{r k}{2}\right)
\Bigg] \nn
\eea
\bea
A_{202-2}^2&=&A_{2022}^{-2}=A_{2220}^2=A_{2-220}^{-2}  \\
&=&-\frac{5}{4r^{6} k^{7}}\sqrt{\frac{3}{2}}\Bigg[2 r k \left(5760 - 72 r^{2} k^{2} + r^{4} k^{4}\right) \nn \\
&\times& \cos\left(\frac{r k}{2}\right) + 4 \left(-5760 + 552 r^{2} k^{2} + r^{4} k^{4}\right) \nn \\
&\times& \sin\left(\frac{r k}{2}\right) + r^{6} k^{6} \mathrm{Si}\left(\frac{r k}{2}\right)
\Bigg] \nn
\eea
\bea
A_{2-12-2}^1&=&A_{2-22-1}^{-1}=-A_{2221}^{1}=-A_{2122}^{-1}  \\
&=&\frac{240}{r^{6} k^{7}}\Bigg[r k \left(-60 + r^{2} k^{2}\right) \cos\left(\frac{r k}{2}\right) \nn \\
&-& 12 \left(-10 + r^{2} k^{2}\right) \sin\left(\frac{r k}{2}\right)
\Bigg] \nn
\eea
\bea
A_{2222}&=&A_{2-22-2}  \\
&=&\frac{15}{8r^{4} k^{5}}\Bigg[2 r k \left(24 + r^{2} k^{2}\right) \cos\left(\frac{r k}{2}\right) \nn \\
&+& 4 \left(-24 + r^{2} k^{2}\right) \sin\left(\frac{r k}{2}\right) + r^{4} k^{4} \mathrm{Si}\left(\frac{r k}{2}\right)
\Bigg] \nn
\eea
\bea
A_{2222}^0&=&A_{2-22-2}^0  \\
&=&\frac{120}{r^{6} k^{7}}\Bigg[r k \left(-60 + r^{2} k^{2}\right) \cos\left(\frac{r k}{2}\right) \nn \\
&-& 12 \left(-10 + r^{2} k^{2}\right) \sin\left(\frac{r k}{2}\right)
\Bigg] \nn
\eea

\subsection{Between hybrids and excited quarkonium states}
\label{ap:AlmHQ}

The angular integrals $A_{l'm'JMl}^{\kappa\mu}(r,k)$ \eqref{eq:Alm_h} for the allowed dipion transitions between hybrids and highly excited quarkonium states depending on the form factors in \eqref{eq:ff_h} are the following. Where $\kappa=0,\pm 1$ and $\mu=0,\pm 1$.

\bea
    A_{00111}^{01}&=&-A_{001-11}^{0-1} =-\frac{1}{\sqrt{2}}A_{10110}^{01}=-\frac{1}{\sqrt{2}}A_{101-10}^{0-1} \nn\\
    &=&-\frac{1}{2r^3k^3}\sqrt{\frac{3}{2}}\bigg[  
    4\sin\left(\frac{rk}{2}\right)  +\pi\mathrm{Si}\!\left(\tfrac{1}{2}(\pi-rk)\right) \nn \\
    &-& \pi\mathrm{Si}\!\left(\tfrac{1}{2}(\pi+rk)\right)
    \bigg] 
\eea

\bea
    A_{00111}^{-10}&=&A_{001-11}^{10} = A_{00101}^{11}=A_{00101}^{-1-1} \nn \\
    &=&\frac{1}{\sqrt{2}}A_{11110}^{11}=\frac{1}{\sqrt{2}}A_{11100}^{10}=
    -\frac{1}{\sqrt{2}}A_{1-1100}^{-10} \nn \\
    &=&-\frac{1}{\sqrt{2}}A_{1-11-10}^{-1-1} \nn \\
    &=&\frac{\sqrt{3}}{4\pi\, r^{3}k^{3}}
    \Bigg[
    4\pi\,\sin\!\left(\frac{rk}{2}\right)
    +(\pi^2-r^2k^2) \nn\\
    &\times&\Big(
    \mathrm{Si}\!\left(\tfrac{1}{2}(\pi-rk)\right)
    -
    \mathrm{Si}\!\left(\tfrac{1}{2}(\pi+rk)\right)
    \Big)
    \Bigg] 
\eea

\bea
    A_{22111}^{10}&=&A_{2-21-11}^{-10}=\sqrt{\frac{5}{6}}A_{11112}^{-1-1}=-\sqrt{\frac{5}{6}}A_{1-11-12}^{11}\nn\\
    &=&-\frac{3}{8\pi\, r^{5}k^{5}}\sqrt{\frac{5}{2}}\Bigg[
    16\pi\, r k\,\cos\!\left(\frac{rk}{2}\right) \nn\\
    &+&4\pi\left(-8+\pi^{2}-r^{2}k^{2}\right)\sin\!\left(\frac{rk}{2}\right)\nn \\
    &+&\left(\pi^{2}-r^{2}k^{2}\right)^{2}\Big(
    \mathrm{Si}\!\left(\tfrac{1}{2}(\pi-rk)\right) \nn \\
    &-&
    \mathrm{Si}\!\left(\tfrac{1}{2}(\pi+rk)\right)
    \Big)
    \Bigg]
\eea

\bea
    A_{20111}^{01}&=&-A_{201-11}^{0-1}=-\sqrt{5}A_{10112}^{01}=-\sqrt{5}A_{101-12}^{0-1} \nn \\
    &=&\frac{1}{4 r^{5}k^{5}}\sqrt{\frac{15}{2}}
    \Bigg[
    -48\,r k\,\cos\!\left(\frac{rk}{2}\right) \nn \\
    &-&4\left(-24+3\pi^{2}+2r^{2}k^{2}\right)\sin\!\left(\frac{rk}{2}\right)
    \nn\\
    &-&\pi\left(3\pi^{2}-r^{2}k^{2}\right)\Big(
    \mathrm{Si}\!\left(\tfrac{1}{2}(\pi-rk)\right) \nn\\
    &-&
    \mathrm{Si}\!\left(\tfrac{1}{2}(\pi+rk)\right)
    \Big)
    \Bigg]
\eea

\bea
    A_{20111}^{-10}&=&A_{201-11}^{10}=A_{20101}^{11}=A_{20101}^{-1-1} \nn \\
    &=&\sqrt{5}A_{11112}^{11}=-\frac{\sqrt{5}}{2}A_{11102}^{10}=\frac{\sqrt{5}}{2}A_{1-1102}^{-10} \nn \\
    &=&-\sqrt{5}A_{1-11-12}^{-1-1} \nn \\
    &=&\frac{\sqrt{15}}{8\pi\, r^{5}k^{5}}
    \Bigg[
    48\pi\, r k\,\cos\!\left(\frac{rk}{2}\right) \nn\\
    &+&4\pi\left(-24+3\pi^{2}-r^{2}k^{2}\right)\sin\!\left(\frac{rk}{2}\right)
    \nn\\
    &+&\left(3\pi^{4}-4\pi^{2}r^{2}k^{2}+r^{4}k^{4}\right)\Big(
    \mathrm{Si}\!\left(\tfrac{1}{2}(\pi-rk)\right) \nn \\
    &-&
    \mathrm{Si}\!\left(\tfrac{1}{2}(\pi+rk)\right)
    \Big)
    \Bigg]
\eea

\bea
    A_{21111}^{11}&=&A_{2-11-11}^{-1-1}=\sqrt{2}A_{21101}^{0-1}= -\sqrt{2}A_{2-1101}^{01}\nn\\
    &=&
    \sqrt{\frac{10}{3}}A_{11102}^{0-1}=\sqrt{\frac{10}{3}}A_{1-1102}^{01} \nn \\
    &=&
    \sqrt{\frac{5}{3}}A_{10112}^{-10}
    =\sqrt{\frac{5}{3}}A_{10102}^{11}\nn\\
    &=&
    -\sqrt{\frac{5}{3}}A_{10102}^{-1-1}=
    -\sqrt{\frac{5}{3}}A_{101-12}^{10} \nn \\
    &=&-\frac{3\sqrt{5}}{4\, r^{5}k^{5}}
    \Bigg[
    16\, r k \cos\!\left(\frac{rk}{2}\right) + 4\left(-8 + \pi^2\right) \nn \\
    &\times&\sin\!\left(\frac{rk}{2}\right)
    + \pi (\pi^2 - r^2k^2) \Big(
    \mathrm{Si}\!\left(\tfrac{1}{2}(\pi-rk)\right) \nn \\
    &-& \mathrm{Si}\!\left(\tfrac{1}{2}(\pi+rk)\right)
    \Big)
    \Bigg]
\eea

\section{Dipion invariant mass spectrum}\label{ap:dw}
In order to obtain a general expression for the dipion invariant mass spectrum in \eqref{eq:dims0}, one can transform the integrals over the 3-momentum of the pions into 4-dimension integrals,
\be
\label{eq:int_corr}
    \int \frac{d^3\Vec{q}}{(2\pi)^3 2E_q}=\int \frac{d^4 q}{(2\pi)^3}\Theta( q^0)\delta(q^2-m_\pi^2) \, ,
\ee
and the same for the integral over $\vec p$. After that, by applying the change of variables $\mathbf{k}_\pm = \mathbf{q}\pm \mathbf{p}$, with the corresponding definitions in \eqref{eq:krE}, and rewriting the Dirac deltas as,
\bea
    &&\delta(q^2-m_\pi^2)\delta(p^2-m_\pi^2)\\
    &&= \, 4\delta(E_+^2 + E_-^2 - k_+^2 - k_-^2 - 4m_\pi^2) \nn \\
    &&\times \, \delta(E_+E_- - k_+k_-\cos\theta_-) \nn  \, ,
\eea
Eq.~\eqref{eq:dims0} reads,
\bea
    \frac{d\Gamma}{dm_{\pi\pi}^2}&=&\frac{1}{2^7\pi^5}\int d^4 k_+ \, d^4 k_- \, \Theta(E_+ + E_-) \\
    & \times& \Theta(E_+ - E_-) \lvert \mathcal{M}(k_+,E_+,k_-,E_-,\theta_-,\phi_-)\rvert^2 \nn\\
    &\times& \delta(E_+ + \Delta E)\delta(E_+^2 + E_-^2 - k_+^2 - k_-^2 - 4m_\pi^2) \nn \\
    &\times& \delta(E_+E_- - k_+k_-\cos\theta_-)\delta\left(m_{\pi\pi}^2- E_+^2 +k_+^2\right) \,. \nn
\eea

Since we take $\mathbf{k}_+$ as the $z$-direction to carry out the remaining angular integrals, the amplitude does not depend on the $\mathbf{k}_+$ angular variables, and hence this angular integral can be carried out trivially.  
The $E_+$ integration is also direct from $\delta(E_++\Delta E)$ with $\Delta E<0$,
\bea
    \frac{d\Gamma}{dm_{\pi\pi}^2}&=&\frac{1}{2^{5}\pi^{ 4}}\int dk_+ \, d^4 k_-\, k_+^2  \\
    &\times &  \Theta(- E_- - \Delta E)\Theta(E_- - \Delta E)\nn\\
    &\times & 
    \lvert \mathcal{M}(k_+,-\Delta E,k_-,E_-,\theta_-,\phi_-)\rvert^2 \nn \\
    &\times& \delta(\Delta E^2 + E_-^2 - k_+^2 - k_-^2 - 4m_\pi^2) \nn \\
    &\times& \delta(-\Delta EE_- - k_+k_-\cos\theta_-) \nn \\
    &\times& \delta\left(m_{\pi\pi}^2- \Delta E^2 +k_+^2\right) \,, \nn
\eea
The $E_-$ integration  gives two contributions that differ in the sign of the square root, 
\bea
    &&\frac{d\Gamma}{dm_{\pi\pi}^2}\bigg|_\pm = \frac{1}{2^{ 6}\pi^{ 4}}\int dk_+ \, d k_-\, d\Omega_- \, k_+^2k_-^2  \\
    &&\times \, {\Big| \mathcal{M}\Big(k_+,-\Delta E,k_-,\pm\sqrt{k_+^2 + k_-^2 + 4m_\pi^2 - \Delta E^2},{\theta_-,\phi_-}\Big)\Big|^2}  \nn \\
    &&\times \, \Theta\Big(-\sqrt{k_+^2 + k_-^2 + 4m_\pi^2 - \Delta E^2} - \Delta E\Big) \nn \\
    &&\times \, \frac{ \Theta (k_+^2 + k_-^2 + 4m_\pi^2 - \Delta E^2)}{\sqrt{k_+^2 + k_-^2 + 4m_\pi^2 - \Delta E^2}} \delta\left(m_{\pi\pi}^2- \Delta E^2 +k_+^2\right) \nn \\
    &&\times \, \delta(\mp\Delta E\sqrt{k_+^2 + k_-^2 + 4m_\pi^2 - \Delta E^2} - k_+k_-\cos\theta_-) \, . \nn
\eea

After doing the $\theta_-$  integral we obtain,
\bea
\label{eq:common_dw}
    &&\frac{d\Gamma}{dm_{\pi\pi}^2}{\bigg|_\pm} =\frac{1}{2^{ 6}\pi^{ 4}}\int dk_+ \, d k_-\, d\phi_- \, k_+k_-   \\
    &&\times  \, | \mathcal{ M_\pm}|^2  \Theta\Big(-\sqrt{k_+^2 + k_-^2 + 4m_\pi^2 - \Delta E^2} - \Delta E\Big) \nn \\
    &&\times \, \frac{ \Theta (k_+^2 + k_-^2 + 4m_\pi^2 - \Delta E^2)}{\sqrt{k_+^2 + k_-^2 + 4m_\pi^2 - \Delta E^2}} \delta\left(m_{\pi\pi}^2- \Delta E^2 +k_+^2\right)\nn \\
    &&\times \, \Theta \Bigg( 1 + \frac{\Delta E \sqrt{k_+^2 + k_-^2 + 4m_\pi^2 - \Delta E^2}}{k_+k_-} \Bigg)\nn \, ,
\eea
with 
${ \mathcal{M}_\pm}=\mathcal{M}\Big(k_+,-\Delta E,k_-,\pm\sqrt{k_+^2 + k_-^2 + 4m_\pi^2 - \Delta E^2},$ $\arccos\Big({\pm\frac{\Delta E \sqrt{k_+^2 + k_-^2 + 4m_\pi^2 - \Delta E^2}}{k_+k_-}}\Big),\phi_-\Big)$. By integrating $k_+$ one ends up with a general expression,

\bea
\label{eq:invmass}
    &&\frac{d\Gamma}{dm_{\pi\pi}^2}{\bigg|_\pm} =\frac{1}{2^{ 7}\pi^{ 4}}\int d k_-\, d\phi_- \, \frac{k_- |{ \mathcal{M}_\pm}|^2 }{\sqrt{ k_-^2 + 4m_\pi^2 - m_{\pi\pi}^2}}  \nn  \\
    &&\times  \, \Theta\Big(-\sqrt{k_-^2 + 4m_\pi^2 - m_{\pi\pi}^2} - \Delta E\Big) \Theta(\Delta E^2 -m_{\pi\pi}^2)\nn \\
    &&\times \, \Theta \Bigg( 1 + \frac{\Delta E \sqrt{ k_-^2 + 4m_\pi^2 - m_{\pi\pi}^2}}{k_-\sqrt{\Delta E^2 -m_{\pi\pi}^2}} \Bigg) \nn
    \\
    &&\times \, 
    \Theta (k_-^2 + 4m_\pi^2 - m_{\pi\pi}^2) \, .
\eea
The $\phi_-$ integral is immediate from the expressions of the amplitude \eqref{eq:Mq} and \eqref{eq:Mh} and leads to the same result for the $+$ and $-$ contributions,
\be
    \frac{d\Gamma}{dm_{\pi\pi}^2}= \frac{d\Gamma}{dm_{\pi\pi}^2}{\bigg|_+} +\frac{d\Gamma}{dm_{\pi\pi}^2}{\bigg|_-}=2\frac{d\Gamma}{dm_{\pi\pi}^2}{\bigg|_+}\,.
\ee
The
$k_-$ integration can be analytically done with integration limits obtained from the last two $\Theta$-functions
in \eqref{eq:invmass}.
\bea
    k_-\big|_\text{max} &=& -\Delta E \sqrt{1-\frac{4m_\pi^2}{m_{\pi\pi}^2}}\nn \\
    &&\\
    k_-\big|_\text{min} &=& \sqrt{m_{\pi\pi}^2 - 4m_\pi^2} \,.\nn
\eea

With this procedure we obtain Eq.~\eqref{eq:dipion_q} and \eqref{eq:dipion_h}.

\section{Spin average for the spin 1 case} \label{ap:spin1}
We define $|T(l'm',lm)|^2$ as the squared modulus of the spin 0 amplitude integrated
\bea
    &&|T(l'm',lm)|^2 = \\
    &&\int d \phi_- |\mathcal{M}(\vec P n l m\to {\vec P}' n' l' m'\,\pi(q) \pi(p))|^2 \nn \, ,
\eea
and write them in terms of the radial integrals $R^\kappa\equiv R_{n'l'm,nlm}^\kappa(k)$ depending if they are involved in transitions between $m'=m$, $m'=m\pm1$ or $m'=m\pm2$.
\bea
    &&|T(l'm,lm)|^2=\frac{8\pi}{f_\pi^4}\Bigg[  4R \big( c_\pi m_\pi^2 + c_{\pi\pi}m_{\pi\pi}^2 \big)   \nn \\
    &&+ \, c_E \Big[ 4{E_+}^2 \big( R - R^0 \big) + 4R^0 m_{\pi\pi}^2 +\big( R^0 -3R \big) {k_-}^2 \Big] \nn \\
    && - \, c_E \big( R - 3R^0 \big){k_-}^2 \cos{(2\theta_{-})}\Bigg]^2
\eea
\be
    |T(l'm\pm1,lm)|^2=\frac{32\pi}{f_\pi^4}c_E^2 \big[R^{\pm 1}\big]^2{k_{-}}^4 \sin^2{(2\theta_-)}
\ee
\be
    | T(l'm\pm2,lm)|^2= \frac{8\pi}{f_\pi^4} c_E^2 \big[R^{\pm 2}\big]^2 {k_-}^4 \sin^4{\theta_-}
\ee

For a transition between initial (final) state with quantum numbers $J^{PC}$ ($J'^{P'C'}$) with $J=l \, { \otimes} \, S$ ($J'=l' \, { \otimes} \, S$) total angular momentum, we define the spin average as,
\bea
    &&|\overline{T}(l'J'^{P'C'},lJ^{PC})|^2=\frac{1}{2J+1}\sum_{J=-M}^M\sum_{J'=-M'}^{M'}\sum_{\mu=-1}^{1}\nn \\
    && \, | C(l'1 J',M'-\mu \mu)C(l 1 J,M-\mu \mu) \nn\\
    && \times \,  T(l' M'-\mu,l M-\mu)|^2\,.
\eea

The spin averages for the relevant transitions between highly excited quarkonium states presented in this work, read:

\be
    |\overline{T}(01^{--},0 1^{--})|^2= |T(00,00)|^2
\ee
\bea
    |\overline{T}(21^{--},0 1^{--})|^2 &=&\frac{1}{5} \Big( |T(20,00)|^2 + 2|T(21,00)|^2 \nn \\
    & +& 2|T(22,00)|^2 \Big)
\eea
\bea
    |\overline{T}(22^{--},0 1^{--})|^2 &=&\frac{1}{3} \Big( |T(20,00)|^2 +2|T(21,00)|^2 \nn \\
    &+& 2|T(22,00)|^2 \Big)
\eea
\bea
    |\overline{T}(23^{--},0 1^{--})|^2 &=&\frac{7}{15} \Big( |T(20,00)|^2 +2|T(21,00)|^2 \nn \\
    &+& 2|T(22,00)|^2 \Big)
\eea
\bea
    |\overline{T}(01^{--},2 1^{--})|^2 &=&\frac{1}{5} \Big( |T(20,00)|^2  +2|T(21,00)|^2 \nn \\
    &+& 2|T(22,00)|^2 \Big)
\eea
\bea
    |\overline{T}(01^{--},2 2^{--})|^2 &=& \frac{1}{5} \Big( |T(20,00)|^2 +2|T(21,00)|^2 \nn \\
    &+& 2|T(22,00)|^2 \Big)
\eea
\bea
    |\overline{T}(01^{--},2 3^{--})|^2 &=& \frac{1}{5} \Big( |T(20,00)|^2 +2|T(21,00)|^2 \nn \\
    &+& 2|T(22,00)|^2 \Big)
\eea
\be
    |\overline{T}(10^{++},1 0^{++})|^2 = \frac{1}{9} | 2T(11,11) +T(10,10) |^2
\ee
\bea
    |\overline{T}(11^{++},1 1^{++})|^2 &=& \frac{1}{6} \Big( |T(10,10) + T(11,11)|^2  \nn \\
    &+& 2|T(11,11)|^2 + 2|T(11,10)|^2 \nn \\
    &+& |T(11,1-1)|^2 \Big)
\eea
\bea
    |\overline{T}(12^{++},1 2^{++})|^2 &=&\frac{1}{5} \bigg( \frac{1}{2}|T(10,10)+T(11,11)|^2 \nn \\
    &+& \frac{7}{6}|T(11,1-1)|^2 +  \frac{7}{3}|T(11,10)|^2 \nn \\
    &+&  \frac{1}{9}|2T(10,10)+T(11,11)|^2 \nn \\
    &+& 2|T(11,11)|^2 
    \bigg)   
\eea

In order to obtain the expressions above, we have also used the fact that $T(l'm',lm)$ obey the same relations as $A^\kappa_{l'm'lm}$ with respect to the $l'm'lm$ indices independently of $\kappa$. These relations are displayed in Appendix~\ref{ap:AlmQQ}.

\section{Full calculation of the dipion transitions between hybrids and highly excited quarkonium states} \label{ap:NLO}

In Section~\ref{sec:htq} , the dipion invariant mass spectrum was derived under the assumption that $r \ll 1/\Delta E$. This approximation isolates the term $\frac{\pi}{r}(E_q \mathrm{p} + E_p \mathrm{q})$ as the sole relevant contribution in Eq.~\eqref{eq:matchh}. This approach was highly advantageous, as this term maps exactly onto the BOEFT Lagrangian, yielding the analytical expression for the dipion invariant mass spectrum presented in Eq.~\eqref{eq:dipion_h}. 
However, as discussed in Sec.~\ref{sec:ff}, for an initial hybrid state with a $1^{+-}$ LDF, three operators become relevant at the same order in the chiral and string expansions. Consequently, if the strict condition $r \ll 1/\Delta E$ is relaxed, all three operators must be included.
This inclusion significantly complicates the analysis; employing the notation introduced in App.~\ref{ap:ff_h}, the full amplitude in \eqref{eq:M_HQ} becomes

\begin{widetext}
    \begin{equation}
    \begin{aligned}
    \label{eq:M_HQ_full}
    &\mathcal{M}(\vec P n J M l \to {\vec P}' n' l' m'\,\pi(q) \pi(p))=\int dr r^2 R^{l'}_{n'}(r)R^l_{nJ}(r) \int \Omega_+ Y^{m'\ast}_{l'}   \left(\frac{i4\sqrt{2\pi}}{f_\pi^2 \sqrt{\sigma}r}\right)\frac{\cos{\left[k_+\cos\theta_+\frac{r}{2}\right]}}{k_+^2\cos^2\theta_+ - \frac{\pi^2}{r^2}} \\
    &\times \sum_{\mu ,\nu= 0,\pm1} C(l1J,M-\mu \mu)Y_l^{M-\mu}(-1)^\nu \mathrm{r}_\nu C(111,\mu-\nu) \Bigg\{ \left( \frac{ \eta \pi  }{ r}  \right)\big( E_+ k_+^k - E_- k_-^k \big) 
    +\eta   k_+\cos\theta_+\hat{r}^l \big(k_{+l}k_+^k - k_{-l} k_+^k \big)  \\
    &+ \, k_+^k\Bigg[ \frac{\lambda}{2}m_{\pi\pi}^2 + \left(-\lambda-\eta+\frac{\lambda'}{2B_0}\right)m_\pi^2 
    -\frac{\eta}{4}\Big[ {k_+}^2\cos^2\theta_+  
    - {k_-}^2\left(\cos\theta_+\cos\theta_- + \cos{(\phi_+-\phi_-)\sin\theta_+\sin\theta_-}\right)^2 \Big]  \\
    &+ \,  \frac{\eta}{2}\left[ {E_+}^2 - \frac{1}{2}\left({k_+}^2 + {k_-}^2\right) \right] \Bigg] \Bigg\} \chi_{\mu -\nu, k}  \, ,
\end{aligned}
\end{equation}

\end{widetext}
with $\mathrm{r}_{\nu=\pm1}=\mp \frac{1}{\sqrt{2}}\sin\theta_+ e^{\pm i \phi_+}$, $\mathrm{r}_{\nu=0}=\cos\theta_+$ and $\vec{\chi}_{\mu}$ defined in \eqref{eq:chi}. If the decaying hybrid is in a $p_0$ state, the amplitude is suppressed because this state has only $\Sigma_u^-$ component, see Appendix~\ref{ap:stringdecay}. Then,
the allowed low-lying dipion transitions are still restricted to $(s/d)_1 \to p + \pi\pi$, $p_1 \to s + \pi\pi$, and $p_1 \to d + \pi\pi$. However, the amplitude given in Eq.~\eqref{eq:M_HQ_full} allows for transitions to $m'$ states that previously vanished for the same $l'$.
Finally, by applying the procedure outlined in App.~\ref{ap:dw} to specific transitions, we obtain analytical, albeit highly non-trivial, expressions for the dipion invariant mass spectrum of each transition. Deriving a generalized expression valid for arbitrary transitions, such as the one presented in \eqref{eq:dipion_h}, analytically proved intractable.

\begin{figure*}
\begin{center}
\includegraphics[height=0.25\textheight,width=0.48\textwidth]{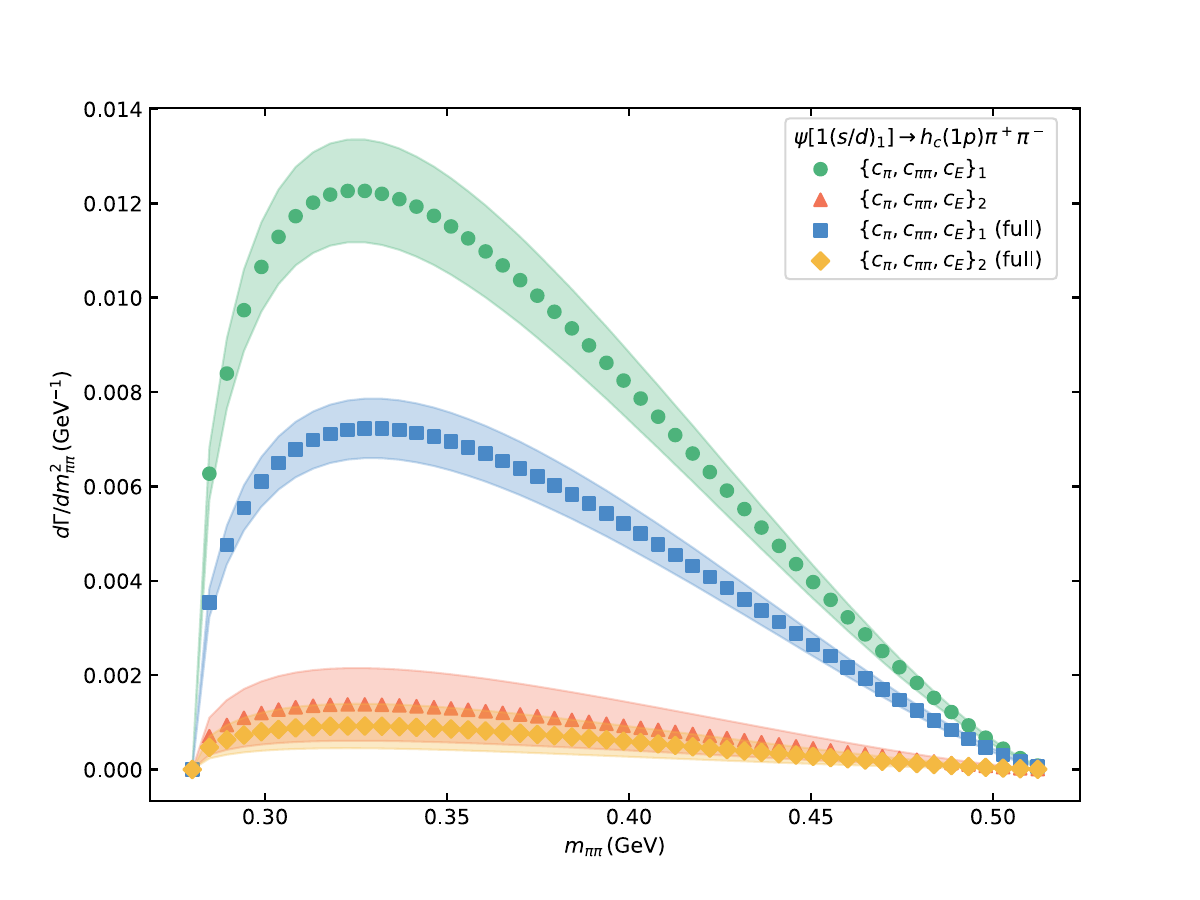}
\includegraphics[height=0.25\textheight,width=0.48\textwidth]{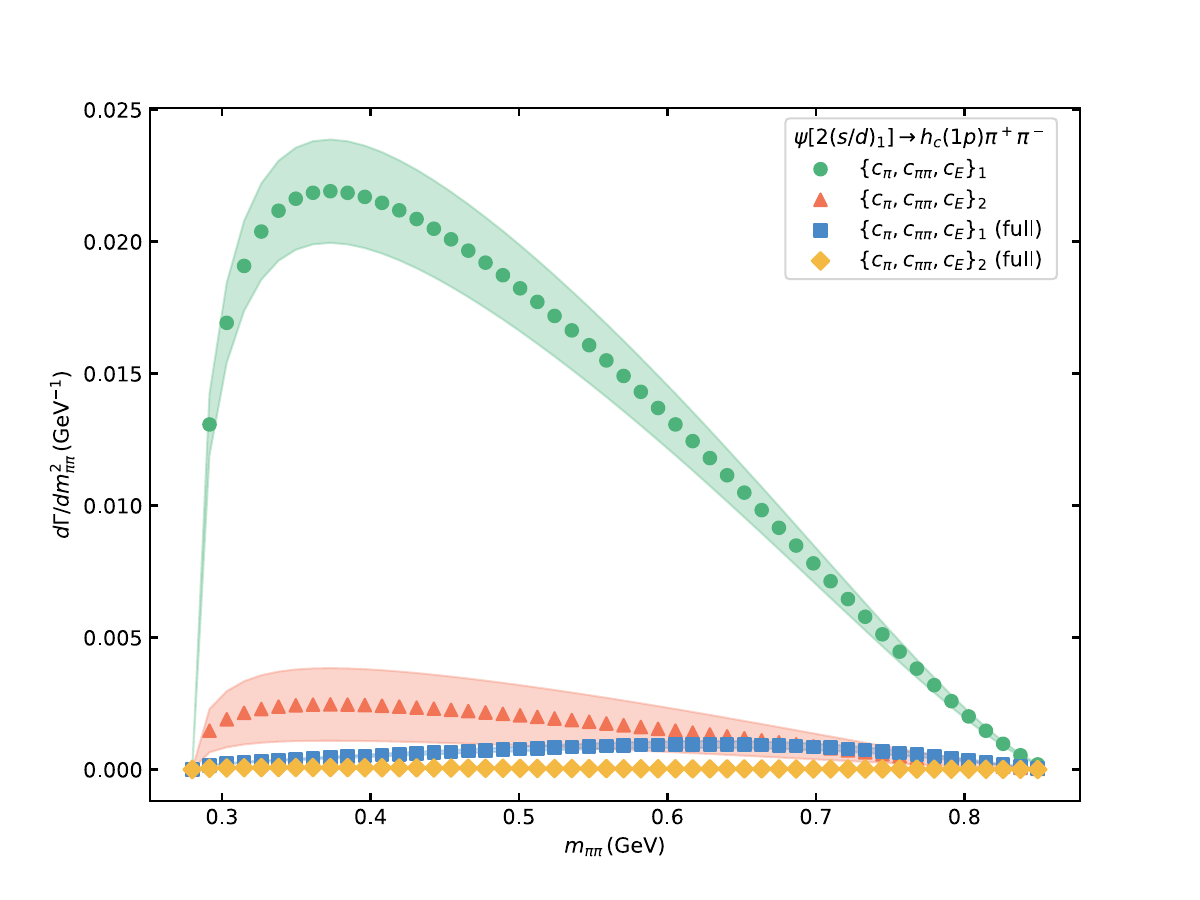} 
\caption{\justifying Comparison between the dipion invariant mass spectrum obtained for charmonium spin-0 transitions from $n(s/d)_1$ ($n=1,2$) hybrid states to $1p$ quarkonium state using \eqref{eq:dipion_h} (green circular and red triangular points) and using the full expression derived from \eqref{eq:matchh} (blue squared and orange diamond points). 
The sets $\{\,c_\pi,\; c_{\pi\pi},\; c_E\,\}_1$ and $\{\,c_\pi,\; c_{\pi\pi},\; c_E\,\}_2$  in the inset are given in \eqref{eq:param1} and \eqref{eq:param2} respectively. The green circular and red triangular points are the ones plotted in Fig. \eqref{fig:SDtoP-c}. 
}
\label{fig:SDtoP-c-full}
\end{center}
\end{figure*}

\begin{figure*}
\begin{center}
\includegraphics[height=0.25\textheight,width=0.48\textwidth]{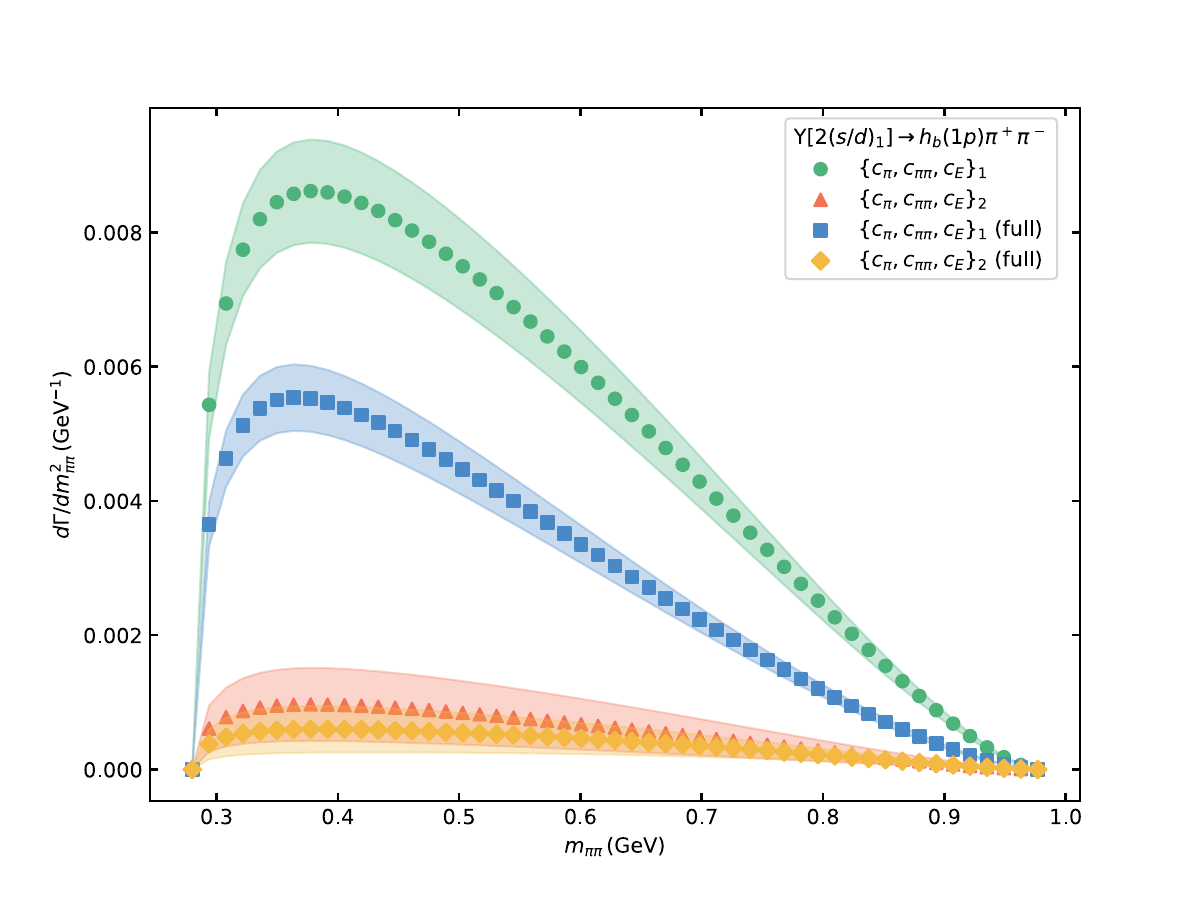}
\includegraphics[height=0.25\textheight,width=0.48\textwidth]{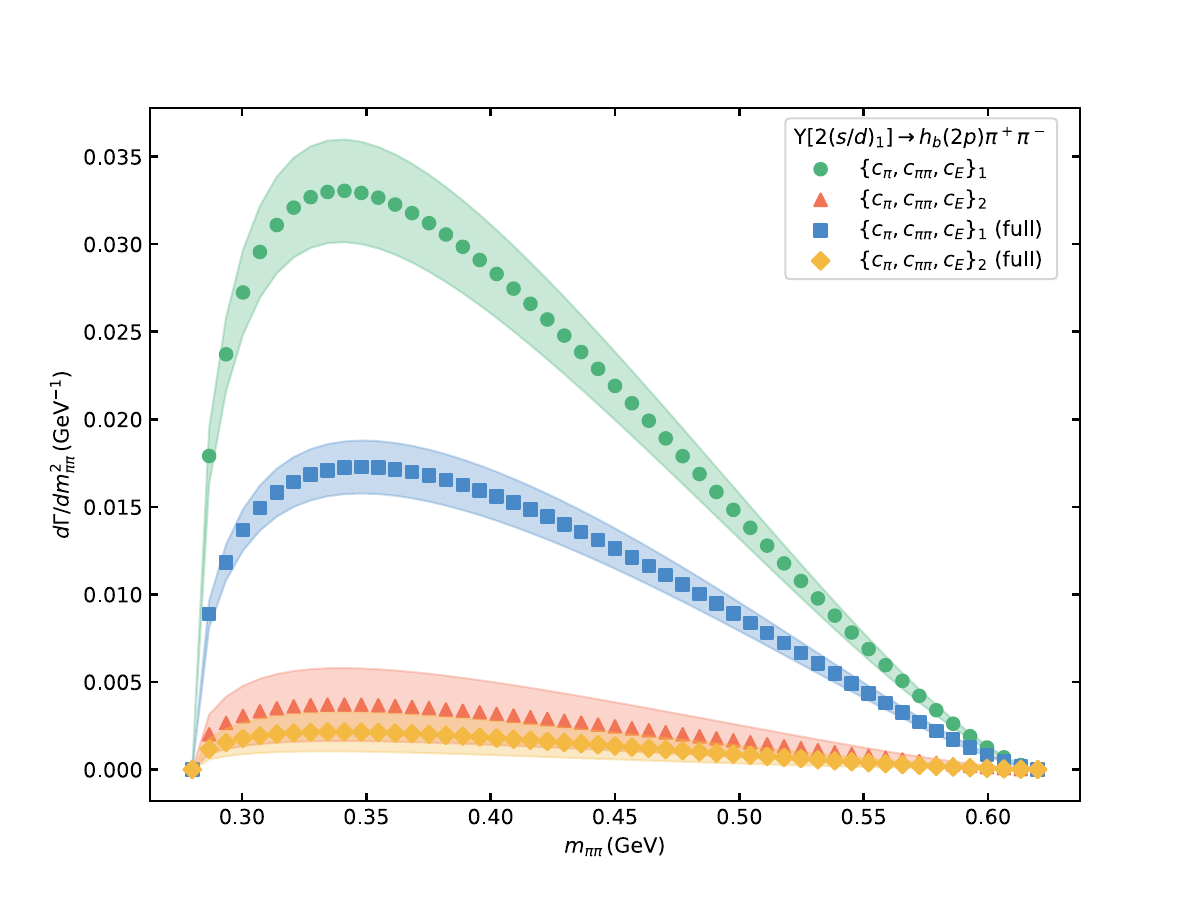} 
\caption{\justifying Comparison between the dipion invariant mass spectrum obtained for bottomonium spin-0 transitions from $2(s/d)_1$ hybrid states to $np$ ($n=1,2$) quarkonium states using \eqref{eq:dipion_h} (green circular and red triangular points) and using the full expression derived from \eqref{eq:matchh} (blue squared and orange diamond points). 
The sets $\{\,c_\pi,\; c_{\pi\pi},\; c_E\,\}_1$ and $\{\,c_\pi,\; c_{\pi\pi},\; c_E\,\}_2$  in the inset are given in \eqref{eq:param1} and \eqref{eq:param2} respectively. The green circular and red triangular points are the ones plotted in Fig. \eqref{fig:SDtoP-b}}
\label{fig:SDtoP-b-full}
\end{center}
\end{figure*}

\begin{figure}[htbp]
\includegraphics[height=0.25\textheight,width=0.48\textwidth]{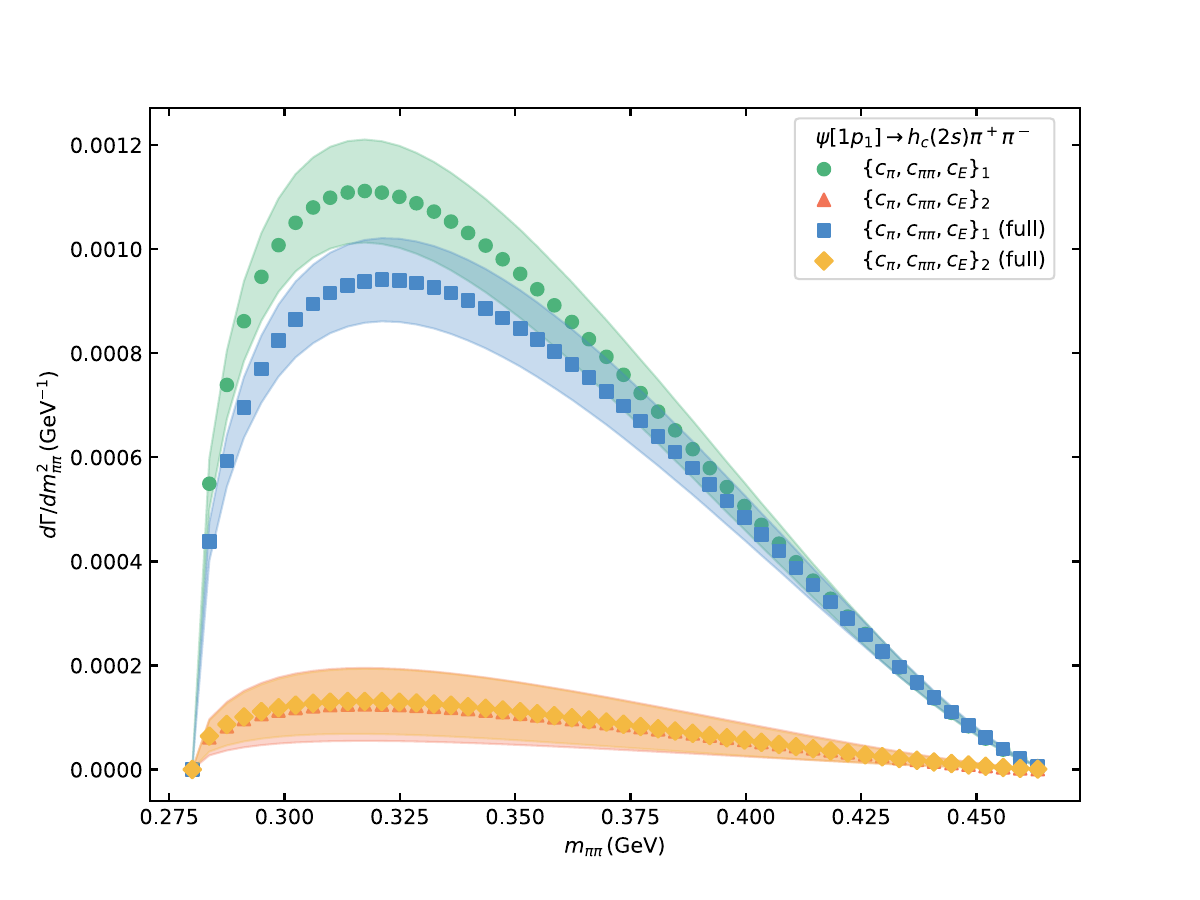} 
\caption{\justifying Comparison between the dipion invariant mass spectrum obtained for charmonium spin-0 $1p_1\to 2s + \pi^+\pi^-$ transition using \eqref{eq:dipion_h} (green circular and red triangular points) and using the full expression derived from \eqref{eq:matchh} (blue squared and orange diamond points). 
The sets $\{\,c_\pi,\; c_{\pi\pi},\; c_E\,\}_1$ and $\{\,c_\pi,\; c_{\pi\pi},\; c_E\,\}_2$  in the inset are given in \eqref{eq:param1} and \eqref{eq:param2} respectively. The green circular and red triangular points are the ones plotted in Fig. \eqref{fig:P1toS}}
\label{fig:P1toS-full}
\end{figure}

In this section, we explore numerically how the dipion invariant mass spectrum and decay width change when all the terms in Eq.~\eqref{eq:matchh} are included. We reproduce Figs.~\ref{fig:SDtoP-c}, \ref{fig:SDtoP-b}, \ref{fig:P1toS} and Table~\ref{t:dwhybrid}, and obtain Figs.~\ref{fig:SDtoP-c-full}, \ref{fig:SDtoP-b-full}, \ref{fig:P1toS-full} and Table~\ref{t:dwhybrid-full}. 
We observe that including these additional terms lowers the values of the decay widths and the dipion invariant mass spectrum by less than a factor of $2$ in all transitions but the charmonium $2(s/d)_1\to 1p$, in which they drop by an order of magnitude.

\begin{table}[htbp]
\centering
\begin{tabular}{ |c||c|c| } 
 \hline
 \multicolumn{3}{|c|}{$\Gamma_i$ (keV)} \\
 \hline
 Transitions + $\pi^+\pi^-$ & $\{\,c_\pi,\; c_{\pi\pi},\; c_E\,\}_1$ & $\{\,c_\pi,\; c_{\pi\pi},\; c_E\,\}_2$ \\
 \hline
 $\psi \left[ 1(s/d)_1 \right] \to h_c(1p)$   & $746 \pm 66$    & $89 \pm 47$ \\
 $\psi\left[ 1p_1 \right] \to \eta_c(2s)$   & $73.6 \pm 6.4$    & $9.6 \pm 4.8$ \\
 $\psi \left[ 2(s/d)_1\right] \to h_c(1p)  $   & $415 \pm 53 $    & $19 \pm 11 $ \\
 \hline
 $\Upsilon \left[2(s/d)_1 \right] \to h_b(1p)$   & $2032 \pm 170$    & $276 \pm 140$ \\  
 $\Upsilon \left[ 2(s/d)_1 \right] \to h_b(2p)$   & $2872 \pm 260$    & $320 \pm 180$ \\  
 \hline
\end{tabular}
\caption{\justifying Full calculation of the dipion decay widths for relevant transitions between charmonium hybrids (up) and bottomonium hybrids (down) and highly excited quarkonium states using the two sets of parameters \eqref{eq:param1} and \eqref{eq:param2} and the full dipion invariant mass spectrum obtained from \eqref{eq:matchh}. We include the uncertainties generated by the errors on the parameters $\{\,c_\pi,\; c_{\pi\pi},\; c_E\,\}$. These figures are to be compared with those in Table \ref{t:dwhybrid}. }
\label{t:dwhybrid-full}
\end{table}

\section{Estimation of the non-resonant amplitude}\label{ap:non-resonant}
The Belle Collaboration observed that the decay rates for $\Upsilon(5s)\to\Upsilon(ns)\pi^+\pi^-$ $(n=1,2,3)$, are unexpectedly high, exceeding those of transitions between lower $\Upsilon(ns)$ by approximately two orders of magnitude \cite{Belle:2007xek,PhysRevLett.108.122001}. Such a significant enhancement strongly indicates the involvement of exotic intermediate states in the $\Upsilon(5s)$ decay chain. Consequently, a formal procedure to determine the decay width for $\Upsilon(5s)\to\Upsilon(3s)\pi^+\pi^-$ is required. 
Following \cite{Voloshin:2007dx,Voloshin:2006ce} the non-resonant (NR) amplitude for a quarkonium transition $A\to B+\pi(q)\pi(p)$ is parametrized as

\be
    \mathcal{A}^{\rm NR}= a_1^{\rm NR} e^{i\delta_1^{\rm NR}} + a_2^{\rm NR} e^{i\delta_2^{\rm NR}}m_{\pi\pi}^2(q,p) \, .
\ee

This parameterization is much simpler that the one we obtain in \eqref{eq:MQQ}, and even simpler than the one that emanates from hadronic theories \eqref{eq:MBonn} \cite{Chen:2016mjn,Baru:2020ywb}, since it does not depend on the kinematic variable $k_-$. Hence, we are only going to use the decay with that follows from it.
Comparing the decay width from this amplitude

\bea
    \Gamma &=&\frac{1}{2m_A}\int \frac{d^3 \vec{p}}{(2\pi)^3 2E_p}\frac{d^3 \vec{q}}{(2\pi)^3 2E_q}\frac{d^3 \vec{p}_B}{(2\pi)^3 2E_B} \nn\\
    && \times \, |\mathcal{A}|^2(2\pi)^4 \delta(-E_A + E_B+E_p+E_q) \nn \\
    &&\times \, \delta\left(\vec{0}+\vec{p}_B +{\vec{p}}+\vec{q}\right) \nn \\ 
    &=&\frac{1}{4m_Am_B}\int \frac{d^3 \vec{p}}{(2\pi)^3 2E_p}\frac{d^3 \vec{q}}{(2\pi)^3 2E_q} \nn\\
    && \times \, |\mathcal{A}|^2 (2\pi) \delta(\Delta E +E_p+E_q) \, ,
\eea
with \eqref{eq:dims0}, relates the non-resonant relativistic amplitude $\mathcal{A}^{\rm NR}$ to our  non-relativistic amplitude as follows: $|\mathcal{M}^{\rm NR}|=\frac{|\mathcal{A}^{\rm NR}|}{ 2\sqrt{m_Am_B}}$. 
Hence,

\bea
    &&|\mathcal{M}^{\rm NR}|^2= \frac{1}{4m_Am_B} \bigg[ \left(a_1^{\rm NR} \right)^2 + \left(a_2^{\rm NR} \right)^2\left(m_{\pi\pi}^2(q,p)\right)^2  \nn \\
    &&+ \, 2 a_1^{\rm NR} a_2^{\rm NR}\cos\left(\delta_2^{\rm NR} -\delta_
    1^{\rm NR}\right)m_{\pi\pi}^2(q,p)  \bigg] \, .
\eea

Using $|\mathcal{M}^{\rm NR}|^2$, we obtain $\Gamma^{\rm NR}$ following the procedure in App.~\ref{ap:dw} up to Eq.~\eqref{eq:common_dw} where the integral over $\phi_-$ is trivially $2\pi$ but we do not have $\delta\left(m_{\pi\pi}^2-\Delta E^2 + k_+^2\right)$ to integrate $k_+$. However, $k_+$ and $k_-$ can be easily integrated in polar coordinates by defining $k_+ = \rho \cos \omega$ and $k_- = \rho \sin \omega$. The integration limits of $\rho$ and $\omega$ are given by the $\Theta$-functions,
\bea
    \rho\big|_\text{max} &=& \sqrt{2\Delta E^2 + 4\Delta E m_{\pi}}\nn \\
    \rho\big|_\text{min} &=& \sqrt{\Delta E^2 - 4m_{\pi}^2}   \, ,
\eea
\bea
    \omega\big|_\text{max} &=& \frac{\pi}{2}- \frac{1}{2}\arcsin S(\rho) \nn \\
    \omega\big|_\text{min} &=& \frac{1}{2}\arcsin S(\rho)  \, .
\eea
with $\Delta E\equiv E_B-E_A$ and $S(\rho)\equiv\frac{-2\Delta E\sqrt{\rho^2+4m_\pi^2-\Delta E^2}}{\rho^2} $.

Reference \cite{Belle:2014vzn} provides data for several models used to fit the transition; for this study, we adopt Model-0 (designated as the nominal model). This model is selected because it sets the amplitudes of potential contributions from the $f_0(980)$ scalar and $f_2(1270)$ tensor states to zero. 
The parameters of this model for the $\Upsilon(5s)\to\Upsilon(3s)\pi^+\pi^-$ transition are $a_1^{\rm NR}=1.03\pm 0.69$, $\delta_1^{\rm NR}= 73\pm 13$ $^{\circ}$, $a_2^{\rm NR}=10.7\pm3.5$ GeV$^{-2}$, and $\delta_2^{\rm NR}=196\pm14$ $^{\circ}$.
With those, we obtain  $\Gamma\left(\Upsilon(5s) \to \Upsilon(3s)\pi^+\pi^-\right)|_{\rm non-resonant}=0.2316^{+0.1791}_{-0.1161}$ keV.

\section{Comparison with hadronic effective theories }\label{ap:comparison}

Hadronic theories involving heavy hadrons are often used for phenomenological analysis (see \cite{Casalbuoni:1996pg} for a review). The BOEFT provides the unknown low energy constant for hadronic theories with fields describing particles with two heavy quarks in terms of expectation values of suitable LEF. Let us illustrate how \eqref{boeftq} reduces to a hadronic theory in the case of $S$-wave quarkonium. We simply write 
\be 
    S(\mathbf{R},\mathbf{r},t)=\sum_{n} \psi_{n00}(\mathbf{r}) J_n(\mathbf{R},t)\,,
\label{BOtohadron}
\ee
where $\psi_{nlm}(\mathbf{r})$ is the wave function of the heavy quarkonium and $J_n(\mathbf{R},t)$ is the heavy quark symmetry multiplet containing the $\Upsilon (nS)$ and the $\eta_b (nS)$ fields, $J_n(\mathbf{R},t) =(\eta_{b\, n}(\mathbf{R},t)\mathbb{I}_2 +  \Upsilon_n^j(\mathbf{R},t)\sigma^j)/\sqrt{2}$. We have restricted ourselves to $S$-wave states by limiting the sum to zero orbital angular momentum ($l=0$, $m=0$). Then,  we have from \eqref{boeftq},
\bea
    \left.{\rm L}_{\rm int}\right|_{S-{\rm wave}}&=&\int d^3 \mathbf{R} \sum_{n m}\,\mathrm{Tr}\Big[J_m^\dagger(\mathbf{R},t)\\
    &\times& \big( g_0^{n m}\partial_0 U^\dagger\partial^0 U + g_1^{n m}\partial_i U^\dagger\partial^i U
    \nn \\
    &+&g_3^{n m}\left(U^\dagger\mathcal{M}+\mathcal{M}^\dagger U\right)\big)J_n(\mathbf{R},t)\Big] \nn \,,
\label{boefthadron}
\eea
where $U=U(\mathbf{R},t)$ and
\bea
    g_k^{n m}&=&\int d^3 \mathbf{r}\psi_{m00}^\ast(\mathbf{r}) g_k(r) \psi_{n00}(\mathbf{r}) \, ,\quad   k=0,3 \\
    g_1^{n m}&=&\int d^3 \mathbf{r}\psi_{m00}^\ast(\mathbf{r}) \left(g_1(r)-\frac{g_2(r)}{3}\right) \psi_{n00}(\mathbf{r})\,.\nn
\eea
$g_1^{n m}$ corresponds to $c_1$ and $g_0^{n m}$ to $c_1 + c_2$ in the notation of Refs. \cite{Chen:2016mjn,Baru:2020ywb}. $g_3^{n m}$ is omitted in those references, although it is taken into account in more recent work \cite{Chen:2025jip}. The term proportional to $g_3^{n m}$ breaks chiral symmetry but it is of the same chiral order as $g_0^{n m}$ and $g_1^{n m}$. In the following we  compare our results with those obtained with the $c_1$ and $c_2$ of Refs. \cite{Chen:2016mjn,Baru:2020ywb}.

Working the Lagrangian from \cite{Chen:2016mjn} we obtain the following amplitude which using our notation in \eqref{eq:krE} reads
\bea
\label{eq:MBonn}
    \mathcal{M}&=&\left(-\frac{4}{f_\pi^2}\right)\Big[\left(\frac{c_1}{2}+\frac{c_2}{4}\right)m_{\pi\pi}^2 - (c_1+c_2)m_\pi^2 \nn \\
    &+& \frac{c_2}{4}E_+^2 - \frac{c_2}{4}k_-^2  \Big] \, .
\eea
This has the form of our amplitude \eqref{eq:MQQ} in the BO limit $r\Delta E\ll 1$, but it depends on two parameters, $c_1$ and $c_2$, rather than three because of the omission of the chiral symmetry breaking operator mentioned above. 
Table~\ref{t:ci} shows the values of $c_i$ with $i=1,2$ from \cite{Chen:2016mjn} and \cite{Baru:2020ywb}, which are related to our $c_\pi$, $c_{\pi\pi}$ and $c_E$ in the following way
\bea
\label{eq:syst}
    c_\pi &=& -\frac{c_1+c_2}{\langle r \rangle } \nn \\
    c_{\pi\pi} &=& \frac{1}{2\langle r \rangle }\left( c_1 + \frac{c_2}{4} \right)\\
    c_E &=& \frac{3}{8}\frac{c_2}{\langle r \rangle} \nn \, ,
\eea
where $\langle r \rangle$ is computed with our wave functions and gives for each transition 
\bea 
\label{eq:<r>}
    \langle \Upsilon(2s) | \,  r \, | \Upsilon(4s) \rangle &=& -0.2578 \, {\rm GeV}^{-1}  \nn \\
    \langle \Upsilon(2s) | \,  r \,  | \Upsilon(5s) \rangle &=& -0.1386 \, {\rm GeV}^{-1}  \\
    \langle \Upsilon(3s) | \,  r \,  | \Upsilon(5s) \rangle &=& -0.3248 \, {\rm GeV}^{-1} \, . \nn
\eea 
Results for the $\{\,c_\pi,\; c_{\pi\pi},\; c_E\,\}$ parameters obtained using $c_i$ ($i=1,2$) and the relations above is also shown in Table~\ref{t:ci}. We observe that the values obtained depend on the process rather than being universal, which is probably a consequence of the missing operator mentioned above.

If we follow App.~\ref{ap:dw} the dipion invaraint mass in terms of $c_1$ and $c_2$ reads,
\bea
\label{eq:dipion_c1c2}
    && \frac{d\Gamma}{dm_{\pi\pi}^2}= \frac{\sqrt{\Delta E^2 - m_{\pi\pi}^2}}{480f_\pi^4\pi^3m_{\pi\pi}^4} \sqrt{1-\frac{4m_\pi^2}{m_{\pi\pi}^2}}  \\
    &&\times \, \Bigg\{ m_{\pi\pi}^4 \Big[ 3c_2^2 \big(m_{\pi\pi}^2 - 4m_\pi^2\big)^2  + 60c_1^2\big(m_{\pi\pi}^2 - 2m_\pi^2\big)^2 \nn \\
    && + \, 
    20c_1c_2 \big( 8m_\pi^4 - 6m_\pi^2m_{\pi\pi}^2 + m_{\pi\pi}^4 \big)
    \Big] + 4c_2m_{\pi\pi}^2\Delta E^2 \nn \\
    && \times \,  \Big[ \big(10c_1 +c_2\big)m_{\pi\pi}^4 -8\big(5c_1 + 3c_2\big)m_\pi^4 + 2c_2m_\pi^2 m_{\pi\pi}^2 
    \Big] \nn \\
    &&+ \, 8c_2^2 \big( 6m_\pi^4 + 2m_\pi^2m_{\pi\pi}^2 + m_{\pi\pi}^4 \big)\Delta E^4
    \Bigg\} \, . \nn
\eea
 In Table~\ref{t:dw_ci} we present the values obtained for the decay width for each transition in  \cite{Chen:2016mjn} and \cite{Baru:2020ywb}, using their parameters $c_i$ with $i=1,2$ and \eqref{eq:dipion_c1c2}, and the ones obtained when using our expression for the dipion invariant mass spectrum \eqref{eq:dipion_q} and $\{\,c_\pi,\; c_{\pi\pi},\; c_E\,\}$ in Table~\ref{t:ci}. In the BO limit, both results become equal. Hence, the difference provides an estimate of the effect of the non-localities emanating from the EST, which ranges within $2\%$ and $30\%$ depending on the transition.

\begin{table*}[t] 
\centering
\begin{tabular}{ |c|c|c|c|c|c|c| } 
\hline
 Transitions + $\pi^+\pi^-$   & $c_1$ (GeV$^{-1}$) & $c_2$ (GeV$^{-1}$) & $c_\pi$ & $c_{\pi\pi}$ & $c_E$ & Ref. \\
 \hline
$\Upsilon(4s)\to \Upsilon(2s) $ & $(1.2\pm0.6)\times 10^{-1}$ &  $(-1.0\pm0.6)\times 10^{-1}$ & $0.0776$ & $-0.1843$ & $0.1455$ &  \cite{Chen:2016mjn} \\
\hline
$\Upsilon(5s)\to \Upsilon(2s) $ &  $(21.1\pm0.6)\times 10^{-4}$ &  $(-12.6 \pm 0.5)\times 10^{-4}$ & $0.0061$ & $-0.0065$ & $0.0034$ & \cite{Baru:2020ywb} \\
\hline
$\Upsilon(5s)\to \Upsilon(3s) $ &  $(-17.8\pm 3.1)\times 10^{-4}$ &  $(16.1\pm 3.6)\times 10^{-4}$ & $-0.0005$ & $0.0021$ & $-0.0019$ & \cite{Baru:2020ywb}  \\
\hline
\end{tabular}
\caption{ \justifying For each bottomonium transition it shows the parameters $c_i$ with $i=1,2$ from \cite{Chen:2016mjn} and \cite{Baru:2020ywb} and the values of $c_\pi$, $c_{\pi\pi}$ and $c_E$ obtained using \eqref{eq:syst} and \eqref{eq:<r>}. }
\label{t:ci}
\end{table*}

\begin{table*}[t]
\centering
\begin{tabular}{ |c|c|c| } 
\hline
 \multicolumn{3}{|c|}{$\Gamma_i$ (keV)} \\
\hline
 Transitions + $\pi^+\pi^-$   & Using $\{\,c_1,\;  c_2\,\}$ and \eqref{eq:dipion_c1c2} & Using $\{\,c_\pi,\; c_{\pi\pi},\; c_E\,\}$ and \eqref{eq:dipion_q} \\
 \hline
$\Upsilon(4s)\to \Upsilon(2s) $ & 152.957 & 232.616 \\
\hline
$\Upsilon(5s)\to \Upsilon(2s) $ & 3.254 & 3.684 \\
\hline
$\Upsilon(5s)\to \Upsilon(3s) $ & 0.0187 & 0.0182  \\
\hline
\end{tabular}
\caption{\justifying Comparison of the predictions of the dipion decay width for bottomonium transitions using $c_i$ ($i=1,2$) values in Table~\ref{t:ci} and Eq.~\eqref{eq:dipion_c1c2}; and using the results obtained for $\{\,c_\pi,\; c_{\pi\pi},\; c_E\,\}$ in Table~\ref{t:ci} with Eq.~\eqref{eq:dipion_q}.}
\label{t:dw_ci}
\end{table*}

\bibliographystyle{ieeetr}
\bibliography{bibliography.bib}

\end{document}